\documentclass[11pt,a4paper]{article}
\usepackage{epsfig}
\usepackage[usenames,dvipsnames]{color}
\usepackage{amssymb}
\usepackage[colorlinks=true,urlcolor=blue]{hyperref}
\usepackage[comma,square,numbers,sort&compress]{natbib}
\usepackage{hyperref}
\usepackage{lineno}
\usepackage{xspace}
\usepackage{authblk}

\graphicspath {{fig/}} 

\def\be{\begin{equation}}
\def\ee{\end{equation}}
\newcommand{\ud}{\mathrm{d}}
\newcommand{\eg}{{\it e.g.}}
\newcommand{\ie}{{\it i.e.}}

\def\bea#1\eea{\begin{align}#1\end{align}}

\newcommand{\nn}{\nonumber}

\newcommand{\bef}{\begin{figure}[hbt]\centering}
\newcommand{\eef}{\end{figure}}

\newcommand{\vc}{\bf}
\newcommand{\f}{\frac}
\newcommand{\eps}{\epsilon}
\def \be  {\begin{equation}}
\def \ee  {\end{equation}}
\def \ba  {\begin{eqnarray}}
\def \ea  {\end{eqnarray}}
\def\mD{m_{\mathrm{D}}}

\newcommand{\bk}{\mathbf{k}}
\newcommand{\bq}{\mathbf{q}}

\begin{document}

\title{Extraction of Heavy-Flavor Transport Coefficients in QCD Matter}

\author[1]{R.~Rapp\thanks{Editor}}
\author[2]{P.B.~Gossiaux$^*$}
\author[3,4]{A.~Andronic$^*$}
\author[3]{R.~Averbeck$^*$}
\author[3]{S.~Masciocchi$^*$}
\author[5]{A.~Beraudo}
\author[3,6]{E.~Bratkovskaya}
\author[3,7]{P.~Braun-Munzinger}
\author[8]{S.~Cao}
\author[9]{A.~Dainese}
\author[10,11]{S.K.~Das}
\author[12]{M.~Djordjevic}
\author[11,13]{V.~Greco}
\author[14]{M.~He}
\author[6]{H.~van Hees}
\author[3,6,15,16]{G.~Inghirami}
\author[17,18]{O.~Kaczmarek}
\author[19]{Y.-J.~Lee}
\author[20]{J.~Liao}
\author[1]{S.Y.F.~Liu}
\author[21]{G.~Moore}
\author[2]{M.~Nahrgang}
\author[22]{J.~Pawlowski}
\author[23]{P.~Petreczky}
\author[11]{S.~Plumari}
\author[5]{F.~Prino}
\author[20]{S.~Shi} 
\author[24]{T.~Song}
\author[7]{J.~Stachel}
\author[25]{I.~Vitev}
\author[26,18]{X.-N.~Wang}

\affil[1]{Cyclotron Institute and Department of Physics and Astronomy, Texas A\&M University, College Station, USA}
\affil[2]{SUBATECH, IMT Atlantique, Universit\'e de Nantes, CNRS-IN2P3, Nantes, France}
\affil[3]{Research Division and EMMI, GSI Helmholtzzentrum f\"ur Schwerionenforschung, Darmstadt, Germany}
\affil[4]{Institut f\"ur Kernphysik, Westf\"alische Wilhelms-Universit\"at M\"unster, M\"unster, Germany}
\affil[5]{INFN, Sezione di Torino, Torino, Italy}
\affil[6]{Institut f\"ur Theoretische Physik, J.W.~Goethe-Universit\"at, Frankfurt am Main, Germany}
\affil[7]{Physikalisches Institut, Ruprecht-Karls-Universit\"at Heidelberg, Heidelberg, Germany}
\affil[8]{Department of Physics and Astronomy, Wayne State University, Detroit, USA}
\affil[9]{INFN, Sezione di Padova, Padova, Italy}
\affil[10]{School of Nuclear Science and Technology, Lanzhou University, Lanzhou, China}
\affil[11]{Department of Physics and Astronomy, University of Catania, Italy}
\affil[12]{Institute of Physics, University of Belgrade, Belgrade, Serbia}
\affil[13]{Laboratori Nazionali del Sud, INFN-LNS, Catania, Italy}
\affil[14]{Department of Applied Physics, Nanjing University of Science and Technology, Nanjing, China}
\affil[15]{Frankfurt Institute for Advanced Studies, Frankfurt am Main, Germany}
\affil[16]{J.~von~Neumann Institute for Computing, J\"ulich, Germany}
\affil[17]{Fakult\"at f\"ur Physik, Universit\"at Bielefeld, Bielefeld, Germany}
\affil[18]{Key Laboratory of Quark and Lepton Physics and IPP, Central China Normal University, Wuhan, China}
\affil[19]{Department of Physics, Massachusetts Institute of Technology, Cambridge, USA}
\affil[20]{Department of Physics, Indiana University, Bloomington, USA}
\affil[21]{Technische Universit\"at Darmstadt, Darmstadt, Germany}
\affil[22]{Institut f\"ur Theoretische Physik, Ruprecht-Karls-Universit\"at Heidelberg, Heidelberg, Germany}
\affil[23]{Physics Department, Brookhaven National Laboratory, Upton, USA}
\affil[24]{Institut f\"{u}r Theoretische Physik, Universit\"{a}t Gie\ss en, Gie\ss en, Germany}
\affil[25]{Theoretical Division, Los Alamos National Laboratory, Los Alamos, USA}
\affil[26]{Nuclear Science Division, Lawrence Berkeley National Laboratory, Berkeley, USA}

\date{\today}

\maketitle

\newpage

\begin{abstract}
We report on broadly based systematic investigations of the modeling components for open heavy-flavor 
diffusion and energy loss in strongly interacting matter in their application to heavy-flavor 
observables in high-energy heavy-ion collisions, conducted within an EMMI Rapid Reaction Task Force 
framework. Initial spectra including cold-nuclear-matter 
effects, a wide variety of space-time evolution models, heavy-flavor transport coefficients, 
and hadronization mechanisms are scrutinized in an effort to quantify pertinent uncertainties 
in the calculations of nuclear modification factors and elliptic flow of open 
heavy-flavor particles in nuclear collisions. We develop procedures for error assessments and
criteria for common model components to improve quantitative estimates for the (low-momentum) 
heavy-flavor diffusion coefficient as a long-wavelength characteristic of QCD matter as a 
function of temperature, and for energy loss coefficients of high-momentum heavy-flavor 
particles.
\end{abstract}

\newpage

{\bf Task Force Formation and Acknowledgment}: \\
The article presents calculations done in the context of the Heavy-Flavor EMMI RRTF effort, 
as agreed among the participants. The publication is signed by the people who actively 
participated in the RRTF meetings and by those who contributed explicitly to the follow-up 
activities leading to this article. We acknowledge that the results presented here would not 
have been possible without the work by all authors of the models and computation approaches 
utilized in this comparison. Therefore we explicitly thank all colleagues who contributed to 
the creation of the different models, independently from the RRTF effort; these are, in
alphabetical order of the model acronyms:
Catania: F.~Scardina; CUJET: M.~Gyulassy and J.~Xu; Duke: S.A.~Bass and Y.~Xu; 
LBL/CCNU: G.-Y.~Qin and T.~Luo; 
Nantes: J.~Aichelin and K.~Werner; PHSD: H.~Berrehrah, D.~Cabrera, W.~Cassing, L.~Tolos 
and J.M.~Torres-Rincon; 
POWLANG: A.~De Pace, M.~Monteno and M.~Nardi;
TAMU: R.J.~Fries, K.~Huggins and F.~Riek; UrQMD: M.~Bleicher, T.~Lang and J.~Steinheimer. \\
We also thank Z.~Conesa del Valle for providing the software used to compare $D$ meson cross 
sections with FONLL, G.E.~Bruno for providing the calculation of the $c$- quark 
$R_{AA}$ with FONLL+EPPS16, and C.~Shen for information on the OSU hydro code; we also thank 
E.~Bruna, C.~Greiner, and J.~Bielcik for discussions in the early stages of the task force
activities.

\newpage

\tableofcontents

\newpage

\section{Introduction} 
\label{sec_intro}
The characterization of the properties of matter can be carried out at various levels,
utilizing different ways of testing its response to external excitations.
Bulk properties are encoded in the equation of state, $\epsilon(P)$, which characterizes 
how a system responds to changes in its pressure. Transport coefficients characterize how 
small perturbations from equilibrium, often associated with conserved quantities, are 
transmitted through the medium. 
In quantum field theory, transport coefficients can be formulated as the zero-energy 
and long-wavelength limit of correlation functions. This, in particular, allows to 
establish connections between microscopic calculations of spectral (or correlation) 
functions and their underlying transport coefficients.   
 
High-energy collisions of atomic nuclei have revealed remarkable properties of strongly
interacting matter at high temperature. For example, the ratio of shear viscosity to the 
entropy density, $\eta/s$, of the medium has been inferred to be the smallest of any known 
substance~\cite{Akiba:2015jwa}. However, the extraction of this quantity, including its 
temperature dependence,
from fitting viscous hydrodynamic simulations of the fireball to final-state hadron spectra, 
is rather indirect, involving the entire system evolution. Progress has been made in 
controlling basic features of the fireball evolution~\cite{Niemi:2015qia}, but significant 
uncertainties persist, \eg, in the initial conditions and pre-equilibrium evolution
of the quark-gluon plasma (QGP). 
Furthermore, the microscopic origin of the small $\eta/s$, \ie, how it emerges from the 
fundamental interactions of Quantum Chromodynamics (QCD) in the medium, remains a central 
question that calls for additional observables and methods. 
In particular, non-monotonic features of transport coefficients in the vicinity of the 
(pseudo-) critical transition temperature are of key relevance to our understanding of
the phase structure of QCD.

The diffusion of heavy quarks in QCD matter at not too high temperatures has long been 
recognized as a promising concept and phenomenological tool to diagnose the medium 
produced in heavy-ion collisions (HICs)~\cite{Svetitsky:1987gq}. The basic realization is 
that the heavy-quark (HQ) mass, $m_Q$, is parametrically large compared to the scales that 
characterize the QCD medium produced in experiment, \ie, its typical temperature, 
including the pseudo-critical transition temperature, $T_{\rm pc}$, which is ultimately
related to the QCD scale parameter, $\Lambda_{\rm QCD}$ and, to a lesser extent, to the
precise values of the light-quark masses. 
This realization entails a sequence of benefits, both phenomenologically and theoretically, 
for using heavy-flavor (HF) particles as a probe of the medium, namely that 
(a) the production of heavy quarks is reasonably well controlled as a hard initial-state 
process, 
(b) the propagation of HF particles through the medium is, at low momenta, of a diffusive 
``Brownian motion" type and thus characterized by well-defined transport coefficients, 
most notably the spatial diffusion coefficient ${\cal D}_s$,  
(c) heavy quarks can remain good quasi-particles (\ie, their collisional width is much
smaller than their energy, $\Gamma_Q \ll E_Q$) in a QGP with large interaction strength
where light partons are already dissolved (\ie, their widths are comparable or larger then
their energies, $\Gamma_q \gtrsim E_q$);
(d) the `'identity" of heavy quarks is preserved in the hadronization process thus 
providing tests of it's microscopic mechanisms, 
(e) interactions of low-momentum heavy quarks with the medium are of potential-type, 
\ie, elastic collisions with small energy transfer.   
Furthermore, the thermalization time of heavy quarks is delayed relative to the light 
partons of the bulk medium, parametrically by a factor of order $\sim$$M_Q/T$, which 
renders it comparable to the lifetime of the QGP fireballs in HICs. Thus, HF particles 
are not expected to fully thermalize and therefore preserve a memory of their interaction 
history which can serve as a gauge of their interaction strength with the medium. The 
HF diffusion coefficient, ${\cal D}_s$, arguably provides the most direct 
window on the in-medium QCD force in HICs, and thus on the coupling strength of the medium. 
To the extent that the same in-medium interactions are operative in the transport of 
different quantities, \eg, energy-momentum or electric charge, one expects the pertinent 
transport coefficients, scaled to dimensionless units, to relate to each other, \eg, 
$\eta/s \sim {\cal D}_s (2\pi T) \sim \sigma_{\rm EM}/T$ \ (where $\sigma_{\rm EM}$ denotes
the electric conductivity).    

At high momentum HQ production in heavy-ion collisions can be understood as a part 
of the in-medium parton shower evolution. The large HQ mass affects the splitting 
functions which encode the many-body physics of parton branching in the QGP, including
the suppression of forward radiation and interference effects. The HQ soft-gluon emission 
energy loss limit, extensively used in jet quenching phenomenology, 
is connected to the general high-energy parton shower picture. The scale separation 
$T\ll M_Q\ll E$ allows for the use of heavy quarks as independent probes of, \eg,  
the Debye screening mass ($m_D$) or the parton mean-free-path ($\lambda$), encoded in 
transport parameters such as $\hat{q} \sim m_D^2/\lambda$.

Over the last decade, open HF observables, \ie, transverse-momentum ($p_T$) spectra and 
elliptic flow ($v_2$) of particles containing a single charm ($c$) or bottom ($b$) quark 
(or their decay products), have much advanced and are now at the verge of becoming a 
precision probe of QCD matter. This has triggered intense theoretical activity aimed at 
understanding the intriguing experimental results on how HF spectra are modified when 
going from elementary proton-proton collisions to reactions with heavy nuclei, see, \eg, 
Refs.~\cite{Beraudo:2014iva,Andronic:2015wma,Prino:2016cni} for recent reviews. 
The modeling efforts have reached a critical stage. Several approaches have accomplished 
a qualitative or semi-quantitative agreement with (some of the) existing data, but it 
seems fair to say that no single approach is yet able to quantitatively describe all 
available HF measurements from the Relativistic Heavy-Ion Collider (RHIC) and the Large 
Hadron Collider (LHC), from low to high $p_T$. 
At the same time, a fundamental understanding of the underlying processes employed in the 
phenomenological modeling efforts requires the latter to be firmly rooted in QCD. The 
complexity of the problem likely involves different prevalent mechanisms not only as a 
function of $p_T$ (\eg, collisional vs. radiative and/or perturbative vs. non-perturbative 
interactions), but also as a function of temperature. In addition, the accuracy of the 
information extracted from HF observables also hinges on a realistic space-time evolution 
for the bulk medium, \eg, hydrodynamic or transport models, as well as initial conditions 
for the HQ spectra, presumably modified by ``cold-nuclear-matter" (CNM) effects in the 
incoming nuclei prior to the collision. Indeed, the individual models currently in use 
employ rather different ingredients for each of the modeling components, including a 
varying degree of fit parameters (\eg, $K$ factors for the transport coefficients). At 
this point, it becomes compelling to go beyond incremental improvements of individual 
approaches and launch a broader effort supported by active researchers in the field. 
For this purpose an EMMI Rapid Reaction Task Force was initiated, approved and formed,
and two onsite meetings convened at GSI Darmstadt (Germany) in July and December of 
2016~\cite{rrtf:2016}.   
The key open questions and objectives in the open HF problem that were identified and
addressed during the meeting are:
\begin{itemize} 
\item[1)] 
How do the conceptual underpinnings of the current theoretical models compare and constrain 
their applicability in various regions of phase space and temperature? Do these uncertainties 
provide a sufficiently robust basis for systematic uncertainty evaluations of the 
extracted transport coefficients? Can quantitative connections to jet quenching in 
the light-flavor sector be made?
\item[2)] 
What is the impact of the available implementations of hadronization, in particular HQ 
coalescence, on $D$-meson spectra, and how can they be seamlessly connected to the QGP and 
hadronic diffusion processes?
\item[3)] 
What are the benefits and limitations for Boltzmann vs. Langevin implementations of the 
HF transport in an evolving medium?
\item[4)] 
What is the role of the different medium evolution models, and how do different predictions 
for the temperature- and momentum-dependent transport coefficients in current model 
calculations manifest themselves in observables?
\item[5)] 
What are future precision requirements on existing observables, and are there other ways to 
analyze data (new observables), to improve current accuracies, and to what extent? In 
particular, in what ways are the upcoming data from RHIC and the LHC instrumental in 
extending our knowledge of deconfined matter?
\end{itemize}

The present effort is a first step in these directions by scrutinizing the different
components in the modeling efforts of various research groups and performing targeted 
calculations to unravel, and ultimately quantify, how pertinent uncertainties impact  
the extraction of HF transport coefficients. 
This document is organized as follows. 
In Sec.~\ref{sec_initial}, we start by developing a common baseline for the initial HQ 
and $D$-meson transverse-momentum spectra from $pp$ collisions, explore uncertainties in
the implementation of CNM effects and suggest a standardized approach to account for them. 
In Sec.~\ref{sec_bulk} we investigate the role of bulk evolution models, by evaluating 
the outcomes for the nuclear modification factor and elliptic flow of charm quarks 
in Pb-Pb(2.76\,TeV) collisions by using a common pre-defined HQ interaction in the QGP 
within various hydrodynamic and transport simulations as used in current phenomenological
applications. 
In Sec.~\ref{sec_hadro} we study the differences in the treatment of HQ hadronization, 
in particular different schemes of HQ recombination with light partons in the quark-hadron
transition of the bulk medium, as well as fragmentation mechanisms.  
In Sec.~\ref{sec_trans} we focus on the interactions of heavy quarks at low and 
intermediate momenta in the QGP and their transport implementation; transport coefficients 
as used by different groups are scrutinized, a publicly accessible repository of transport 
coefficients~\cite{emmi-repo} is provided, the outcome of different charm-quark transport 
coefficients from Langevin simulations in a common hydrodynamic evolution is studied, 
insights from perturbative QCD (pQCD) and lattice QCD (lQCD) are discussed and put into context, 
and two common schemes (Boltzmann and Fokker-Planck Langevin) for implementing HF transport 
into bulk evolution models are compared.   
In Sec.~\ref{sec_eloss} the transport of high-momentum heavy quarks is discussed, 
starting with basic definitions of momentum broadening and the pertinent transport coefficient 
($\hat{q}$), its dependence on jet energy, dynamical effects in energy loss, next-to-leading 
order calculations and the problem of non-locality in radiative processes. 
A summary with an outlook for future developments is given in Sec.~\ref{sec_sum}. 

Concerning notation, we will use $Q=c,b$ as a generic for the heavy quarks charm and bottom,
$q=u,d,s$ for light quarks, $p_t$ for quark and $p_T$ for hadron transverse-momenta. A listing
of the model approaches involved in the studies reported in the following is given in 
App.~\ref{app_models}.

\section{Initial Heavy-Flavor Spectra}
\label{sec_initial}

\subsection{Baseline for $p_t$ distributions of $c$ quarks and fragmentation 
functions to $D$ mesons}
\label{ssec_baseline}

In this section we describe the construction of a common baseline for the initial 
$p_t$-differential cross sections of $c$ quarks which are required as an input to simulations 
for their transport through a QGP formed in ultrarelativistic heavy-ion collisions (we will use
the notation $p_t$ and $p_T$ for quark and hadron transverse momenta, respectively). The initial 
$p_t$ spectra will be based on FONLL calculations~\cite{Cacciari:1998it,Cacciari:2012ny} and
will be supplemented by fragmentation functions to $D$ mesons based on the BCFY 
framework~\cite{Braaten:1994bz}, also used by the FONLL authors, to enable quantitative
constraints from experimental measurements in proton-proton ($pp$) collisions and also 
to serve as a baseline
for hadronization (at least at high $p_t$) in the heavy-ion environment.
A common baseline for the input to transport calculations will reduce uncertainties in the 
comparison of results obtained with various transport for heavy-ion observables such as the
nuclear modification factor, $R_{AA}$, and elliptic flow, $v_2$, of heavy-flavor (HF) 
particles (or their decay products). 
Indeed, different shapes of the initial $p_t$ distribution lead to different values of 
$R_{AA}$ and $v_2$ for the same QGP parameters (energy loss or diffusion coefficients), as
will be discussed in Sec.~\ref{ssec_pp-uncert}. This can be easily seen for the simple case of 
a power-law $p_t$ distribution $\propto 1/p_t^n$ and constant energy loss $\Delta E$, 
giving a quark-level $R_{AA}\sim p_t^n/(p_t+\Delta E)^n$, which clearly depends on $n$. 
In the same way, different fragmentation functions can lead to different values for the 
hadron-level $R_{AA}(p_T)$ for the same quark-level $R_{AA}(p_t)$. Similar arguments apply 
for $v_2$.

Fixed-Order Next-to-Leading-Log (FONLL) calculations are widely used to obtain the initial 
heavy-quark (HQ) $p_t$-differential cross sections that serve as an input for the transport and 
energy loss models. This is a perturbative-QCD (pQCD) calculation in which the HQ production 
cross section, also denoted as partonic cross section $\hat\sigma$,  is obtained through an 
expansion in powers of the strong coupling constant, $\alpha_s$. In particular, 
in FONLL the partonic cross section is calculated to next-to-leading order (NLO) in 
$\alpha_{\rm s}$ (with the terms proportional to $\alpha_s^2$ and $\alpha_s^3$) and  
with an all-order resummation of logarithms of $p_t/m_Q$, where $p_t$ and $m_Q$ are the HQ 
transverse momentum and mass, respectively. 
The HF hadron cross section is factorized as a convolution of the parton distribution 
functions (PDFs) of the incoming partons, of the partonic cross sections 
and of a nonperturbative fragmentation function that encodes the probability for a heavy 
quark with momentum $p$ to fragment into a HF hadron with 
momentum $z\cdot p$, with $0<z<1$. FONLL uses the collinear factorization scheme, in which 
the factorization variable is related to the squared momentum transfer in the hard process, 
$Q^2$. In particular, the PDFs, the partonic cross section and the fragmentation function 
are evaluated at the same scale $\mu_F$ (factorization scale), 
which is taken to be proportional to the transverse mass of the produced heavy quark 
$\mu_0=\sqrt{m_Q^2+p_t^2}$. The strong coupling constant $\alpha_s$ is evaluated at the 
renormalization scale, $\mu_R$, also taken to be proportional to $\mu_0$. The central values 
of the perturbative scales are taken as $\mu_F=\mu_R=\mu_0$.
The uncertainties of the $c$-quark cross section are estimated using three values of 
$m_c$,  1.3, 1.5 and 1.7~GeV$/c^2$, and, for the central value of $m_c$, seven sets of 
values of $\mu_F$ and $\mu_R$ defined by 
$(\mu_F/\mu_0,\mu_R/\mu_0)=(1,1)$, (0.5,0.5), (0.5,1), (1,0.5), (2,2), (1,2), (2,1). 
The bands defined by the envelope of the minimum and maximum cross sections obtained from 
the mass variation and from the scale variations are summed in quadrature. In addition, 
the envelope obtained by varying the PDFs within their uncertainties is also added in 
quadrature. The PDF set used in recent FONLL calculations is CTEQ6.6~\cite{Pumplin:2002vw}.

\begin{figure}[t]
\begin{center}
\includegraphics[width=0.51\textwidth]{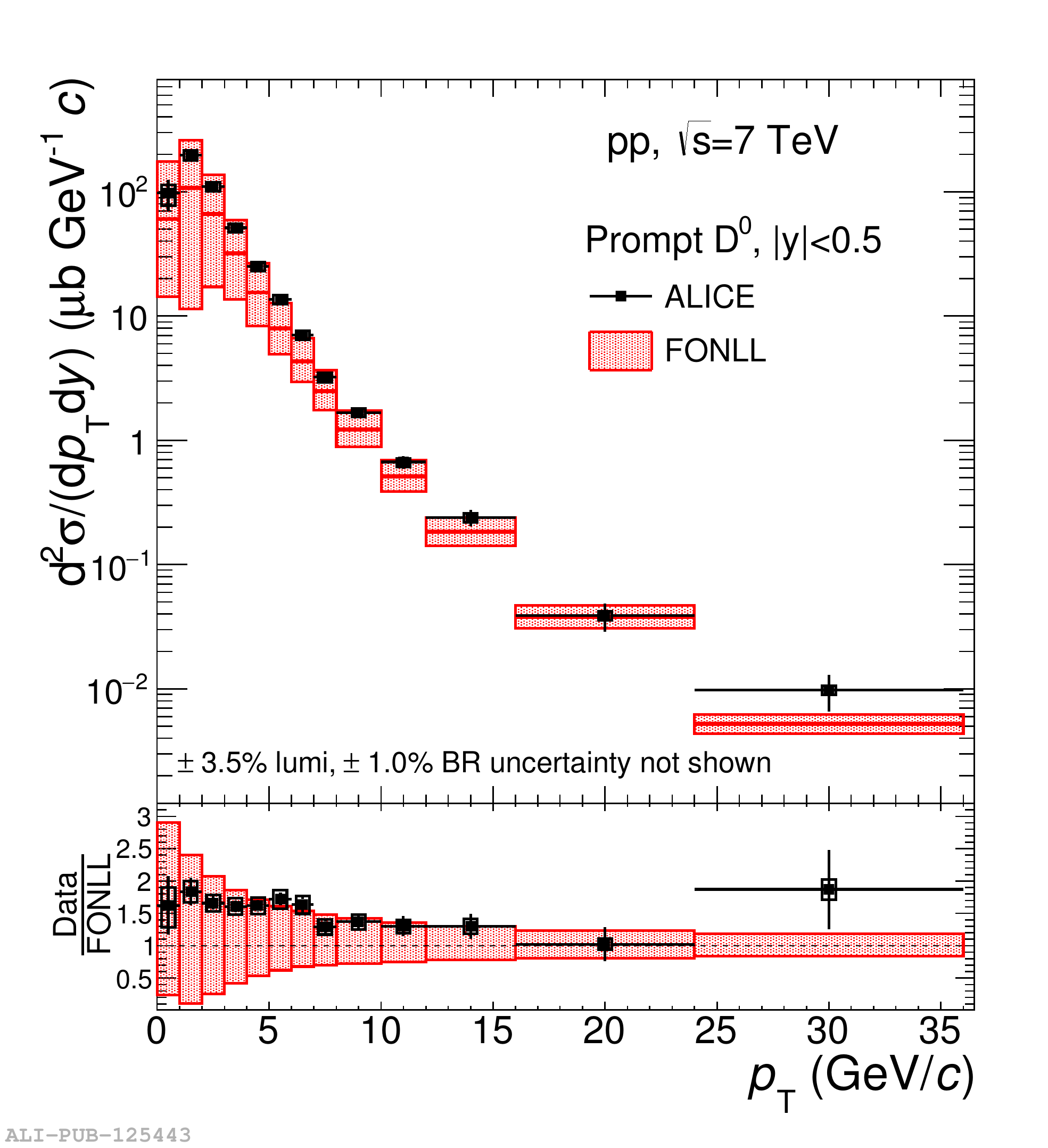}
\includegraphics[width=0.48\textwidth]{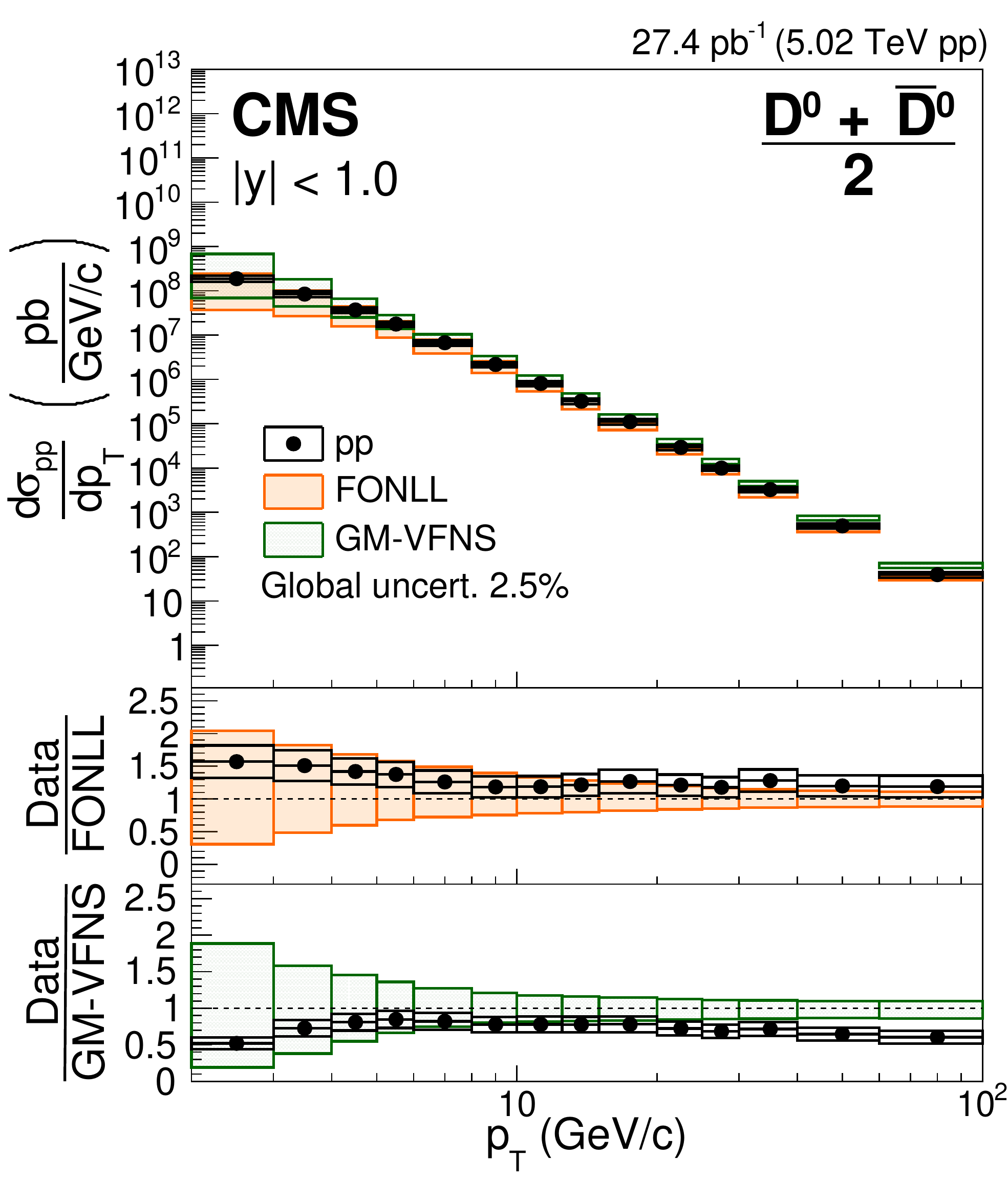}
\caption{Cross sections for $D^0$ meson production at central rapidity in $pp$ collisions at 
$\sqrt s=7$ and 5.02~TeV, measured by ALICE~\cite{Acharya:2017jgo} and CMS~\cite{Sirunyan:2017xss}, 
respectively, compared with the FONLL and GM-VFNS calculations.}
\label{fig:ALICECMSD0}
\end{center}
\end{figure}

The FONLL calculation provides a good description of the production cross sections of 
$D$ and $B$ mesons in $pp$ (and $p\bar p$) collisions at center-of-mass energies from 
0.2 to 13~TeV over a wide $p_T$ range at both central and forward rapidities 
(see, \eg Ref.~\cite{Andronic:2015wma} and references therein). In the case of $D$-meson 
production, two general observations can be made, which are clearly illustrated by the 
$D^0$-meson cross sections at $\sqrt s=7~$TeV and 5.02~TeV measured by 
ALICE~\cite{Acharya:2017jgo} and CMS~\cite{Sirunyan:2017xss}, respectively, 
and shown in Fig.~\ref{fig:ALICECMSD0}:
\begin{itemize}
\item the central value of the FONLL calculation ($m_c=1.5$~GeV$/c^2$, $\mu_F=\mu_R=\mu_0$) 
yields spectra that lie below the experimental data, and the ratio data/FONLL depends on $p_T$, 
increasing a bit towards low $p_T$; therefore, the central value of FONLL does not provide an 
optimal description of the shape of the $p_T$ distribution of $D$ mesons (we recall that it is 
the shape of the $p_T$ distribution that affects $R_{AA}$ and $v_2$); 
\item the uncertainties of the theoretical calculation, dominated by the perturbative scale 
setting, are significantly larger than the experimental 
ones~\cite{Adamczyk:2012af,Acosta:2003ax,Aad:2015zix,ALICE:2011aa,Abelev:2012vra,Abelev:2012tca,Adam:2016ich,Acharya:2017jgo,Sirunyan:2017xss,Aaij:2013mga,Aaij:2016jht,Aaij:2015bpa}, 
especially for transverse momenta smaller than 10-15~GeV/$c$; therefore, considering the 
total FONLL uncertainties is not the best approach. 
\end{itemize}

We therefore choose to define the baseline $p_t$ distribution using the FONLL cross section that 
best describes the shape of the $D^0$ cross sections at both energies, $\sqrt s=5$~TeV
and 7~TeV (shown in Fig.~\ref{fig:ALICECMSD0}), among the nine FONLL cross section obtained by 
the aforementioned seven scale sets and three $m_c$ values. The uncertainty introduced by the 
choice of the $p_T$ shape can be studied by using the two FONLL cross sections that span the 
maximum shape variation but are still consistent with the shape of the data.
We propose to use the cross sections with these three sets of parameters for all LHC energies, 
in particular $\sqrt{s_{NN}}=2.76$ and $5.02$~TeV for Pb--Pb collisions. We did not use $pp$ 
data at $\sqrt s=2.76$~TeV, because the available $D$-meson cross section measurements at this 
energy at central rapidity~\cite{Abelev:2012vra} have much larger experimental uncertainties 
than the data at the higher energies.

\begin{figure}[t]
\begin{center}
\includegraphics[width=0.32\textwidth]{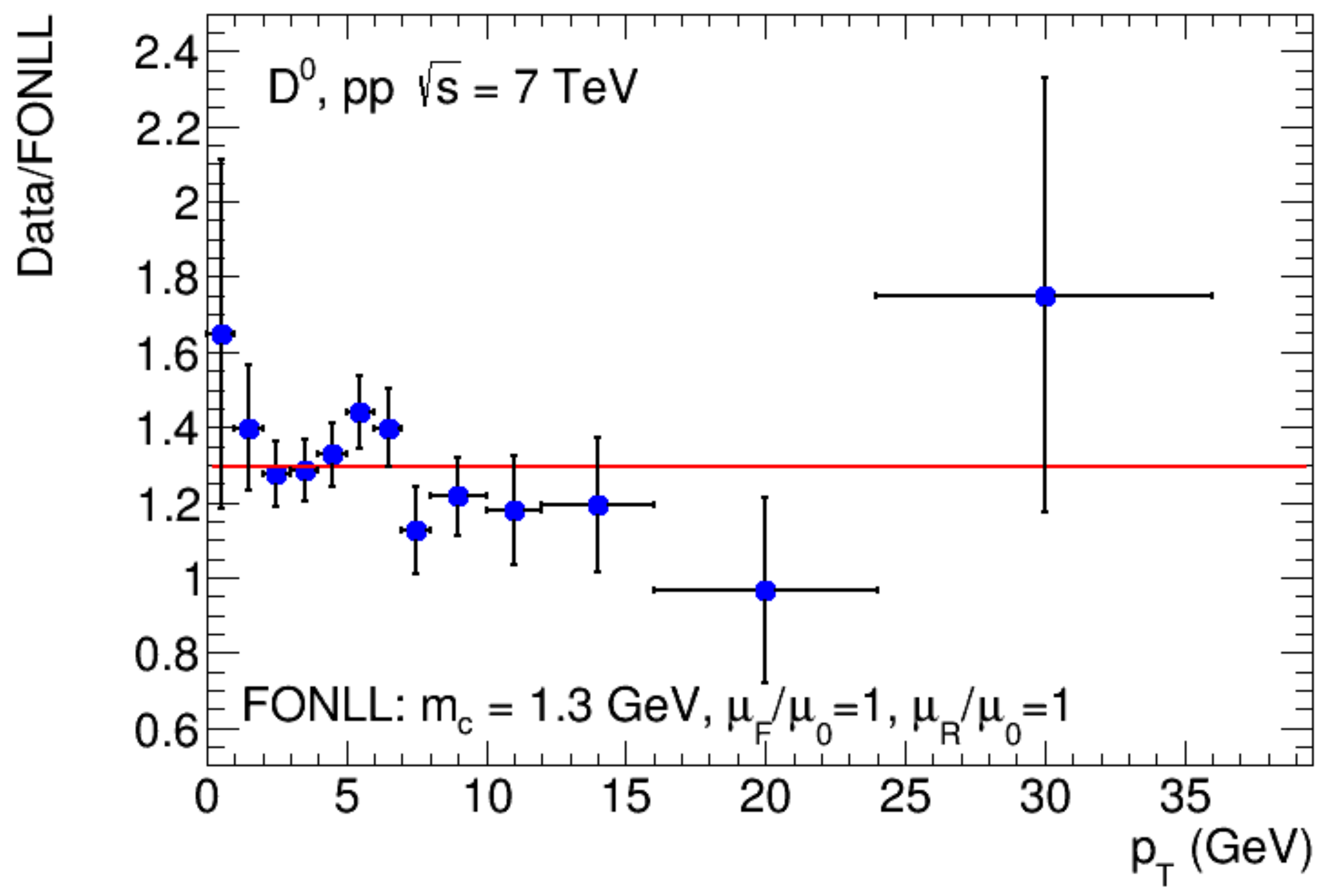}
\includegraphics[width=0.32\textwidth]{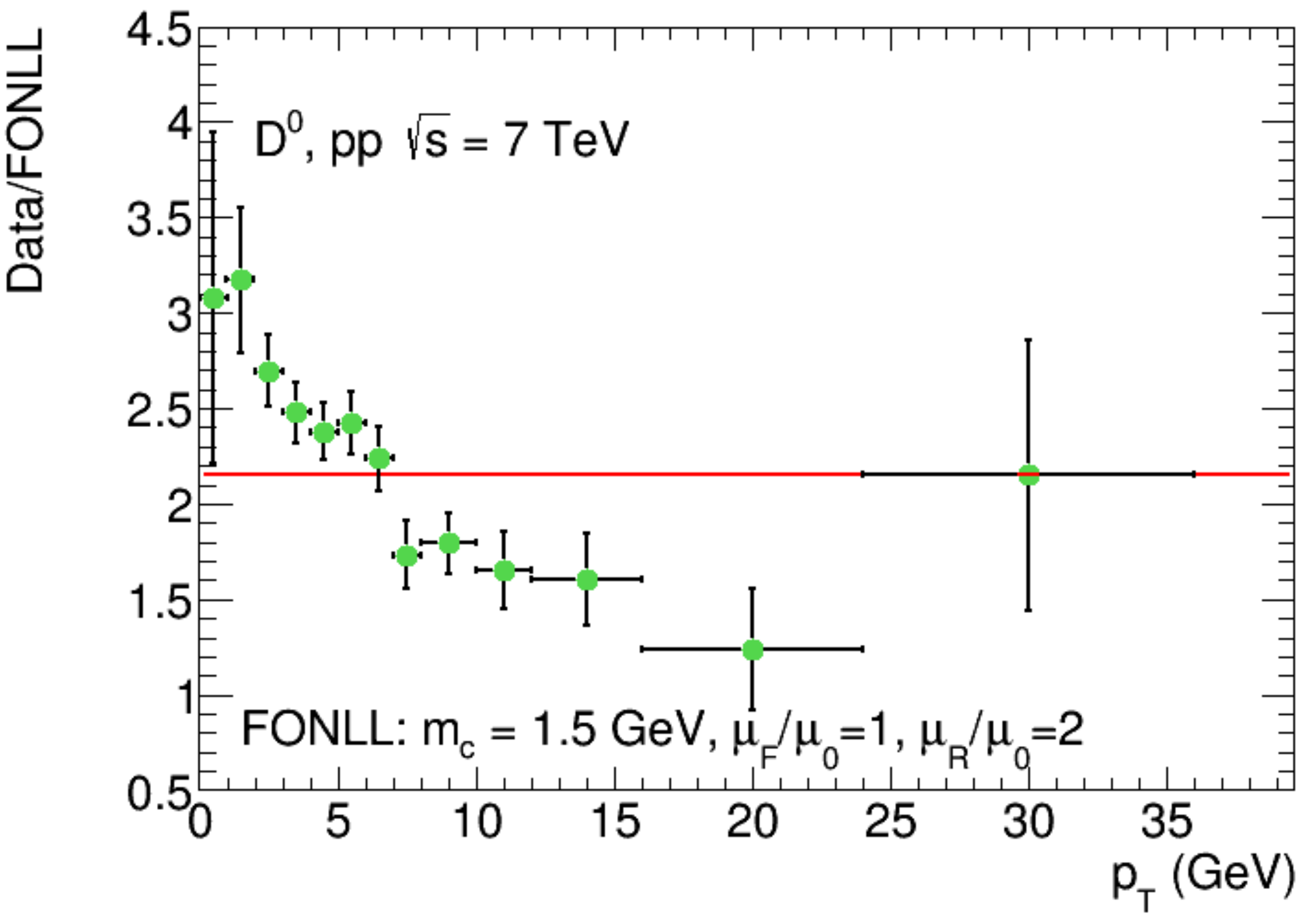}
\includegraphics[width=0.32\textwidth]{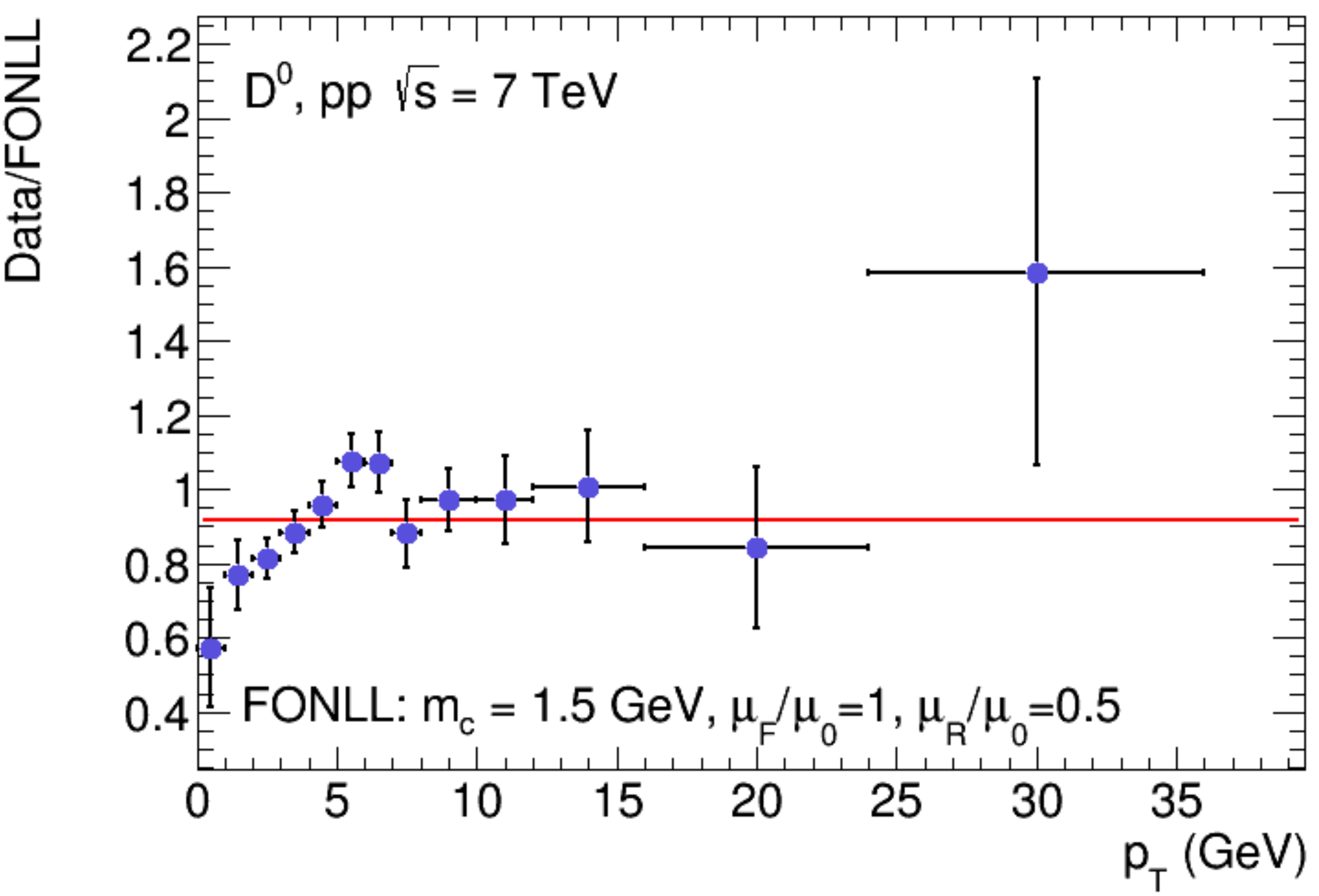}
\includegraphics[width=0.32\textwidth]{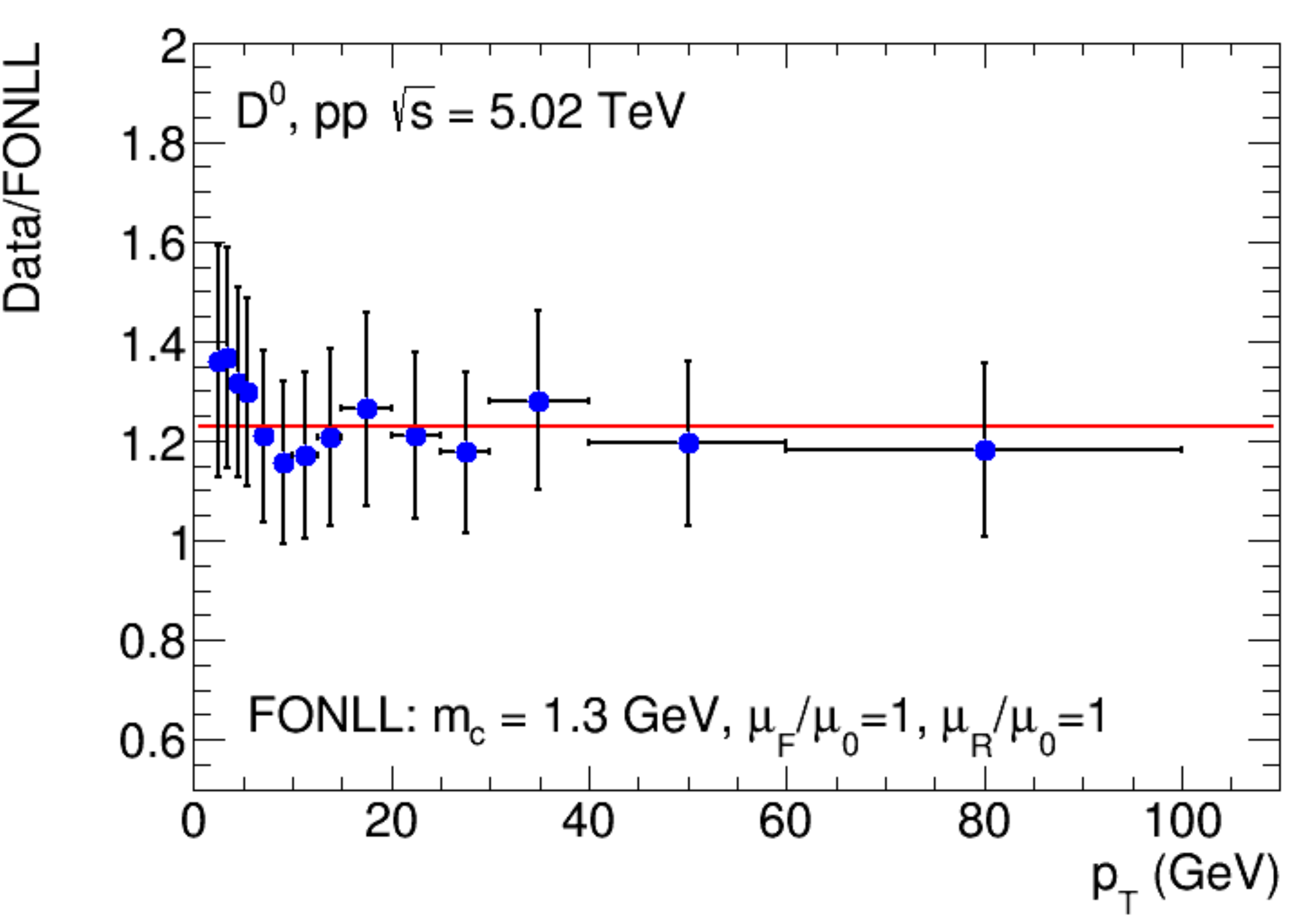}
\includegraphics[width=0.32\textwidth]{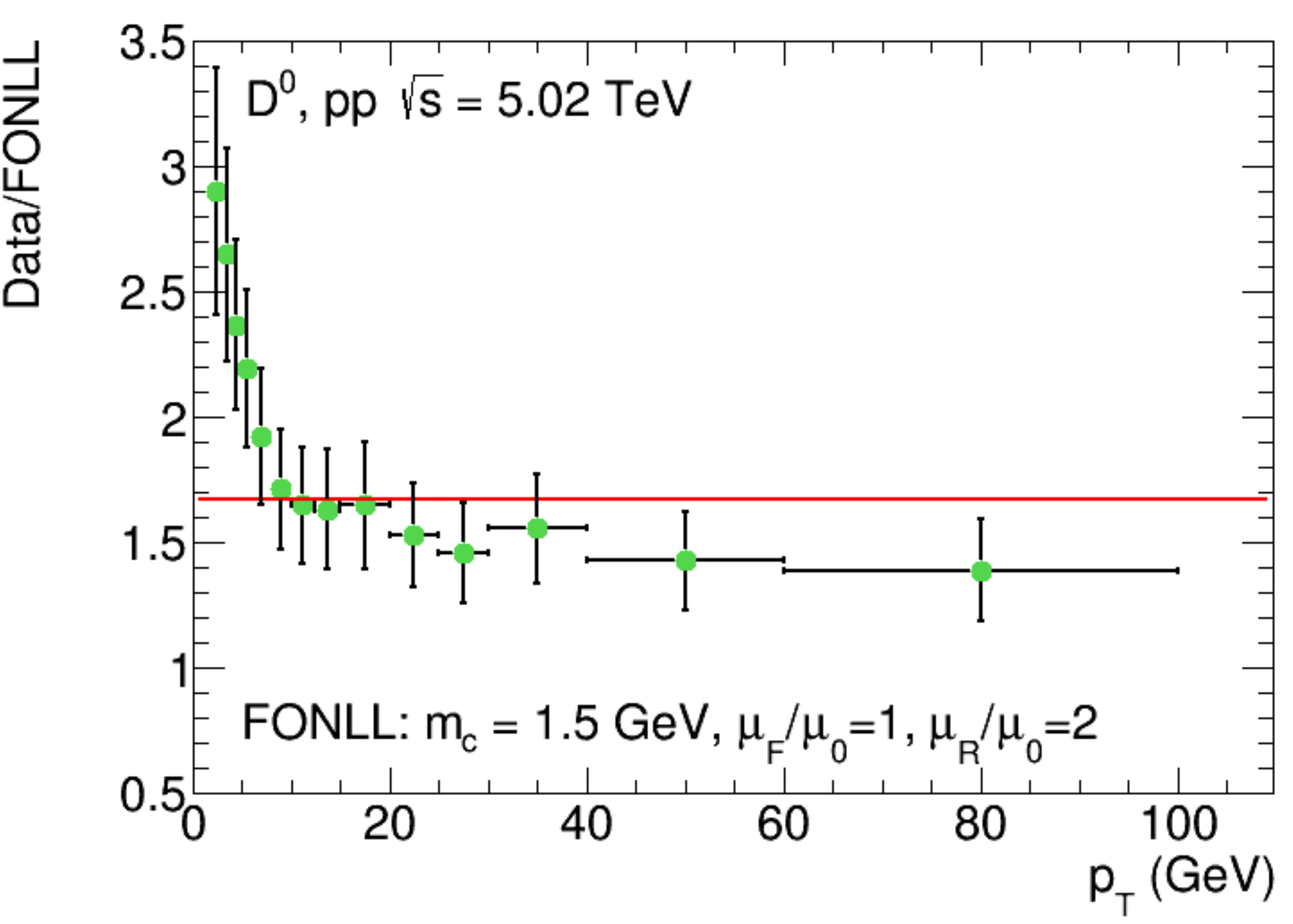}
\includegraphics[width=0.32\textwidth]{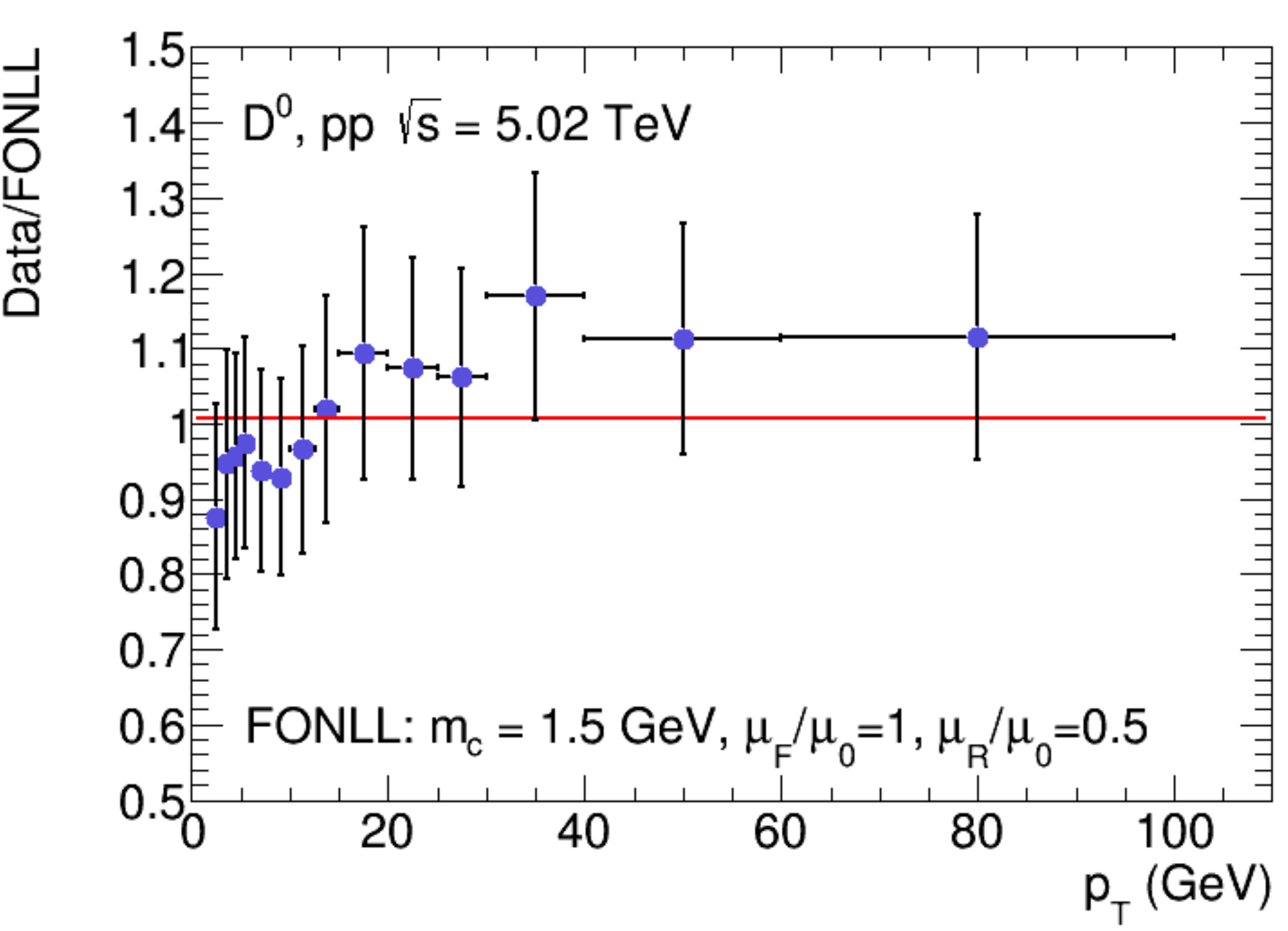}
\caption{Ratios of the cross sections for $D^0$ meson production at central rapidity in $pp$ 
collisions at $\sqrt s=7$ (upper rows) and 5.02~TeV (lower rows), measured by 
ALICE~\cite{Acharya:2017jgo} and CMS~\cite{Sirunyan:2017xss}, respectively, with FONLL cross 
sections for the best-fitting (left) and the two extreme parameter sets (central and right).}
\label{fig:datatoFONLL}
\end{center}
\end{figure}

\begin{figure}[!ht]
\begin{center}
\includegraphics[width=0.45\textwidth]{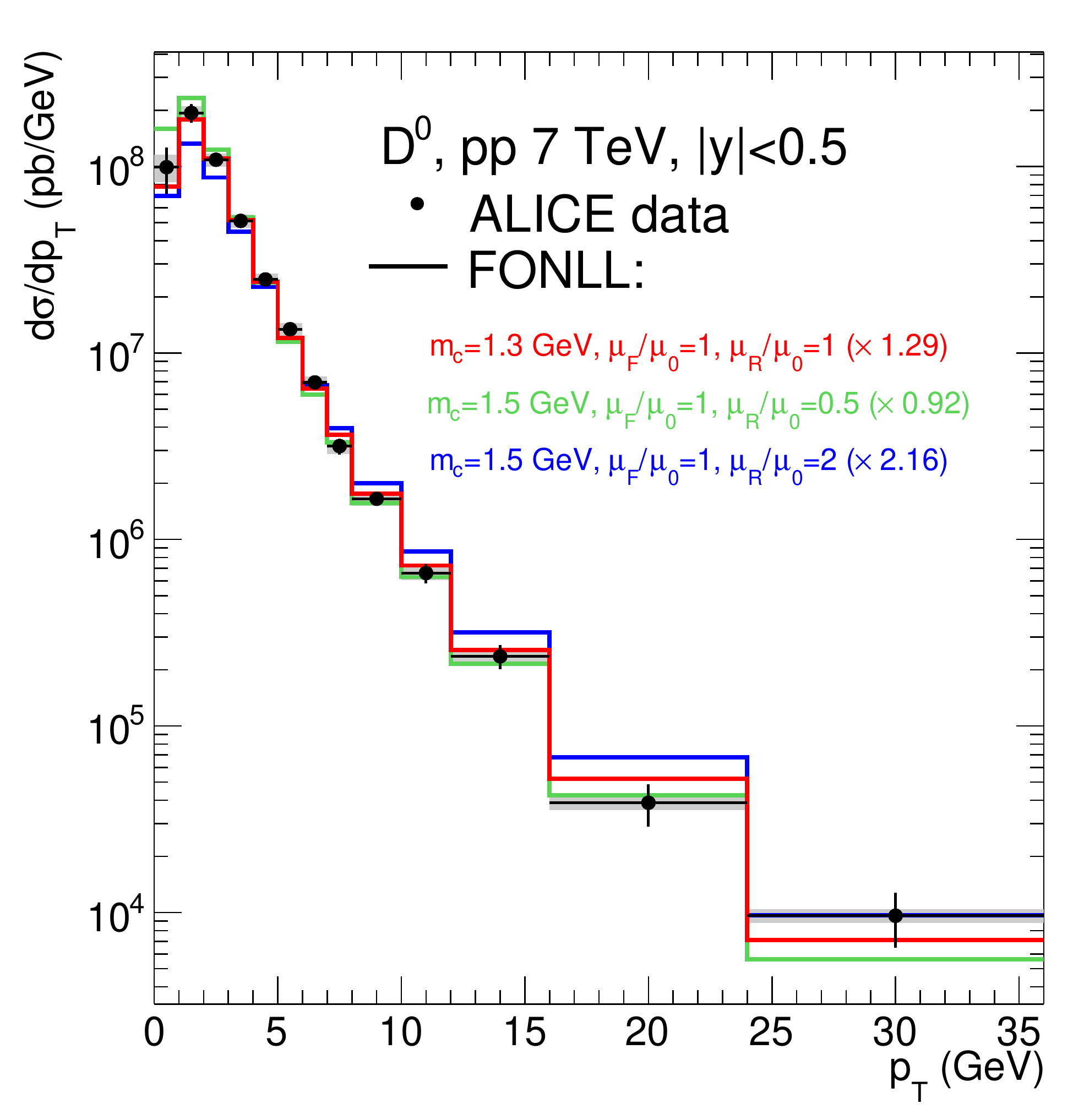}
\includegraphics[width=0.45\textwidth]{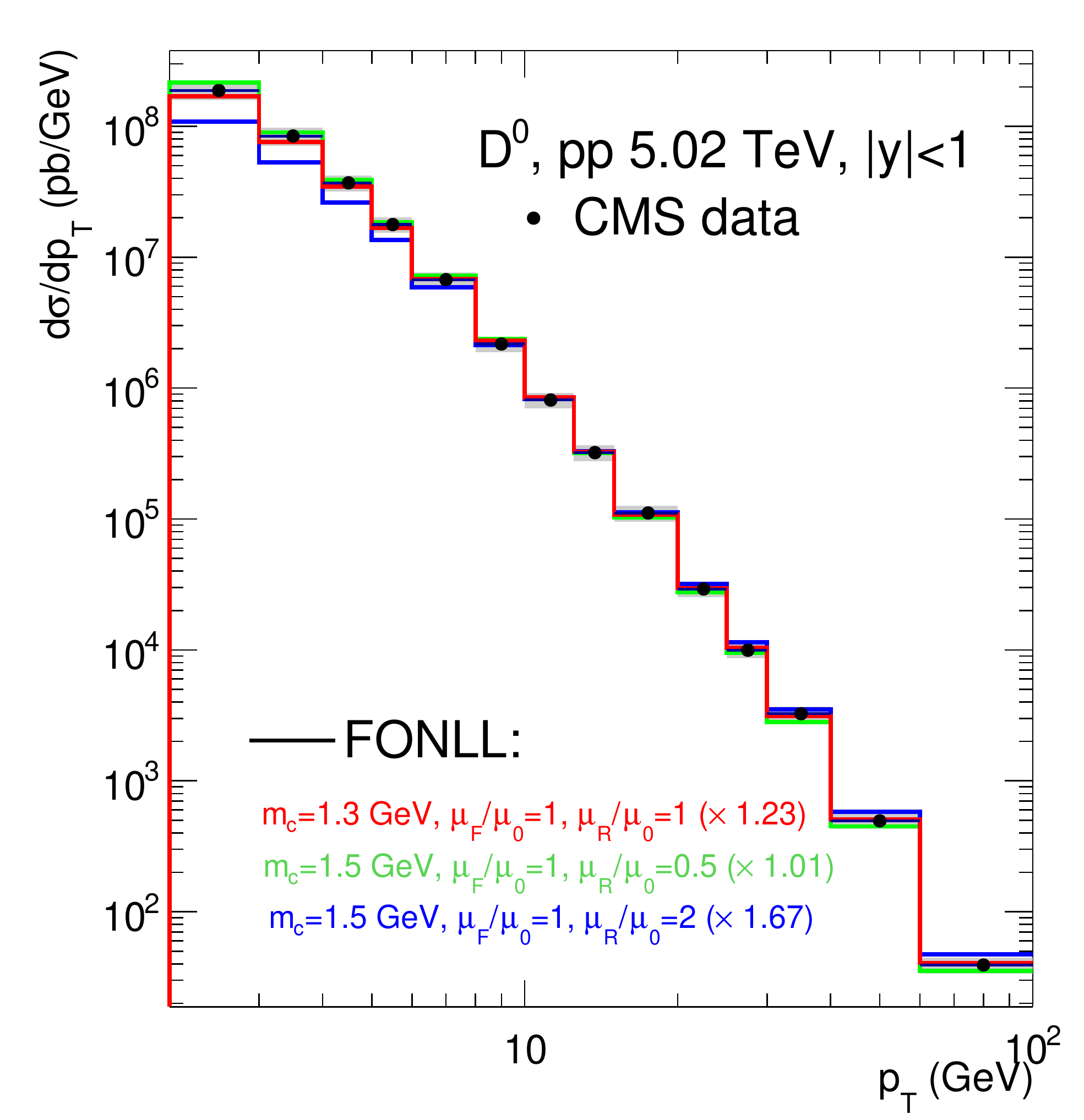}
\includegraphics[width=0.44\textwidth]{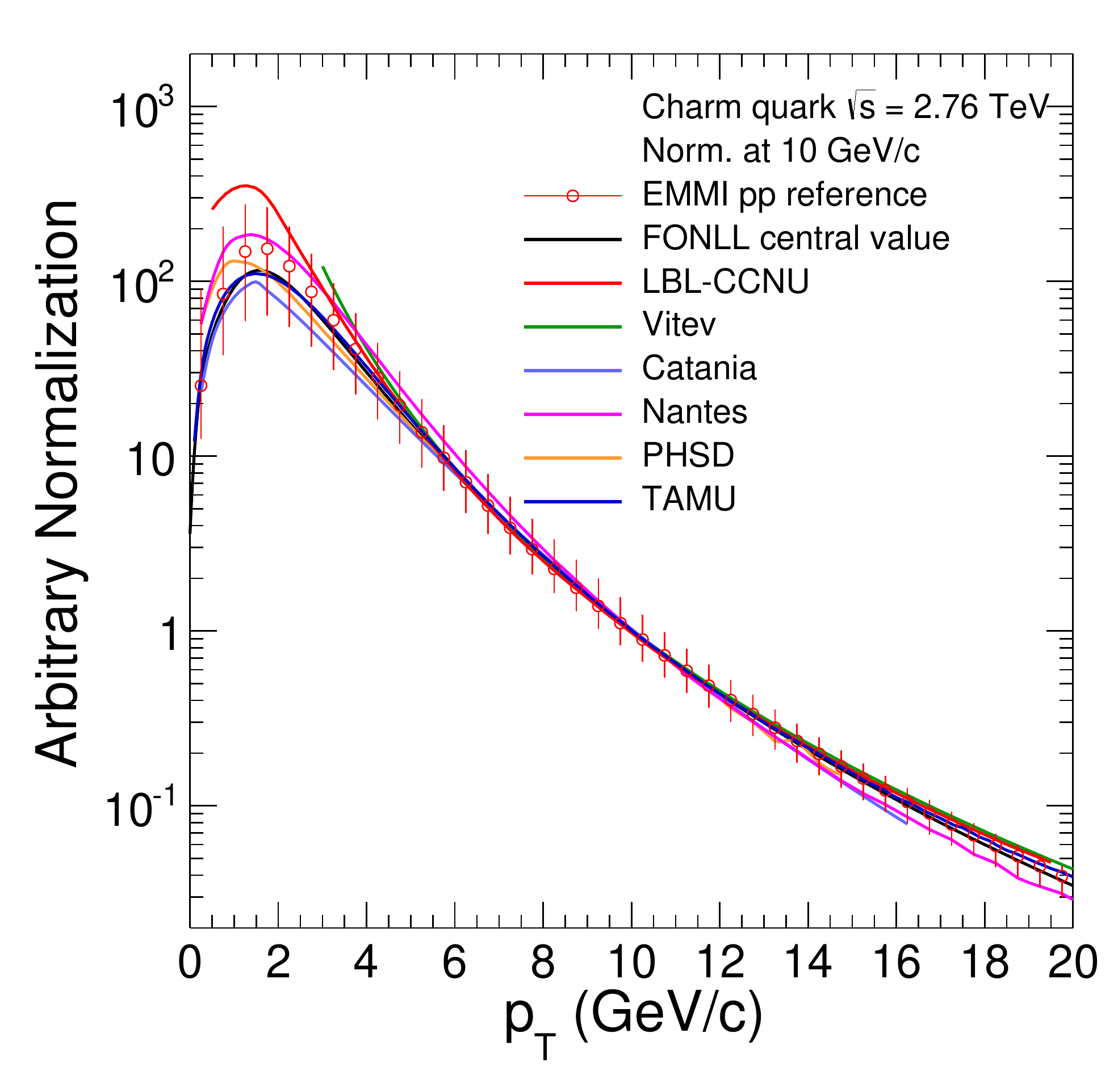}
\caption{Top-left: cross section for $D^0$-meson production at central rapidity in $pp$ 
collisions at $\sqrt s=7$~TeV, measured by ALICE~\cite{Acharya:2017jgo}, compared with the 
best-fitting and extreme FONLL cross sections scaled to match the data.
Top-right: cross section for $D^0$-meson production at central rapidity in $pp$ collisions at 
$\sqrt s=5.02$~TeV, measured by CMS~\cite{Sirunyan:2017xss}, compared with the best-fitting 
and extreme FONLL cross sections scaled to match the data.
Lower panel: our proposed reference for the shape of the $c$-quark $p_t$ distribution in $pp$ 
collisions at $\sqrt s=2.76$~TeV, compared with those used in some of the heavy-ion models; to
readily compare their shapes, all curves are normalized to $p_t$=10\,GeV.}
\label{fig:cmpALICECMS}
\end{center}
\end{figure}

More concretely, we adopt the following procedure.
\begin{enumerate}
\item For the two energies, $\sqrt s=5.02$ and 7~TeV, we constructed a ratio data/FONLL for 
each of the nine parameter sets. The uncertainties on this ratio are the uncertainties of the 
ALICE and CMS data, were we removed the global uncertainties that do not change the shape of 
the cross section.
\item We fitted each ratio with a constant.
\item We defined the `best-fitting' FONLL cross section as the one having the minimum value 
of $(\chi^2/ndf)_{\rm 5.02\,TeV}+(\chi^2/ndf)_{\rm 7\,TeV}$ and the two `extreme' cross sections 
as those with opposite slope in the data/FONLL ratio, largest values of the $\chi^2$ sum, but 
both $(\chi^2/ndf)_{\rm 5.02\,TeV}$ and $(\chi^2/ndf)_{\rm 7\,TeV}$ smaller than 2.
\end{enumerate}
The resulting cross sections are:
\begin{itemize}
\item `best fitting' set: $m_c=1.3$~GeV$/c^2$, $\mu_F=\mu_R=\mu_0$, with 
$(\chi^2/ndf)_{\rm 5.02\,TeV}=0.13$ and fitted constant value of $1.23\pm0.05$ at 
5.02~TeV, $(\chi^2/ndf)_{\rm 7\,TeV}=0.88$ and fitted constant value of $1.29\pm0.03$ at 7~TeV;
\item `extreme' sets: $m_c=1.5$~GeV$/c^2$, $\mu_F=\mu_0$, $\mu_R=0.5\,\mu_0$ and 
$m_c=1.5$~GeV$/c^2$, $\mu_F=\mu_0$, $\mu_R=2\,\mu_0$.
\end{itemize}
The best-fitting cross section uses the minimum value of $m_c$ (1.3~GeV$/c^2$), which is 
consistent with the fact that for $m_c=1.5~$GeV$/c^2$ the ratio data/FONLL increases at low 
$p_T$, where the value of the quark mass has the largest influence on the cross section, 
and a smaller mass value increases the cross section.
The sets with the largest $\chi^2/ndf$ values (up to 4-5) are those with $\mu_F=0.5\,\mu_0$.

Figure~\ref{fig:datatoFONLL} shows the data/FONLL ratios and fits for the best-fitting (left) 
and extreme sets (center and right), at 7 (upper row) and 5.02~TeV (lower row).
Figure~\ref{fig:cmpALICECMS} (upper panels) shows the comparison of the three cross sections, 
scaled by the fitted constants, with the data at $\sqrt s=7$ and 5.02~TeV. The lower panel of 
the same figure compares our proposed reference for $c$ quarks at $\sqrt s=2.76$~TeV  with those 
used as initial conditions in several transport calculations in heavy-ion collisions.

The FONLL $c$-quark $p_t$-differential cross sections, $d\sigma/dp_t$, integrated over 
$|y|<1$ for the three parameters sets at $\sqrt s= 2.76$ and $5.02$~TeV, are provided 
in an online repository~\cite{emmi-repo}. For the Pb-Pb input $p_t$ distributions, these cross 
sections can be multiplied by a nuclear modification factor $R_{AA}$ for $c$ quarks obtained 
from nuclear shadowing according to the EPS09NLO PDF modifications~\cite{Eskola:2009uj}, which 
is discussed in more detail in the following section. 

In order to obtain the $D$-meson $p_T$ distributions, $c$ quarks are fragmented using the BCFY 
function~\cite{Braaten:1994bz} to convert them into pseudoscalar ($D_{Q\to P}(z)$) and vector 
($D_{Q\to V}(z)$) mesons:
\begin{eqnarray}
\label{pqcd1}
D_{Q\to P}(z) &=&  N\, \frac{rz(1-z)^2}{(1-(1-r)z)^6}
\left [ 6 - 18(1-2r)z + (21 -74r+68r^2) z^2 \nonumber  \right. \\
&& \left. -2(1-r)(6-19r+18r^2)z^3  + 3(1-r)^2(1-2r+2r^2)z^4 \right ]\,, \\
\label{pqcd2}
D_{Q\to V}(z) &=& 3N \,\frac{rz(1-z)^2}{(1-(1-r)z)^6}
\left [ 2 - 2(3-2r)z + 3(3 - 2r+ 4r^2) z^2 \nonumber \right. \\
&& \left. -2(1-r)(4-r +2r^2)z^3  + (1-r)^2(3-2r+2r^2)z^4 \right ] \,.
\end{eqnarray}
These functions are the same as used in FONLL calculations. 
The only parameter (apart from the normalization $N$) is $r$, which can be
set to the same values used in FONLL, which were obtained by fitting the 
analytical forms reported above to the $D^\star$ fragmentation function 
measured by ALEPH (see Ref.~\cite{Cacciari:2003zu} for details).
The resulting values are $r=0.06$ for $m_c=1.3$~GeV$/c^2$, which is the mass used 
for the best-fitting cross section,
and $r=0.10$ for $m_c=1.5$~GeV$/c^2$.
These functions should be used for both the Pb-Pb spectra (when considering independent 
fragmentation outside the QGP), \ie, the numerator of the $R_{AA}$, and the $pp$ spectra 
figuring into the denominator of the $R_{AA}$.

\subsection{Cold Nuclear Matter Effects}
\label{ssec_cnm}

Cold-nuclear-matter (CNM) effects modify the yields and kinematic distributions of 
hadrons produced in hard scattering processes in the case of $p$A and AA collisions, see, \eg,
Refs.~\cite{Albacete:2017qng,Albacete:2013ei} for recent analyses at the LHC.
The largest CNM effect at LHC energies is the nuclear modification of parton distribution 
functions, \ie, the fact that the PDFs of nucleons within nuclei are different from the PDFs 
of free protons. We report the expected effects using the EPS09NLO~\cite{Eskola:2009uj} and 
EPPS16~\cite{Eskola:2016oht} parameterizations of the nuclear PDF modifications, which both 
depend on $x$, $Q^2$ and the mass number $A$, and are defined as 
$R^A_i(x,Q^2) = f^A_i(x,Q^2)/f^p_i(x,Q^2)$, where $i$ denotes the parton type (gluon, valence 
quark or sea quark) and $f^A$, $f^p$ are the PDFs of the nucleon in a nucleus of mass $A$ and 
of the free proton, respectively. The features of the modifications that are most relevant for HQ 
production up to momenta of a few tens of GeV$/c$ are the reduction of $R^A_i$ below unity for 
$x$ lower than about $3\cdot 10^{-2}$, namely nuclear shadowing, and the increase above unity 
for $x>3\cdot 10^{-2}$, namely anti-shadowing.
These modifications are larger for small values of $Q^2$.  

\begin{figure}[!ht]
\begin{center}
\includegraphics[width=\textwidth]{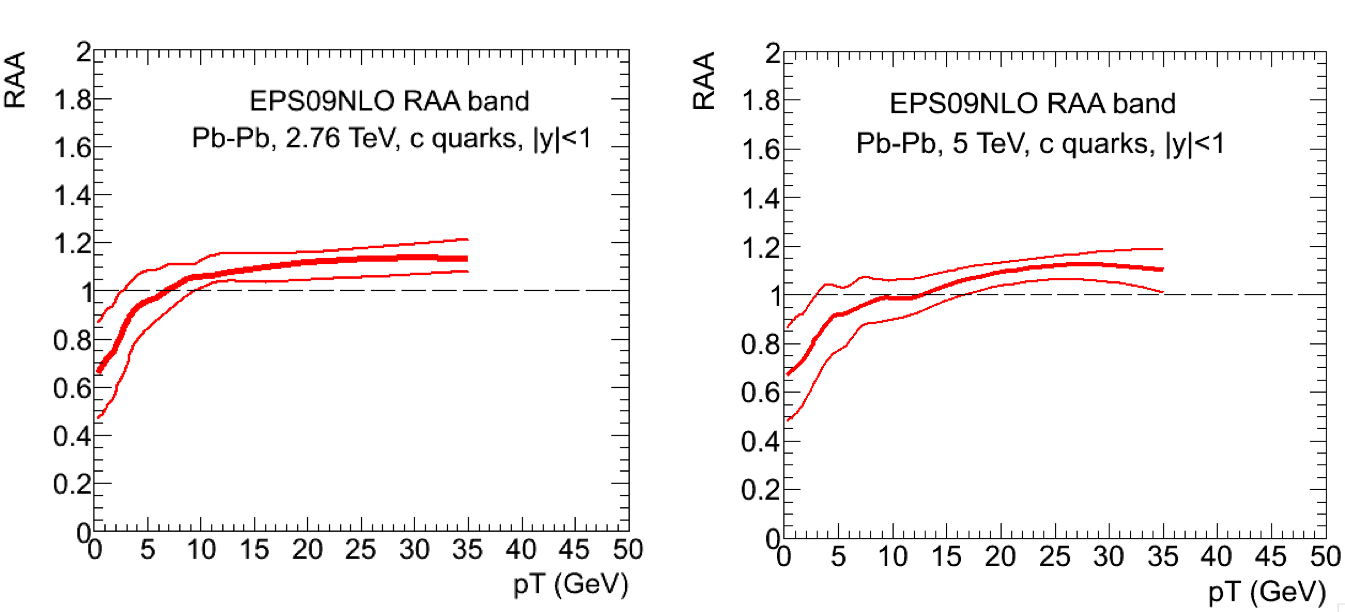}
\includegraphics[width=.5\textwidth]{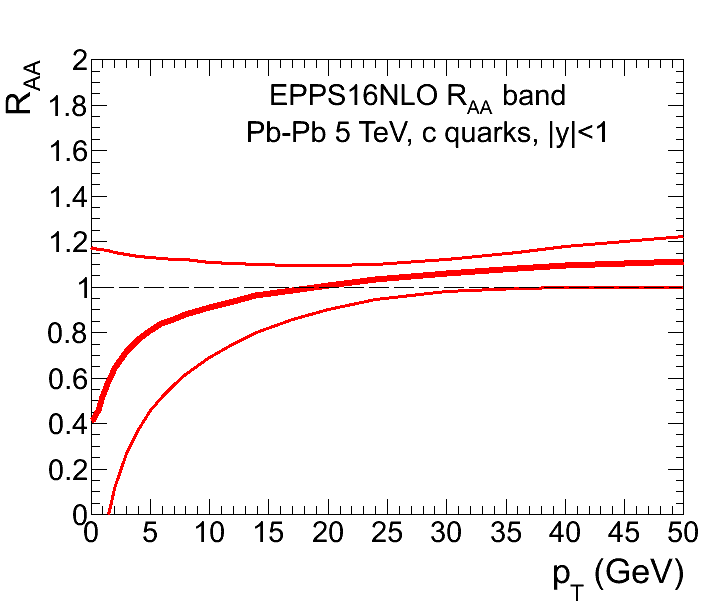}
\caption{Nuclear modification factor of the $c$-quark $p_t$ distribution as obtained using, 
in the two upper panels, the HVQMNR NLO calculation~\cite{Mangano:1991jk} with  CTEQ6M 
PDFs~\cite{Pumplin:2002vw} and EPS09NLO nuclear modification~\cite{Eskola:2009uj},
and, in the lower panel, the FONLL calculation~\cite{Cacciari:1998it,Cacciari:2012ny} with 
CTEQ10 PDFs~\cite{Lai:2010vv} and EPPS16 nuclear modification~\cite{Eskola:2016oht}. 
The uncertainty bands correspond to the EPS09 and EPPS16 uncertainties.}
\label{fig:cRAAeps09}
\end{center}
\end{figure}

Figure~\ref{fig:cRAAeps09} shows the CNM modification of $R_{AA}$ for the $c$-quark $p_t$ 
distribution in Pb-Pb collisions at LHC energies. 
The upper panels show the results for $\sqrt{s_{NN}}= 2.76$ and $5.02$~TeV 
as obtained with the HVQMNR NLO calculation~\cite{Mangano:1991jk}, the 
CTEQ6M PDFs~\cite{Pumplin:2002vw} and the EPS09NLO nuclear modification. 
The lower panel shows the result for $\sqrt{s_{NN}}=5.02$~TeV 
as obtained with the FONLL calculation~\cite{Cacciari:1998it,Cacciari:2012ny}, 
the CTEQ10 PDFs~\cite{Lai:2010vv} and the EPPS16 nuclear modification. 
The uncertainty bands are obtained 
according to the EPS09 and EPPS16 prescriptions, thus representing the uncertainty on the nuclear 
modification of the PDFs. The central values of the $R_{AA}$ values at the two energies are 
similar, with the result given by the EPPS16 set being slightly lower than that of the EPS09 set. 
The uncertainties are significantly larger with the more recent EPPS16 set compared to the EPS09 set, 
because in the EPPS16 analysis the authors allow for additional parameters (which are not constrained 
by the existing data) in the functional form that regulates the $x$ and $Q^2$ dependence of 
$R^A_i$~\cite{Eskola:2016oht}. While we report here results with both sets, we note that the 
nuclear modification factor of $D$ mesons in $p$Pb collisions at the LHC as measured by the
ALICE~\cite{Adam:2016ich} and LHCb~\cite{Aaij:2017gcy} experiments at central and forward 
rapidity, respectively, is described within uncertainties using the EPS09 set. Therefore, EPS09 
could be considered as an acceptable effective implementation of the shadowing effects for charm 
quarks, while EPPS16 could be considered as the most up-to-date implementation of nuclear PDFs 
and their uncertainties.

The $R_{AA}$ values for both sets can be downloaded in numerical format and used as multiplicative 
factor, depending on the $c$-quark $p_t$, to obtain an input $c$-quark $p_t$
distribution for transport simulation in Pb-Pb collisions.

\subsection{Exploration of Uncertainty on $R_{\rm AA}$ and $v_2$ in Pb-Pb(2.76\,TeV)}
\label{ssec_pp-uncert}
\begin{figure}[!t]
\includegraphics[width=0.45\textwidth]{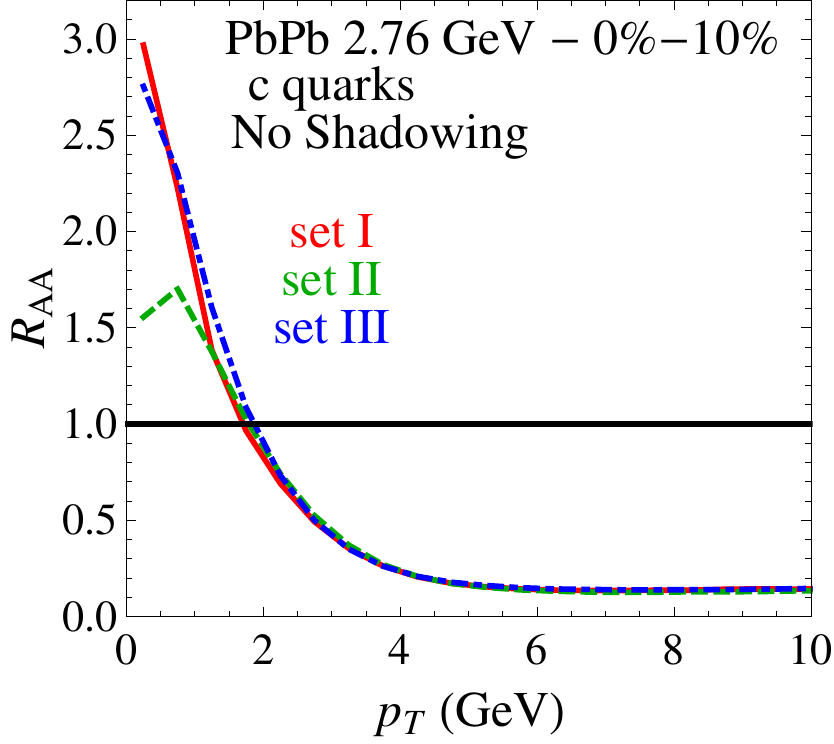}
\includegraphics[width=0.45\textwidth]{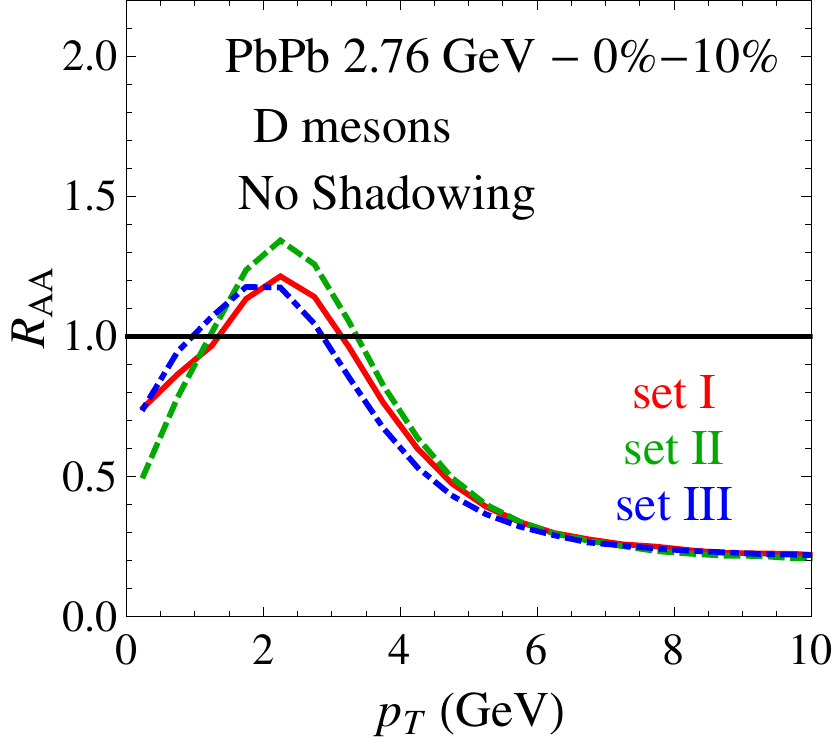}
\includegraphics[width=0.47\textwidth]{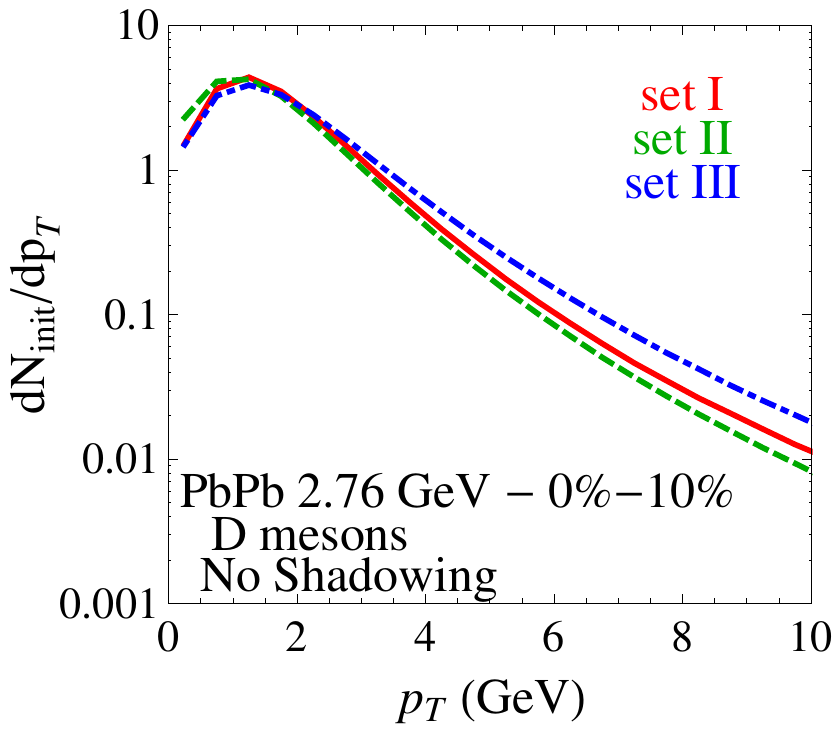}
\includegraphics[width=0.47\textwidth]{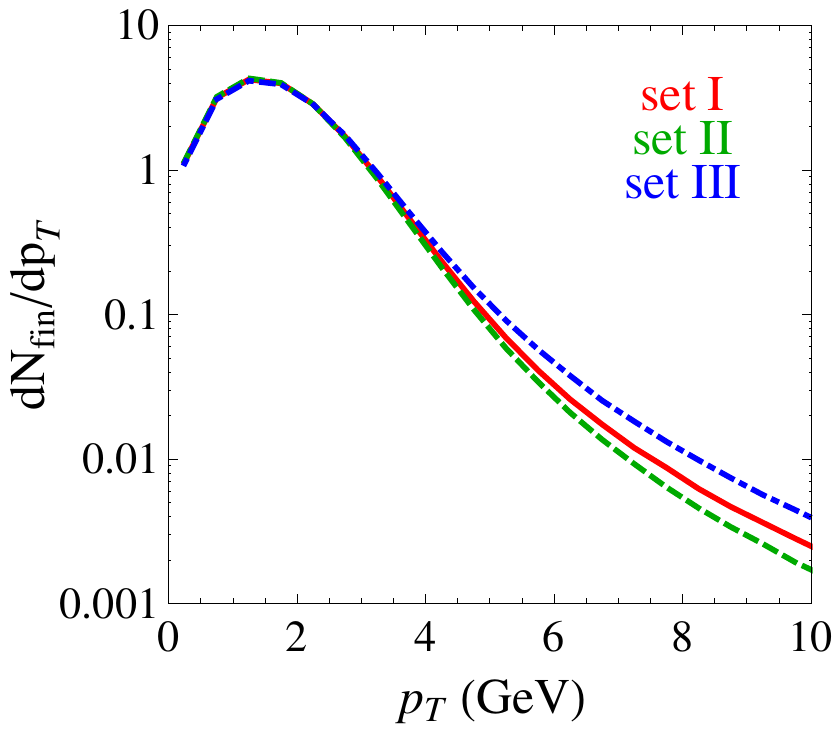}
\caption{Impact of the variation in the initial $c$-quark spectrum from $pp$ collisions 
on the $R_{AA}$ (top row) and transverse-momentum spectra (bottom row) of $c$ quarks and 
$D$ mesons.
Upper left: $R_{AA}$ of $c$ quarks stemming from the three sets for the initial spectrum, 
transported through the QGP in the Nantes model; upper right: same for $D$ mesons;
lower left: spectra of ``initial" $D$ mesons, obtained from the hadronization of initial 
$c$ quarks; lower right: spectra of final $D$ mesons.}
\label{fig:ppspec-uncert}
\end{figure}

In this section we address the consequences of an imprecise knowledge of the initial 
$c$-quark spectrum on the nuclear modification factor in heavy-ion collisions, thereby 
focusing on the impact on the final spectra of $c$ quarks just before hadronization and 
$D$ mesons just after hadronization. We conduct this study within the Nantes transport 
model~\cite{Nahrgang:2013saa}\footnote{We expect that the conclusions do 
not strongly depend on the specific transport model used for this study.} with elastic 
energy loss only (with a $K$ factor of $K=1.5$ for the pQCD* HQ-medium interaction, and 
a coalescence/BCFY fragmentation scheme) in 0-10\,\% Pb-Pb($\sqrt{s_{NN}}$=2.76\,TeV) 
collisions.

We start by focusing on the effects of the $c$-quark spectrum produced in $pp$ (\ie, without
CNM effects) taken as the initial condition for the transport. This is carried out by 
considering the FONLL approach with the 3 parameter sets resulting from the study performed in 
Sec.~\ref{ssec_baseline}:
\begin{itemize}
\item set I: $m_c=1.3$ GeV, $\mu_F/\mu_0=1$, $\mu_R/\mu_0=1$,  scaled by 1.33
\item set II: $m_c=1.5$ GeV, $\mu_F/\mu_0=1$, $\mu_R/\mu_0=0.5$,  scaled by 0.93
\item set III: $m_c=1.5$ GeV, $\mu_F/\mu_0=1$, $\mu_R/\mu_0=2$,  scaled by 2.21
\end{itemize}

In the upper panels of Fig.~\ref{fig:ppspec-uncert}, we show the $R_{AA}$ of $c$ quarks (left) 
and the $R_{AA}$ of $D$ mesons after hadronization (right). We find a clear separation between 
a high-$p_T$ regime 
($p_T\gtrsim 3\,{\rm GeV}$), for which the $R_{AA}$ is essentially independent of the 
chosen set, and a low-$p_T$ regime ($p_T\lesssim 3\,{\rm GeV}$), for which significant differences 
are found in the c-quark $R_{AA}$, which are somewhat tempered in the $D$-meson $R_{AA}$.   

\begin{figure}[!t]
\begin{center}
\includegraphics[width=0.48\textwidth]{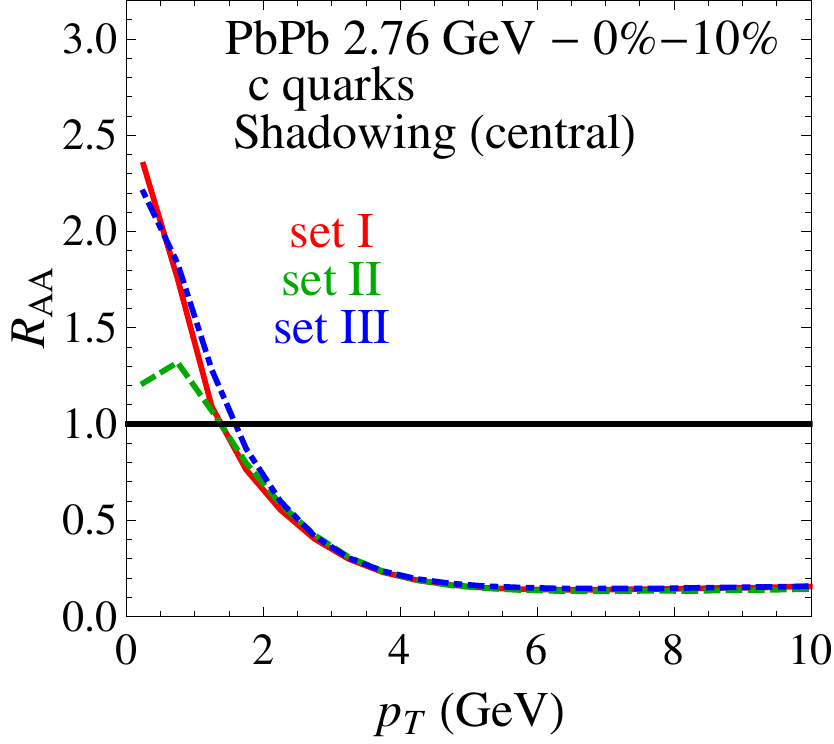}
\includegraphics[width=0.48\textwidth]{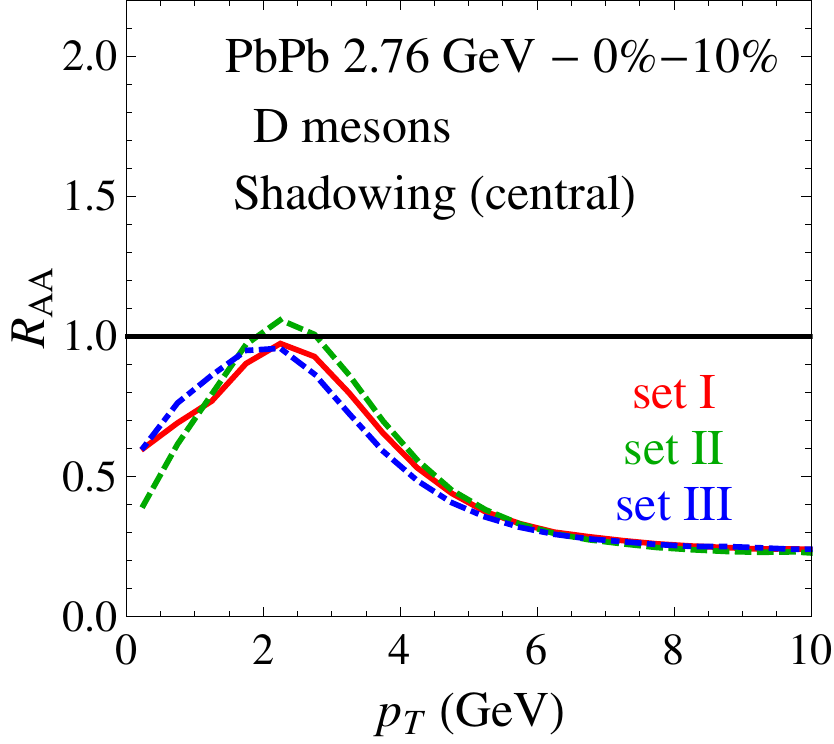}
\caption{Same as upper panels in Fig.~\ref{fig:ppspec-uncert} with CNM effects included in the
initial $c$-quark $p_t$ spectrum (see text for details).}
\label{fig:RAAfromvariousppsetsshadow}
\end{center}
\end{figure}
In order to clarify these observations, we display in the lower panels of Fig.~\ref{fig:ppspec-uncert} 
the spectra of 
``initial" $D$ mesons (left; obtained from the distributions of initial $c$-quarks through 
fragmentation) as well as the spectra of final $D$ mesons (right). In the high-$p_T$ regime, 
we see that the hierarchy between the 3 sets is preserved through the HQ transport, as energy loss 
is the dominant process. This explains the overlaps observed at the $R_{AA}$ level. In the low-$p_T$ 
regime, distributions ensuing from the transport of various sets converge towards a unique profile, 
a feature arising from the rather large degree of thermalization of $c$ quarks in the QGP 
phase\footnote{The degree of $c$-quark thermalization in the QGP can of course vary in 
different transport models.}.        
Hence, one concludes that the differences seen for the $R_{AA}$ in this regime merely result 
from the choice of the initial distribution adopted in the denominator. This could lead to systematic 
differences between various groups that could be easily corrected by adopting a common baseline 
of the type suggested in Sec.~\ref{ssec_baseline}.

Next we investigate the consequences resulting from the CNM effects described in Sec.~\ref{ssec_cnm}. 
In Fig.~\ref{fig:RAAfromvariousppsetsshadow}, we display the same quantities, but with CNM effects taken 
into account by multiplying the various sets evaluated for $pp$ collisions by a reduction factor chosen 
as the central line of Fig.~\ref{fig:cRAAeps09}. All curves in Fig.~\ref{fig:RAAfromvariousppsetsshadow} 
follow the same trend as the corresponding ones in the upper row of Fig.~\ref{fig:ppspec-uncert}, with an 
expected additional suppression at low $p_t$. 

\begin{figure}[!t]
\begin{center}
\includegraphics[width=0.46\textwidth]{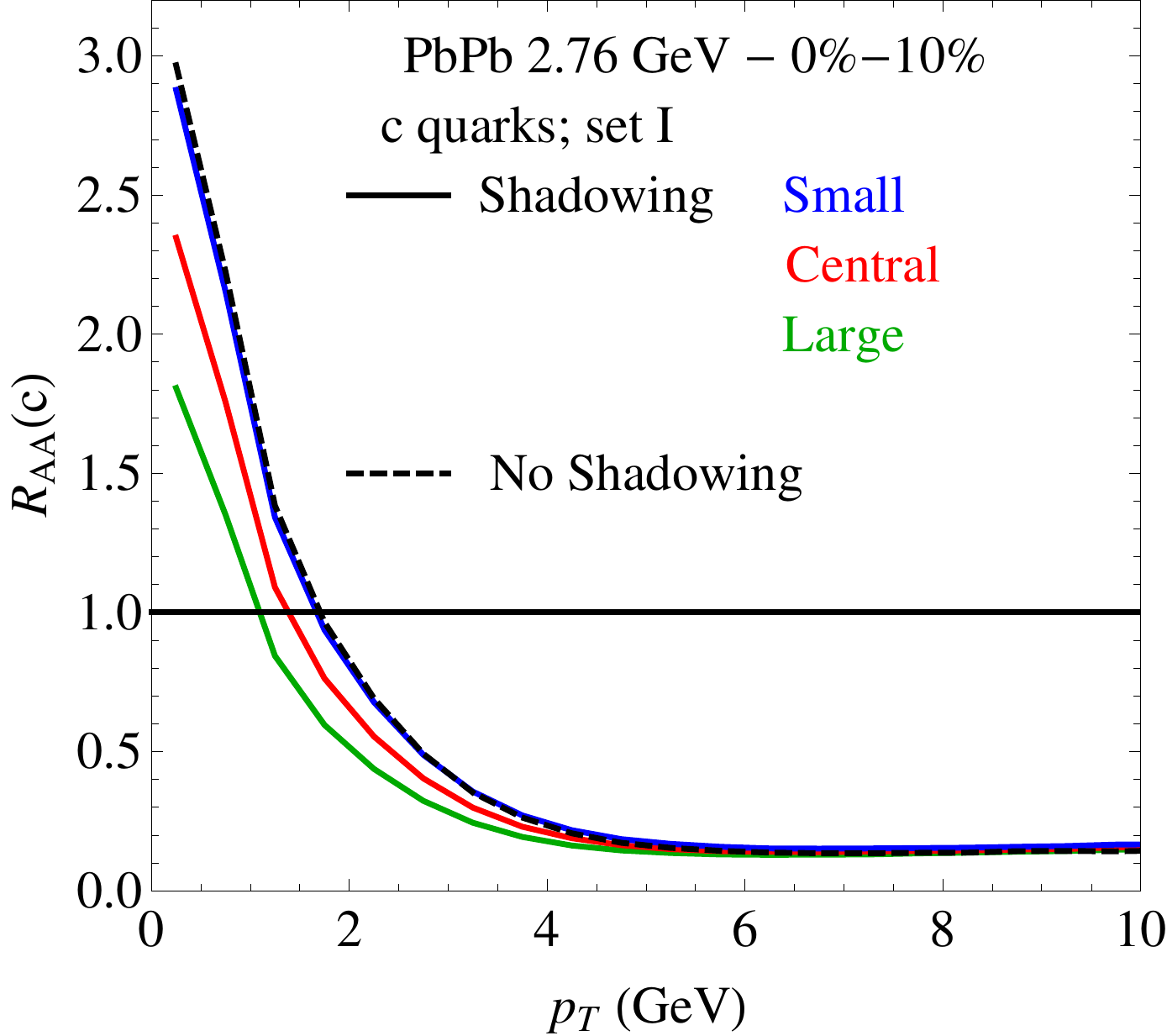}
\includegraphics[width=0.48\textwidth]{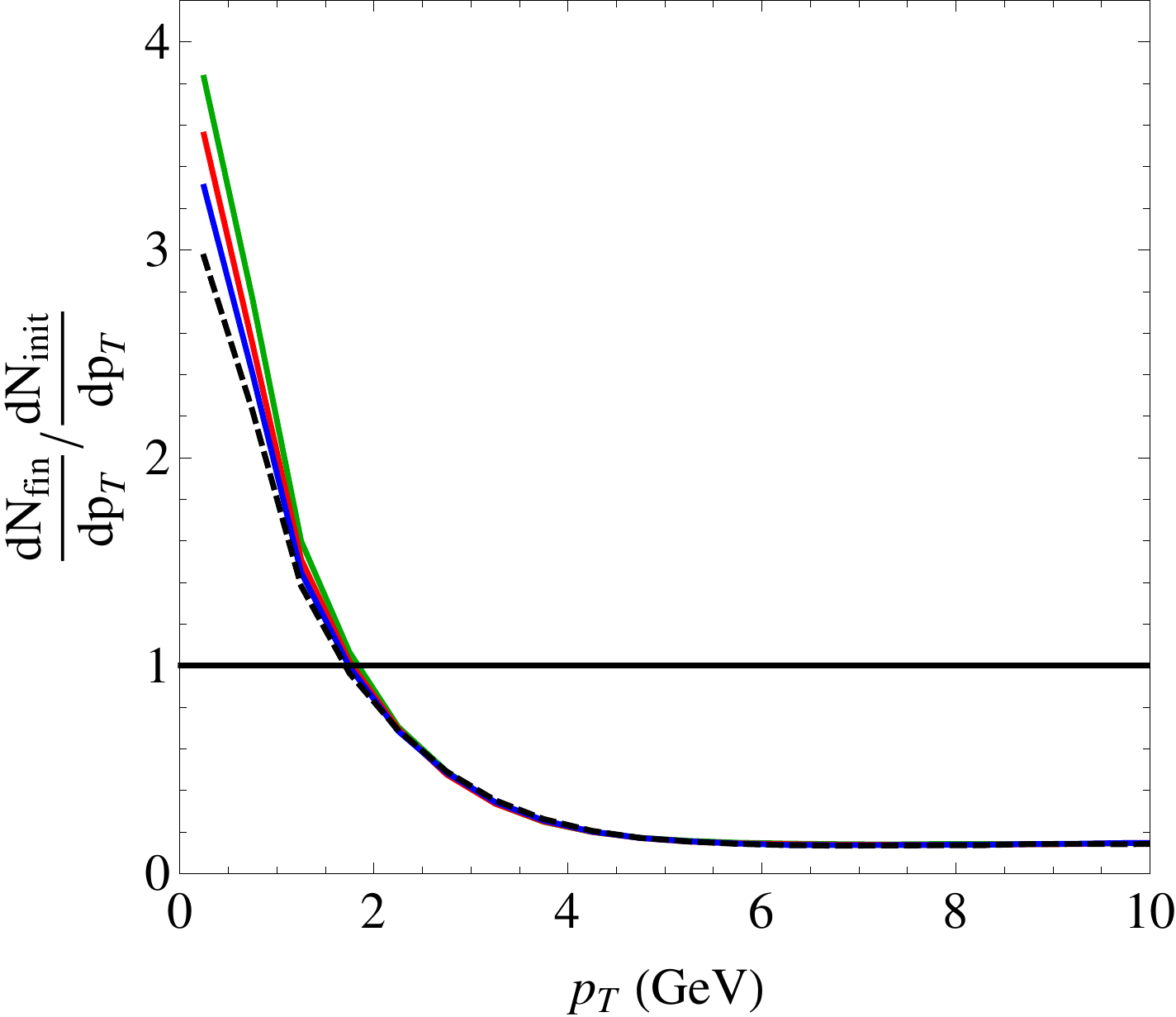}
\includegraphics[width=0.46\textwidth]{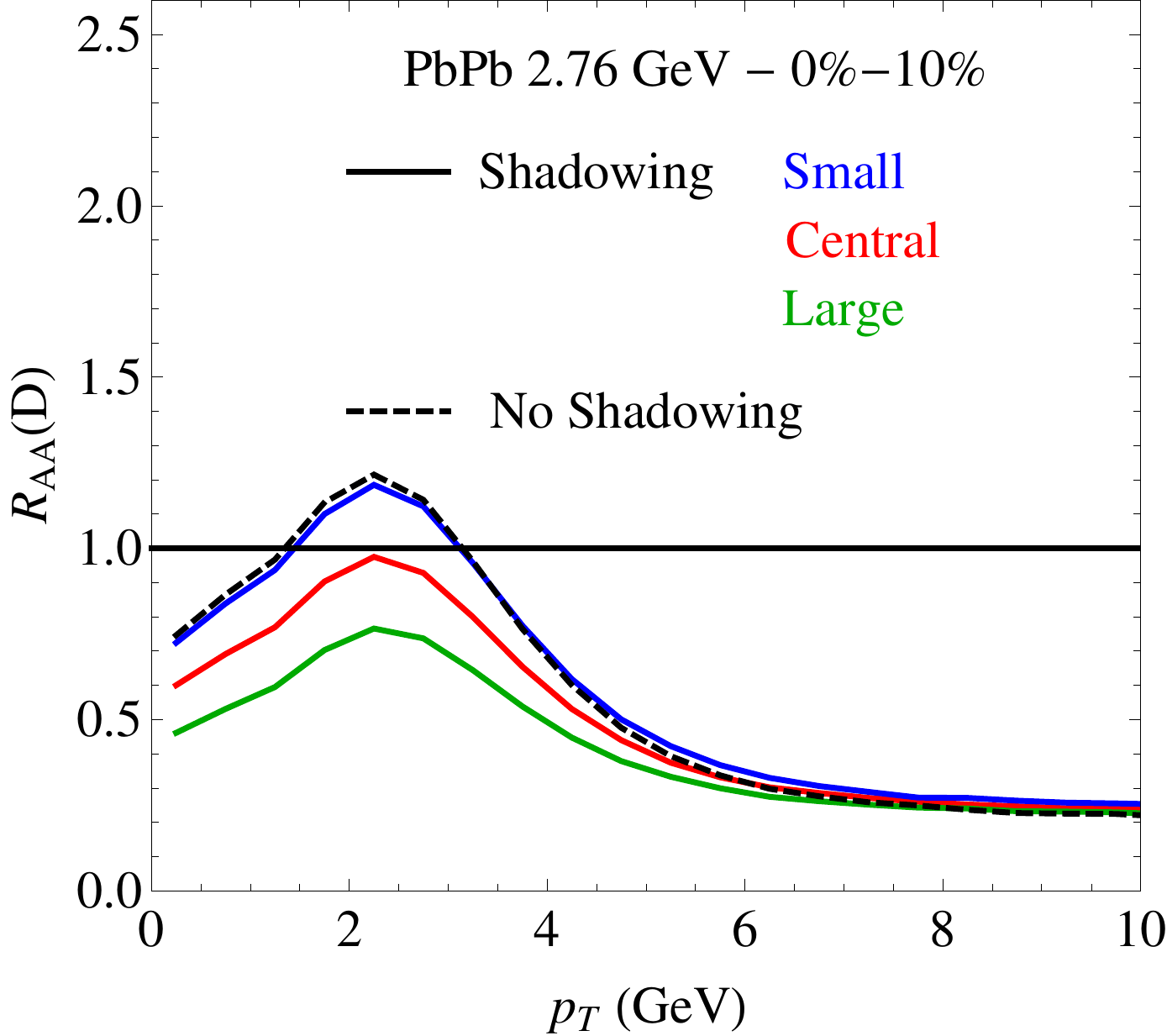}
\includegraphics[width=0.48\textwidth]{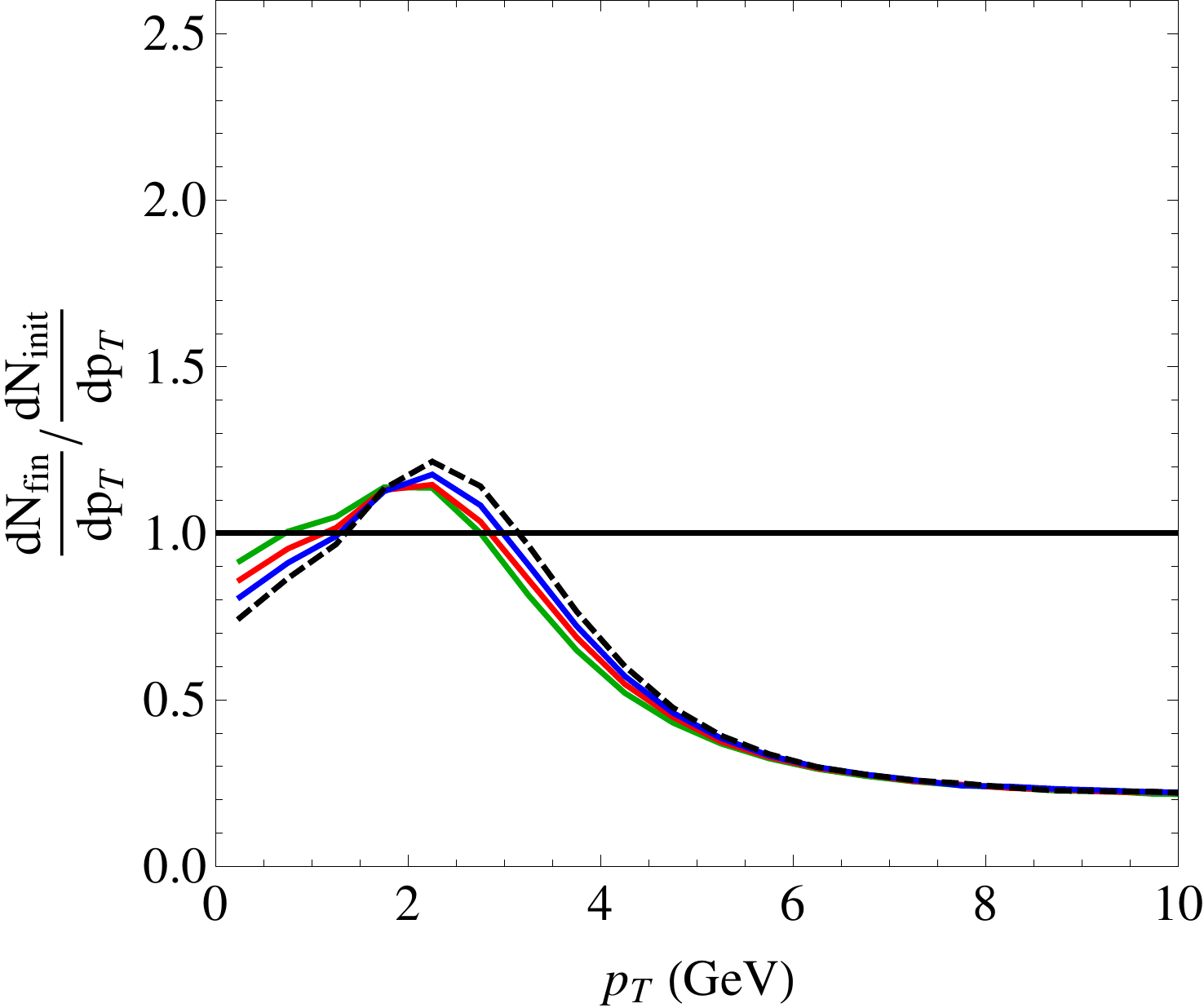}
\caption{Top row: $R_{AA}$ of $c$ quarks (left panel) resulting from initial-set I without 
(dashed line) and with (solid line) multiplication by various CNM prescriptions for the initial 
$c$-quark spectrum as shown in Fig.~\ref{fig:cRAAeps09}; right panel: ratio of final $c$-quark 
$p_t$-spectrum to the initial input spectrum (including CNM effects) to the transport simulation. 
Bottom row: same as top row but for $D$ mesons.}
\label{fig:RAA-shadow}
\end{center}
\end{figure}

In upper left panel of Fig.~\ref{fig:RAA-shadow}, we explore more systematically the CNM effects 
on the $c$-quark $R_{AA}$ by considering various input distributions for the transport obtained 
from set I multiplied by the initial ``$R_{AA}$" corresponding to the various prescriptions shown 
in Fig.~\ref{fig:cRAAeps09}. The case without CNM effect is also shown. It nearly overlaps with 
the ``small" shadowing case, while prescriptions corresponding to larger shadowing naturally lead 
to a stronger suppression at small $p_t$ ($p_t\lesssim 4\,{\rm GeV}$). 

In the upper right panel of Fig.~\ref{fig:RAA-shadow}, we display the ratios between the final 
and the initial spectra (where the latter also {\em include} CNM effects). These ratios reflect 
the genuine modifications in the QGP phase and are seen  
to be not much affected by the precise shape of the initial distribution. Thermalization of $c$ quarks 
at small $p_t$ has the tendency to compensate the depletion in the initial distribution and is the 
reason for the reversal of the hierarchy observed in this range.    

In the bottom row of Fig.~\ref{fig:RAA-shadow}, we show the same quantities as in the top row
but for the $D$ mesons; the same observations as for $c$ quarks apply. 
In the low-$p_T$ regime, we find differences of the order of $\pm 25\%$ 
for small/large CNM effects as compared to the intermediate case. Comparing with 
Fig.~\ref{fig:ppspec-uncert}, we conclude from this study that the uncertainties on the CNM effects 
dominate by far over those on the $c$-quark spectra from $pp$ collisions.

\section{Bulk Evolution Models} 
\label{sec_bulk}
The evolution of the bulk medium in HICs, characterized by the space-time dependence 
of temperature and local flow velocity, provides the link between the interactions of
HF particles with the medium and the time evolution of their spectra. It is therefore 
mandatory to carefully compare the different medium evolutions as employed in HF phenomenology 
in the current literature.  
  
In the following comparison, we have focused on approaches which were primarily designed to
work at low and intermediate charm-quark momenta. Specifically, and very briefly, these are:
\begin{itemize}
\item UrQMD:
HQ Langevin transport in
a 3+1D ideal hydrodynamic evolution, initialized by smeared UrQMD string/energy
density configurations at a starting time of $\tau_0$=0.5\,fm, with a Polyakov-loop
model-based QGP EoS fitted to lQCD with final QGP temperature of
$T_{\rm c}$=160\,MeV~\cite{Steinheimer:2009nn,Petersen:2011sb};
\item TAMU:
HQ Langevin transport in a 2+1D ideal hydrodynamic evolution with smooth initial conditions,
starting time $\tau_0$=0.4\,fm, with lQCD-fitted EoS and final QGP temperature of
$T_{\rm c}$=170\,MeV~\cite{He:2011zx};
\item Nantes:
HQ Boltzmann transport using the EPOS-2 event generator with fluctuating initial conditions and
3+1D ideal hydro starting at $\tau_0$=0.3\,fm, with lQCD-based EoS~\cite{Werner:2010aa} and 
final QGP temperature $T_{\rm c}$=166\,MeV~\cite{Pierog:2010dt};
\item Catania:
HQ Langevin transport using Boltzmann simulations for the bulk evolution with massive
quasiparticles, coarse-grained to obtain local temperatures, starting time $\tau_0$=0.3\,fm
and final QGP temperature of $T_{\rm c}$=170\,MeV~\cite{Plumari:2012ep};
\item LBL-CCNU:
HQ Boltzmann transport assuming massless thermal partons in the VISHNU 2+1D viscous
hydrodynamic evolution (OSU hydro) including event-by-event
initial conditions, starting at $\tau_0$=0.6\,fm, with lQCD-fitted EoS~\cite{Huovinen:2009yb} 
and final QGP temperature of $T_{\rm c}$=165\,MeV~\cite{Song:2007fn,Song:2007ux,Qiu:2011hf}
\item Duke:
HQ Langevin transport with the same hydrodynamic evolution as in LBL-CCNU.
\item CUJET:
HQ energy loss calculation (limited to $p_T$$>$6\,GeV)~\cite{Liao:2008dk,Xu:2014tda,Xu:2015bbz} 
with elastic-only friction coefficient from pQCD*5 using the VISHNU 2+1D viscous hydrodynamic 
model with $T_{\rm c}$=160\,MeV (the radiative processes included in the default CUJET framework 
have been switched off). 
\item POWLANG:
HQ Langevin transport in the ECHO-QGP 3+1D viscous hydrodynamic evolution with lQCD-based
EoS, starting time $\tau_0$=0.6\,fm and final QGP temperature of
$T_{\rm c}$=155\,MeV~\cite{DelZanna:2013eua}.
\item PHSD:
HQ Boltzmann transport in a microscopic off-shell transport model utilizing a dynamical 
quasiparticle model for the QGP EoS fitted to lQCD data with a final QGP energy
density of 0.5\,GeV/fm$^3$ corresponding to a would-be equilibrium temperature of
$T_{\rm c}$$\simeq$160\,MeV~\cite{Bratkovskaya:2011wp,Song:2015sfa,Song:2015ykw}.
\end{itemize}

We note that for the purpose of the present discussion the notion of $T_{\rm c}$ pertains
to the temperature where the hadronization of heavy quarks into HF hadrons is carried out,
which also delineates the partonic and hadronic treatment of the HQ and HF-hadron interactions
in the bulk medium (the latter are not discussed in this section).

\subsection{Bulk Comparisons}
\label{ssec_bulk-comp}
\begin{figure}[!t]
\hspace{-0.2cm}
\includegraphics[width=\linewidth,height=13cm]{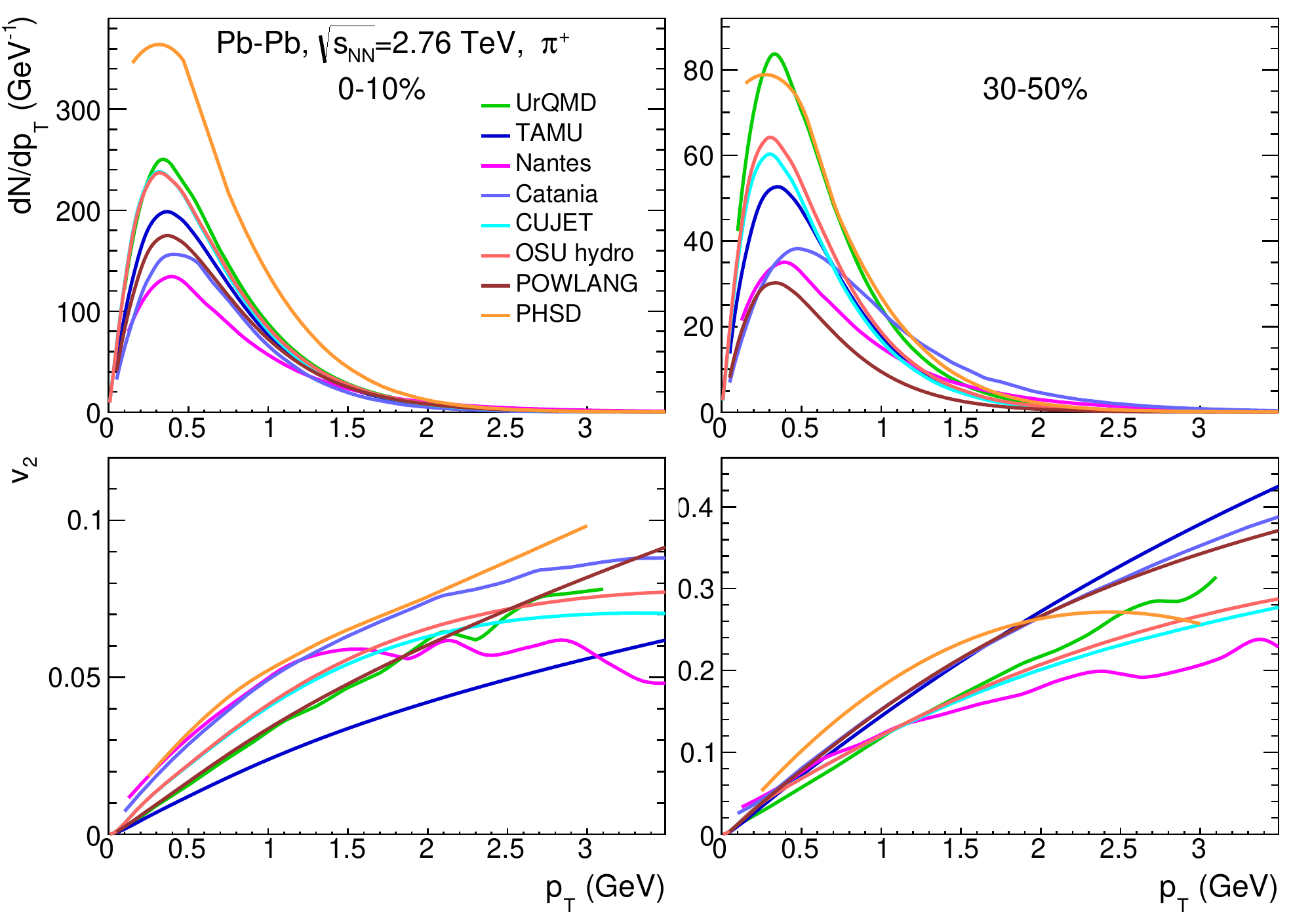}
\caption{Comparison of the $p_T$ spectra (upper panels) and elliptic flow (lower
panels) of direct pions (no feeddown) right after hadronization in 0-10\% (left panels) and
30-50\% (right panels) Pb-Pb(2.76\,TeV) collisions within the different fireball evolution
models used for HF transport (as identified in the legends).
}
\label{fig:pions}
\end{figure}

To prepare for the interpretation of the resulting charm-quark (and later $D$-meson) 
spectra with a common QGP transport interaction, we start by inspecting the results of 
the bulk evolution for their radial and elliptic flow in terms of light-hadron 
production at $T_{\rm c}$. While this does not give a complete picture of the space-time
history of the medium expansion as experienced by the propagating $c$-quarks, it should
nevertheless provide a useful benchmark. To reduce ambiguities in the comparison of the 
collective properties of the bulk across the different evolution models, we perform the 
comparison of $p_T$-spectra and $v_2$ at the hadron level, specifically for direct pions 
and protons, \ie, without any feeddown from resonance decays. This avoids, \eg, 
complications associated with different quark masses in the description of the QGP or 
issues related with gluonic degrees of freedom. 

\begin{figure}[!t]
\hspace{-0.2cm}
\includegraphics[width=\linewidth,height=13cm]{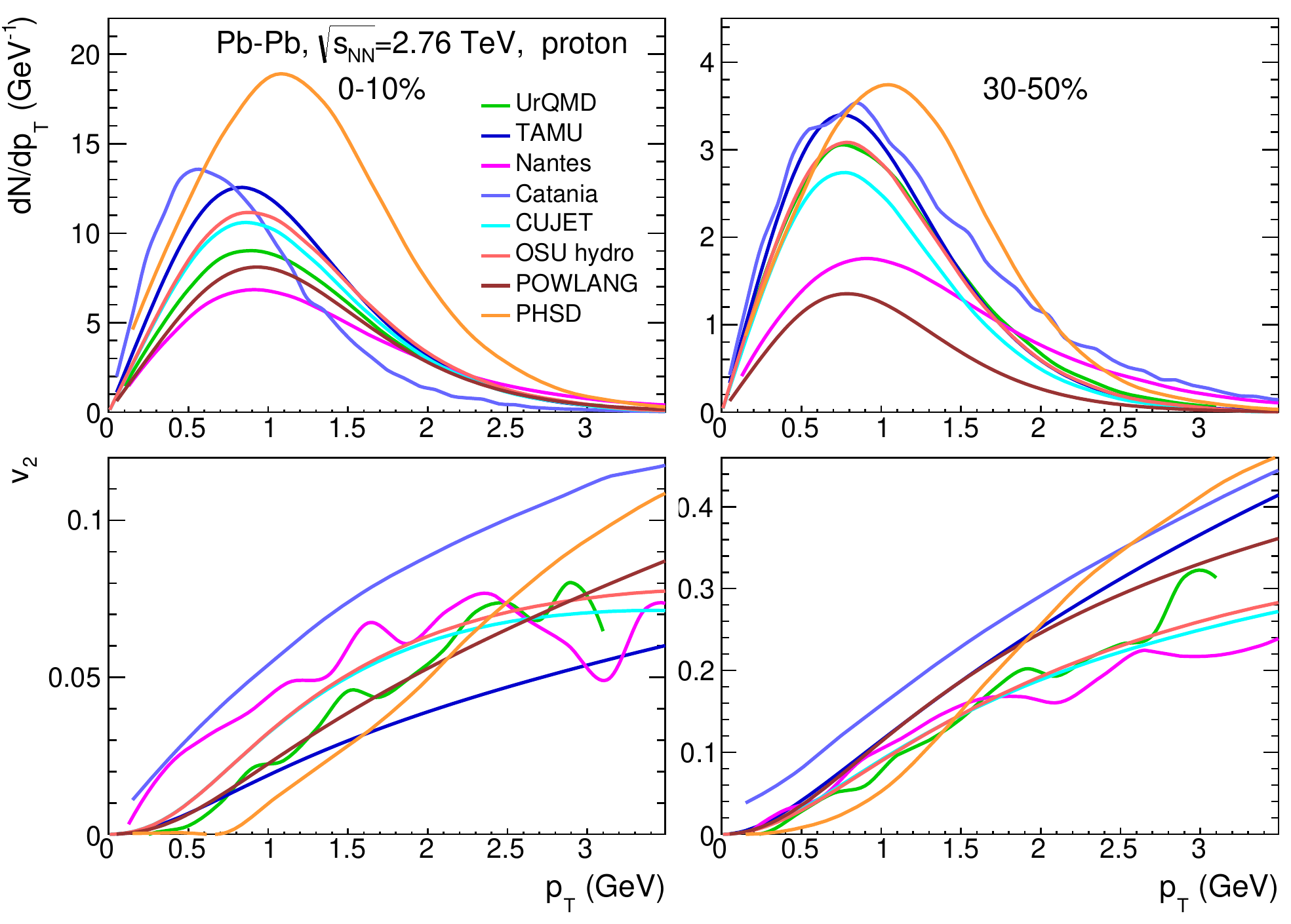}
\caption{Comparison of the $p_T$ spectra (upper panels) and elliptic flow (lower panels) 
of direct protons (no feeddown) right after hadronization in 0-10\% (left panels) and 
30-50\% (right panels) Pb-Pb(2.76\,TeV) collisions within the different fireball evolution 
models used for HF transport (as identified in the legends).
}
\label{fig:protons}
\end{figure}

Direct-pion and -proton $p_T$ spectra and $v_2$ at hadronization are summarized in
Figs.~\ref{fig:pions} and \ref{fig:protons}, respectively, for 0-10\% and 30-50\% 
Pb-Pb(2.76\,TeV) collisions. For both pions and protons all evolution models give
maximum structures in the $p_T$ spectra whose locations are indicative for the size of 
the transverse flow at hadronization. There is an approximate agreement within about 20\%
in both magnitude and shape of the pion and proton spectra for the hydrodynamic models 
used by TAMU, CUJET and the OSU hydro (used by LBNL/CCNU and Duke). The ECHO-QGP model 
used by the Torino group and the EPOS-2 model used by the Nantes group tend to fall somewhat 
off in both the pion and proton multiplicities, especially for the 30-50\% centrality class. 
The PHSD model, where the direct contributions to pions and protons are not readily 
extracted from the underlying off-shell hadronization scheme, shows somewhat larger yields. 
They differ from those of hydro-based models since the hadronization of massive 
quasiparticles in the PHSD goes via the production of resonances/strings,
which includes feeddown from resonance and string decays. 

\begin{table}[!t]
\centering
\begin{tabular}{|c|c|c|c|c|}
\hline
Model  & \multicolumn{2}{c}{$dN_{\pi^+}/dy$ ($dS/d\eta$) } \vline & \multicolumn{2}{c}{$dN_{p}/dy$}  \vline
\\
\hline
   & 0-10\%   & 30-50\%   &  0-10\%    &  30-50\%
\\
\hline
UrQMD & 495 & 152 & 34 & 11
\\
\hline
TAMU & 682 (12400) & 170 (3080) & 58 & 15
\\
\hline
Nantes & 478 & 129 & 38 & 10
\\
\hline
Catania & (14000) & (3700) & &
\\
\hline
LBL-CCNU/Duke & 653 (12600) & 160 (3080) & &
\\
\hline
CUJET & 610 (10820) & 142 (2610)& 45 & 11
\\
\hline
POWLANG & (9100) & (1450) & &
\\
\hline
PHSD & 722 & 148 & 31 & 6
\\
\hline
exp. & 670$\pm$68 & 163$\pm$15 & 31$\pm$4 & 8$\pm$1
\\
\hline
\end{tabular}
\caption{Inclusive $\pi^+$ and proton numbers (\ie, including strong and electromagnetic feeddown)
per unit momentum-space rapidity in Pb-Pb(2.76\,TeV) collisions in the various bulk evolution models.
Also shown in parentheses are the values for the total entropy per unit space-time rapidity at
the end of the QGP phase (as available). As a reference the last row shows experimental values from
Ref.~\cite{Abelev:2013vea}.
}
\label{tab_bulk}
\end{table}

The shape of the pion spectra, \ie, the maximum structure and its location in $p_T$, 
generally agrees quite well across the models. The sensitivity to the transverse flow 
is enhanced in the spectra of protons due to their larger sensitivity to blue shift 
effects.  In this regard, all hydro models, as well as the UrQMD model, show satisfactory 
agreement. The maximum in the proton spectra in PHSD is at slightly higher $p_T$ than in 
the other models, indicative for a stronger radial flow at hadronization, while it is 
at slightly lower $p_T$ in the Catania transport model for 0-10\% centrality.

To shed more light on the somewhat unexpected range in the yields of the direct 
pions (and, to a lesser degree, of the direct protons), we collect in Tab.~\ref{tab_bulk} 
information on the inclusive yields, \ie, including strong and electromagnetic decay 
feeddowns, and/or on the total entropy per unit rapidity as available, at the respective 
ends of the QGP phase in the various bulk evolution models. In the table we also quote 
the experimentally measured values, which are, however, not necessarily to be understood 
as a precise benchmark for the model yields at $T_{\rm c}$ since further 
chemistry-changing processes (\eg, entropy production in both viscous-hydro and 
transport models, inelastic reactions in transport models), or a later chemical 
freezeout (\eg, $T_{\rm c}$=170\,MeV vs. $T_{\rm ch}$=160\,MeV in the TAMU model)
can affect the finally observable yields.  
The TAMU, OSU (used by LBL-CCNU and Duke) and CUJET hydro models are within a 10\% 
range of the inclusive pion numbers, while the POWLANG and Nantes models come out on the 
low end, with a $\sim$25\% smaller total entropy / inclusive pion number, respectively; 
this range is approximately consistent with the comparisons of the direct-pion spectra
discussed in Fig.~\ref{fig:pions}, although the TAMU results for the latter are trending 
somewhat lower given that is has the largest inclusive-pion numbers of the hydro models. 
For the transport models Catania and PHSD, where the direct-pion numbers in 0-10\% defined
the upper and lower ends of the range, the total entropy ($\sim$14000) and inclusive pion
number (722), respectively, are better aligned with the range defined by the hydro models. 
This reiterates that direct-pion numbers in transport models may not be a good bulk measure,
due to different hadronization mechanisms (\eg, coalescence or string fragmentation) which
generally do not result in chemical equilibrium abundances as implied by hydrodynamic models.  
In addition, inelastic processes in the transport through the hadronic phase may further 
modify the hadro-chemistry prior to kinetic freezeout. These considerations suggest that 
the overall discrepancies in the direct-pion numbers are reduced when comparing
more inclusive measures; the spread in the hydro models mostly originates from the
different entropy inputs (which may be the most relevant quantity thus 
far for characterizing the matter content of the QGP), while differences in the
hadro-chemistry (which is also affected by the resonance content of the hadron gas EoS) 
indice some variations when going from direct to inclusive pions. Further studies are 
required to further resolve these discrepancies and reduce the uncertainties related 
to the bulk evolution models.

The linear $y$-scale in the $p_T$ spectra emphasizes that more than 90\% of the 
pion (proton) yields are concentrated at rather low momenta, $p_T\lesssim 1.5(2.5)$\,GeV.
The low-$p_T$ particles are the relevant scattering partners for HF diffusion, and for the 
interpretation of the bulk $v_2$ that we turn to next. 

For the $v_2$ (lower panels in Figs.~\ref{fig:pions} and \ref{fig:protons}), 
the pion and proton curves are reasonably well collimated for semicentral
collisions for $p_T<2$\,GeV; the spread is larger for central collisions where the treatment
of initial-geometry fluctuations and initial-flow fields have a more significant
impact on the evolution of the overall smaller (and thus less robust) spatial eccentricity 
and its conversion into momentum eccentricity. At higher $p_T$, transport calculations and 
hydro models with viscosity exhibit a more pronounced levelling off than ideal hydrodynamic 
models; However, as mentioned above, at these momenta the
phase space density of the medium is suppressed by more than an order of magnitude
and thus not expected to play a significant role for interactions with heavy quarks.

\subsection{Charm-Quark Spectra with Common Transport Coefficient}
\label{ssec_com-trans}
As an initial test of how different bulk medium evolution models as employed by the various 
research groups affect the results for HF observables, calculations were carried out by the 
groups using their own evolution model but with a common pre-defined transport coefficient (for 
Langevin approaches) or pertinent cross section (for Boltzmann approaches). Specifically, pQCD
Born diagrams for elastic charm-quark scattering off thermal quarks, anti-quarks and gluons
were used, where, for example, the basic matrix element of $t$-channel gluon exchange
is given by 
\begin{equation}
{\cal M}_t \propto \frac{\alpha_s}{t-m_D^2}  \ .   
\end{equation}  
The coupling constant has been fixed at $\alpha_s=0.4$ (corresponding to $g$=2.24), the
Debye mass at $m_D=gT$, and thermal parton masses in the heat bath at $m_{\rm th}=gT$, 
assuming 3 light-quark flavors. For the charm-quark mass a constant value of 
$m_{\rm c}$=1.5\,GeV has been used, and an overall $K$ factor of 5 was applied to the 
squared matrix elements (numerical tables for the pertinent HQ transport coefficients 
are available from  the HF-RRTF repository~\cite{emmi-repo}). in the following, we
refer to this interaction as ``pQCD*5".
The resulting  spatial HQ diffusion coefficient amounts to ${\cal D}_s (2\pi T) \simeq 6$ 
at $T$=300\,MeV, with a weak temperature dependence. 
Furthermore, in the Langevin approaches, a uniform implementation of the Einstein
relation was adopted, with friction ($A$) and transverse diffusion ($B_0$) coefficients 
as calculated from the pQCD scattering matrix elements  and the longitudinal one adjusted 
to $B_1=TEA$ to ensure the correct equilibrium limit ($E=\sqrt{m_c^2+p^2}$ is the on-shell
$c$-quark energy).

\begin{figure}[!t]
\hspace{-0.2cm}
\includegraphics[width=\linewidth,height=13cm]{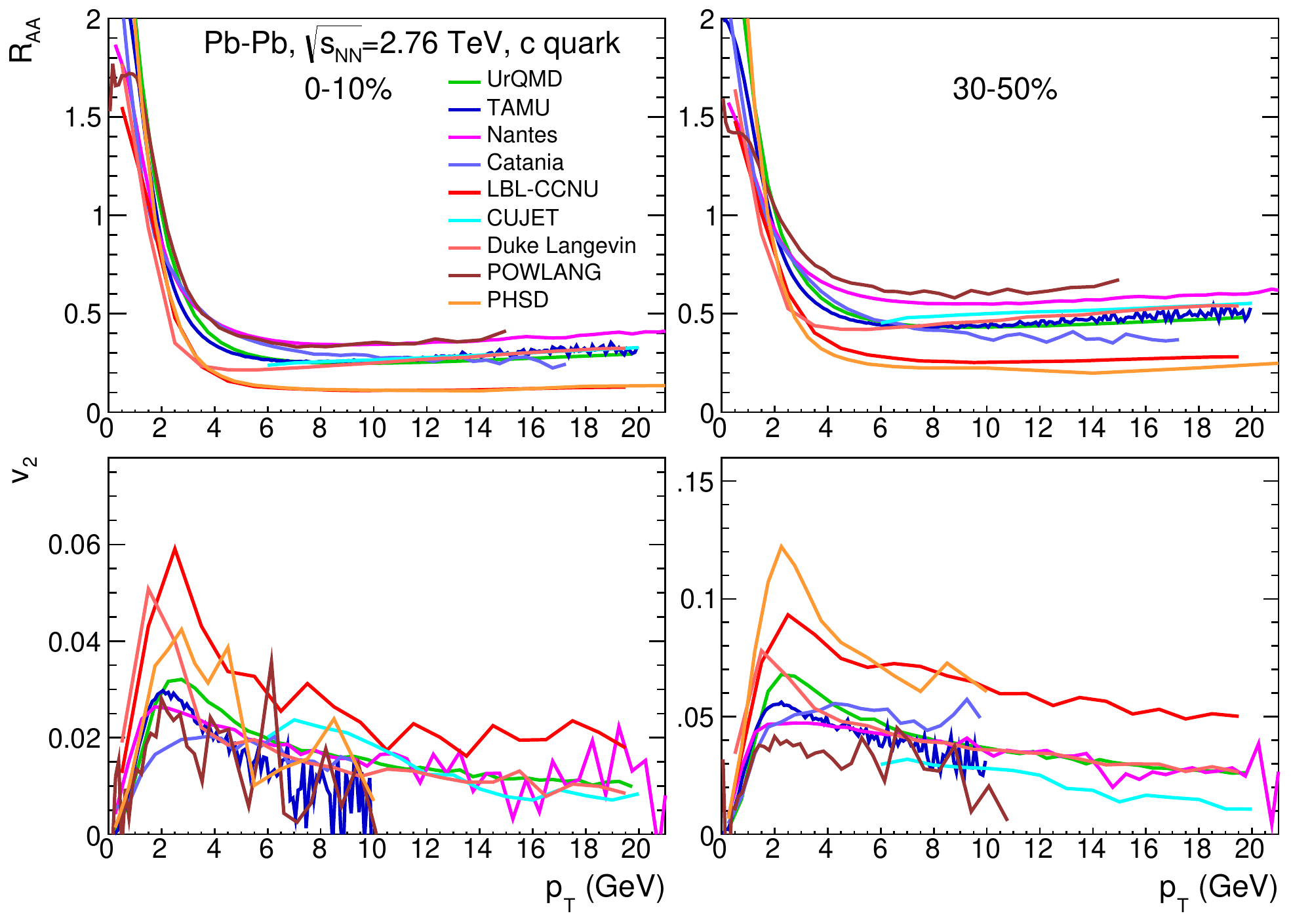}
\caption{Comparison of the nuclear modification factor (upper panels) and elliptic flow (lower 
panels) of charm quarks in 0-10\% (left panels) and 30-50\% (right panels) Pb-Pb(2.76\,TeV) 
collisions using a common charm-quark pQCD*5 interaction in the QGP (transport coefficient 
for Langevin diffusion or cross section for Boltzmann transport implementations). The different
model approaches are identified in the legends and detailed in the text. 
}
\label{fig:c-quark}
\end{figure}
Within the above bulk models, the charm-quark $R_{AA}$ and $v_2$ have been evolved through 
the QGP phase of 0-10\% and 30-50\% Pb-Pb(2.76\,TeV) collisions, with initial charm-quark 
spectra from $pp$ collisions without CNM effects (such as shadowing or Cronin effect). 
The results recorded at the end of the QGP phase are collected in Fig.~\ref{fig:c-quark}. 

All $R_{\rm AA}$'s (upper panels of Fig.~\ref{fig:c-quark}) are reasonably well collimated 
in the fall-off region around $p_t\simeq2$\,GeV where they pass through one; this is largely 
a consequence of charm-quark number conservation, as the total yield is mostly concentrated 
around this value of transverse momentum. This somewhat limits the discrimination power 
of the low-$p_t$ $c$-quark $R_{\rm AA}$
(this situation will much improve at the $D$-meson level).
At higher $p_t$, all calculations level off for $p_t\gtrsim6$\,GeV, for most models in a 
reasonably collimated range of $R_{\rm AA}\simeq$0.3-0.4 and 0.4-0.6 for 0-10\% and 30-50\%
centrality, respectively. Notable outliers are the LBL-CCNU Boltzmann transport model and the 
PHSD transport model. In the former case, this can be understood as being due to the use
of massless thermal partons in the bulk evolution, which implies a significantly larger 
number of scatterers at a given temperature compared to the massive thermal partons used
in the calculation of the pQCD*5 transport coefficients (conversely, if the transport 
coefficients are calculated with massless thermal partons it typically increases the 
low-momentum thermalization rate a factor of up to $\sim$2). In the case of PHSD, we note 
that the pQCD*5 interaction is implemented via charm-quark Born scattering diagrams off the 
bulk partons as used in the transport model (which are not necessarily in chemical and 
thermal equilibrium and generally differ in mass from the value of $gT$~\cite{Xu:2017pna}), 
with a momentum- and temperature-dependent $K$-factor to match 
the pQCD*5 friction coefficient in the equilibrium limit (which may also entail differences
in the transverse and longitudinal diffusion coefficients). Also recall that the PHSD bulk 
evolution is the one with the largest inclusive pion number in central collisions. On the other 
hand, the POWLANG and Nantes results, which are among the ones with the weakest medium 
modifications (\ie, relatively large $c$-quark $R_{\rm AA}$ and small $v_2$), are obtained 
from the hydro evolutions with the smallest entropy content; this qualitative consistency is 
quite encouraging. Also to be kept in mind are the varying treatments of the initial conditions 
(\eg, with or without transverse flow, the initial transverse energy density distribution 
and its possible fluctuations), for which a more in-depth analysis is left to future work. 
For central collisions, the UrQMD, TAMU, Nantes, Catania, CUJET, Duke and POWLANG $R_{\rm AA}$'s 
for $p_t$=10-20\,GeV, are all within a range of $\pm0.05$ around 0.3-0.35, with a slight upward 
trend for most, which is quite encouraging. For semi-central collisions POWLANG and Nantes, on
the one hand, and Catania, on the other hand, lie somewhat above and below, respectively, a 
$\pm$0.05 range of the other models.  
These deviations can at least in part be understood by the factor of $\sim$2.5 difference of
the total entropies in these calculations, being at the low (POWLANG, Nantes) and high 
(Catania) end of the various evolution models, recall Table~\ref{tab_bulk}.

For the $v_2$, displayed in the lower panels of Fig.~\ref{fig:c-quark}, the sensitivity is 
somewhat limited for 0-10\% centrality due to the relatively small signal and appreciable 
statistical fluctuations in the calculations. For the 30-50\% centrality, the stronger impact 
found for the bulk media in the PHSD and LBL-CCNU models in the $R_{\rm AA}$ is also reflected 
in the larger $v_2$ within these approaches, reaching a maximum of more $\ge$9\% while all other 
calculations mostly lie within a $\sim$4-6\%, with DUKE and UrQMD exhibiting slightly more
pronounced (\ie, narrow) maxima near 7\% but subsequently leveling off well within the 
common trend. These systematics proceed out to higher $p_t$, with the main band largely 
bracketed by Catania (upper part) and POWLANG (lower part), correlating well with the 
observations for the $R_{\rm AA}$. 

\begin{figure}[!t]
\begin{tabular}{cc} 
\begin{minipage}{.5\textwidth}
\hspace{-.5cm}{\includegraphics[width=9cm,height=7cm]{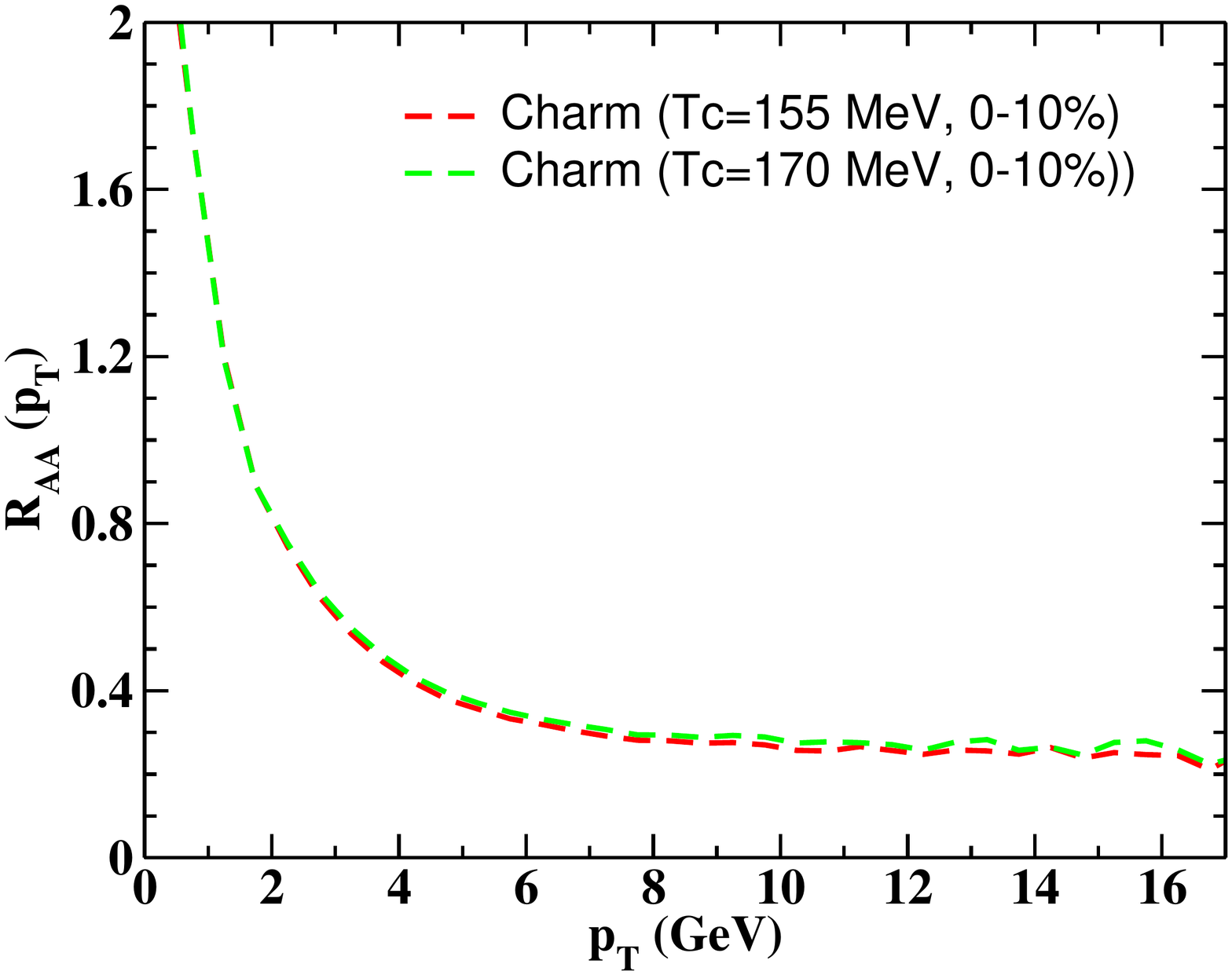}}
\end{minipage} & 
\begin{minipage}{.5\textwidth}
\hspace{-.5cm}{\includegraphics[width=9cm,height=7cm]{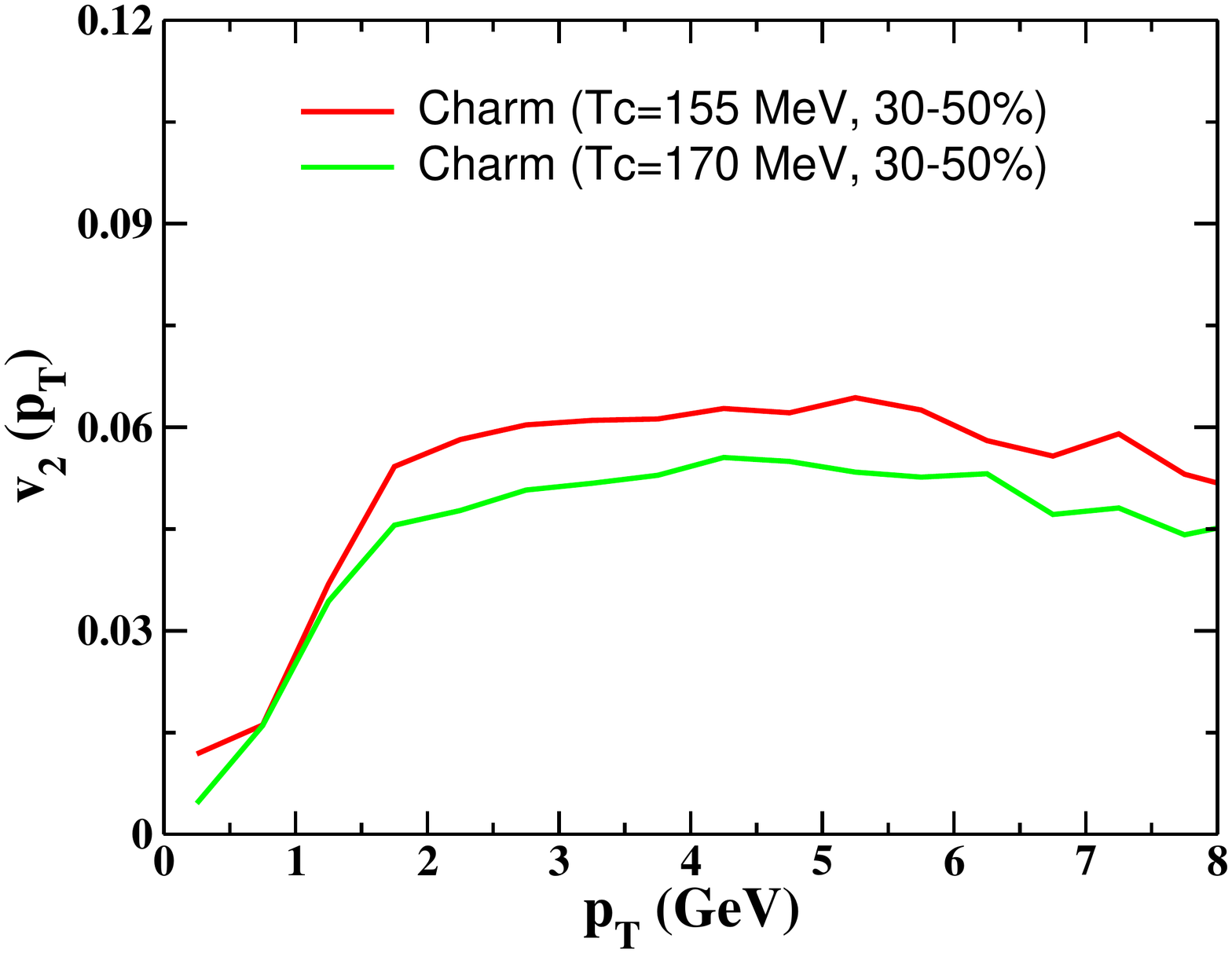}}
\end{minipage} 
\end{tabular}
\caption{Comparison of the charm-quark nuclear modification factor in 0-10\% (left panel) and 
elliptic flow in 30-50\% (right panel) Pb-Pb(2.76\,TeV) collisions computed with the pQCD*5 
interaction within the Catania Langevin transport approach when varying the final temperature of 
the QGP evolution from the default value of $T_{\rm c}$=155\,MeV (red lines) to $T_{\rm c}$=170\,MeV
(green lines).}
\label{fig:Tc-study}
\end{figure}
One of the differences in the bulk evolution models relevant to HF observables is
the temperature $T_{\rm c}$ where the QGP evolution is assumed to end. To illustrate this 
uncertainty, we show in Fig.~\ref{fig:Tc-study} the impact of a variation of this quantity
on the final $c$-quark spectra. Concretely, within the Catania transport approach, the results
for the default value of $T_c$=155\,MeV are compared to terminating the QGP evolution at
$T_{\rm c}$=170\,MeV, as used, \eg, in the TAMU hydro evolution. It turns out that the
$R_{\rm AA}$ is affected little, while the $v_2$ picks up an appreciable contribution
of up to 20\% when the evolution is run to the lower temperature. This result is consistent
with previous studies~\cite{Rapp:2008zq,Das:2015ana} where the suppression figuring in the 
$R_{\rm AA}$ is 
identified a density-driven effect which is most effective in the earliest phases of the 
fireball, while the transfer of $v_2$ from the medium to the heavy quark is most effective when 
the fireball $v_2$ is large which is primarily in the later phases of the evolution, closer 
to $T_{\rm c}$. This effect is further augmented when the coupling strength of the medium 
is largest near $T_{\rm c}$ (including coalescence processes), for which initial evidence was 
already deduced from the first PHENIX HF electron data~\cite{Adare:2006nq,vanHees:2005wb}. The 
pQCD*5 interaction underlying the studies in this section does not feature an enhanced strength
near $T_{\rm c}$, and therefore the increase of the HQ $v_2$ near $T_{\rm c}$ is expected to be
more pronounced for nonperturbative interactions.

\begin{figure}[!t]
\begin{tabular}{cc}
\begin{minipage}{.5\textwidth}
\hspace{-.3cm}{\includegraphics[width=8.5cm,height=6.5cm]{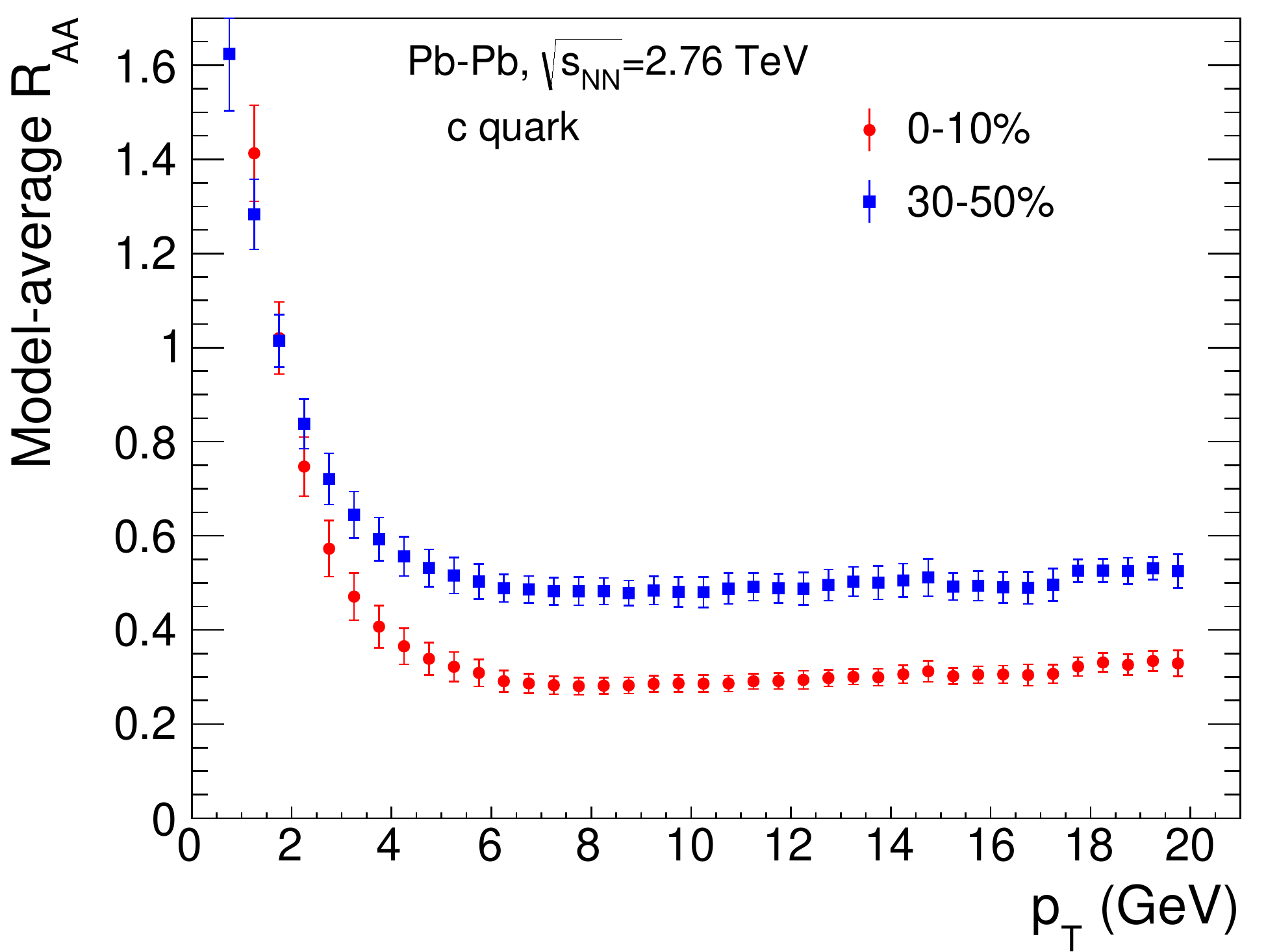}}
\end{minipage} &
\begin{minipage}{.5\textwidth}
\hspace{-.5cm}{\includegraphics[width=8.5cm,height=6.5cm]{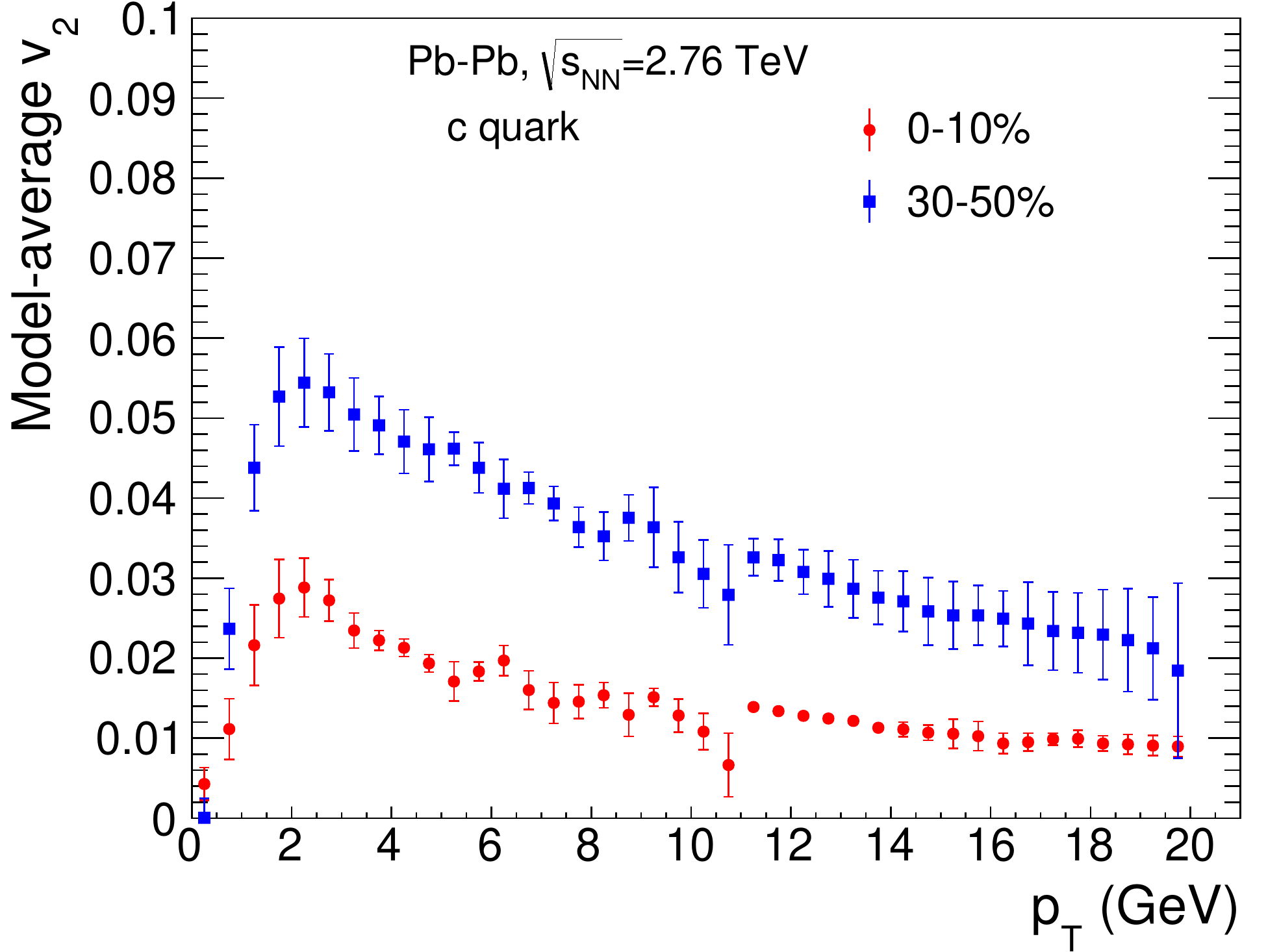}}
\end{minipage}
\end{tabular}
\caption{Statistical average of the transport models calculations using the (elastic) pQCD*5 
interaction for charm-quark nuclear modification factor (left panel) and elliptic flow (right panel) 
at the end of the QGP phase in 0-10\% (red dots) and 30-50\% (blue squares) Pb-Pb(2.76\,TeV) 
collisions.}
\label{fig:c-avg}
\end{figure}
We finish this section by performing a statistical average of the above discussed model
calculations for the $c$-quark $R_{\rm AA}$ and $v_2$, cf.~Fig.~\ref{fig:c-avg} (for reasons 
mentioned earlier we do not include the LBL-CCNU and PHSD results in the averages).
The result of this procedure suggests a roughly $\pm$10\% uncertainty due to the different 
bulk evolution models for the QGP phase (somewhat larger for the high-$p_T$ $v_2$).

\section{Hadronization} 
\label{sec_hadro}

The hadronization mechanism of heavy quarks into heavy mesons and 
baryons~\cite{Lin:2003jy,Greco:2003vf} 
in heavy-ion collisions has been established as an important ingredient to the 
phenomenology of the observed heavy-flavor $R_{\rm AA}$ and $v_2$ at both 
RHIC~\cite{vanHees:2005wb,Adare:2006nq} and the LHC~\cite{Abelev:2014ipa,He:2012df}.
As such it is critical to scrutinize the different theoretical treatments of this modeling 
component. In the following section (\ref{ssec_hadro-com}) we first compare the impact of 
the various hadronization mechanisms from the literature as applied in current model 
approaches to the charm-quark spectra in Pb-Pb(2.76\,TeV) collisions as computed with the 
common pQCD*5 transport coefficient in Sec.~\ref{ssec_com-trans}. We then elaborate on 
different ways of implementing heavy-light quark coalescence by directly comparing several 
approaches applied to the same input charm-quark spectrum within the same bulk medium 
background (temperature and flow field) and critically inspect the effects on the resulting 
$D$-meson $p_T$ spectra and $v_2$ in Sec.~\ref{ssec_hadro-spec}). Finally we discuss an 
alternative for in-medium hadronization based on a fragmentation scheme with surrounding 
medium partons in Sec.~\ref{ssec_in-med-frag}).

\subsection{Comparison of $D$-Meson Spectra from Common Transport Coefficients}
\label{ssec_hadro-com}
Let us start the discussion of charm-quark hadronization by briefly outlining its implementation
by the various groups.
\begin{itemize}
\item UrQMD: 
Instantaneous coalescence model (ICM) in momentum space~\cite{Greco:2003vf} with massive
light quarks ($m_q$=0.37\,GeV), with spatially uniform light-quark distributions but including 
their collective flow, supplemented with Peterson fragmentation for left-over $c$-quarks.
\item TAMU: 
Resonance recombination model (RRM)~\cite{Ravagli:2007xx}, based on a rate 
from the Boltzmann equation with resonant $c+q\to D$ interaction on the hydrodynamic 
hypersurface (including the spatial dependence of flow fields) at $T_{\rm c}$~\cite{He:2011qa} 
with massive light quarks ($m_q$=0.3\,GeV) including effects of hadro-chemistry, 
supplemented by FONLL fragmentation.
\item Nantes: 
ICM with light quarks of mass 
$m_q$=0.1\,GeV including spatial dependence of local flow fields, supplemented
with fragmentation~\cite{Gossiaux:2009mk}.
\item Catania: 
ICM in momentum space~\cite{Greco:2003vf} with massive
light quarks ($m_q$=0.33\,GeV) including global transverse flow, with spatially uniform 
light-quark distributions, supplemented with Peterson fragmentation.
\item LBL-CCNU and Duke:
ICM in momentum space~\cite{Greco:2003vf} with thermal light-quark distributions with 
mass $m_q$=0.3\,GeV and transverse-flow 
effects simulated through an effective temperature, including effects of hadro-chemistry, 
supplemented with PYTHIA fragmentation~\cite{Cao:2013ita}.
\item CUJET: 
Fragmentation only using the perturbative BCFY scheme~\cite{Braaten:1994bz}, 
cf.~eq.~(\ref{pqcd1}), with the input $c$-quark constructed in Sec.~\ref{ssec_baseline}
(in the original CUJET, harder input spectra~\cite{Nelson:2012bc} are combined with 
softer Peterson fragmentation ($\epsilon$=0.06), resulting, however, in very similar 
$D$-meson spectra in $pp$).
\item POWLANG:
In-medium fragmentation which includes string formation with thermal partons in local 
restframe of the expanding medium, followed by PYTHIA fragmentation (cf.~also 
Sec.\ref{ssec_in-med-frag} for further details)~\cite{Beraudo:2014boa,Beraudo:2015wsd}. 
\item PHSD: ICM Wigner functions in coordinate and momentum space, gradually hadronized in time 
based on a classical diffusion argument~\cite{Song:2016lfv}, with stochastic sampling of the local bulk 
environment with thermal parton masses ($m_q$$\simeq$0.31\,GeV) and including higher $D$-meson 
resonance excitations, supplemented by fragmentation~\cite{Song:2015sfa,Song:2015ykw}.
\end{itemize}

\begin{figure}[!t]
\hspace{-0.0cm}{\includegraphics[width=0.99\textwidth]{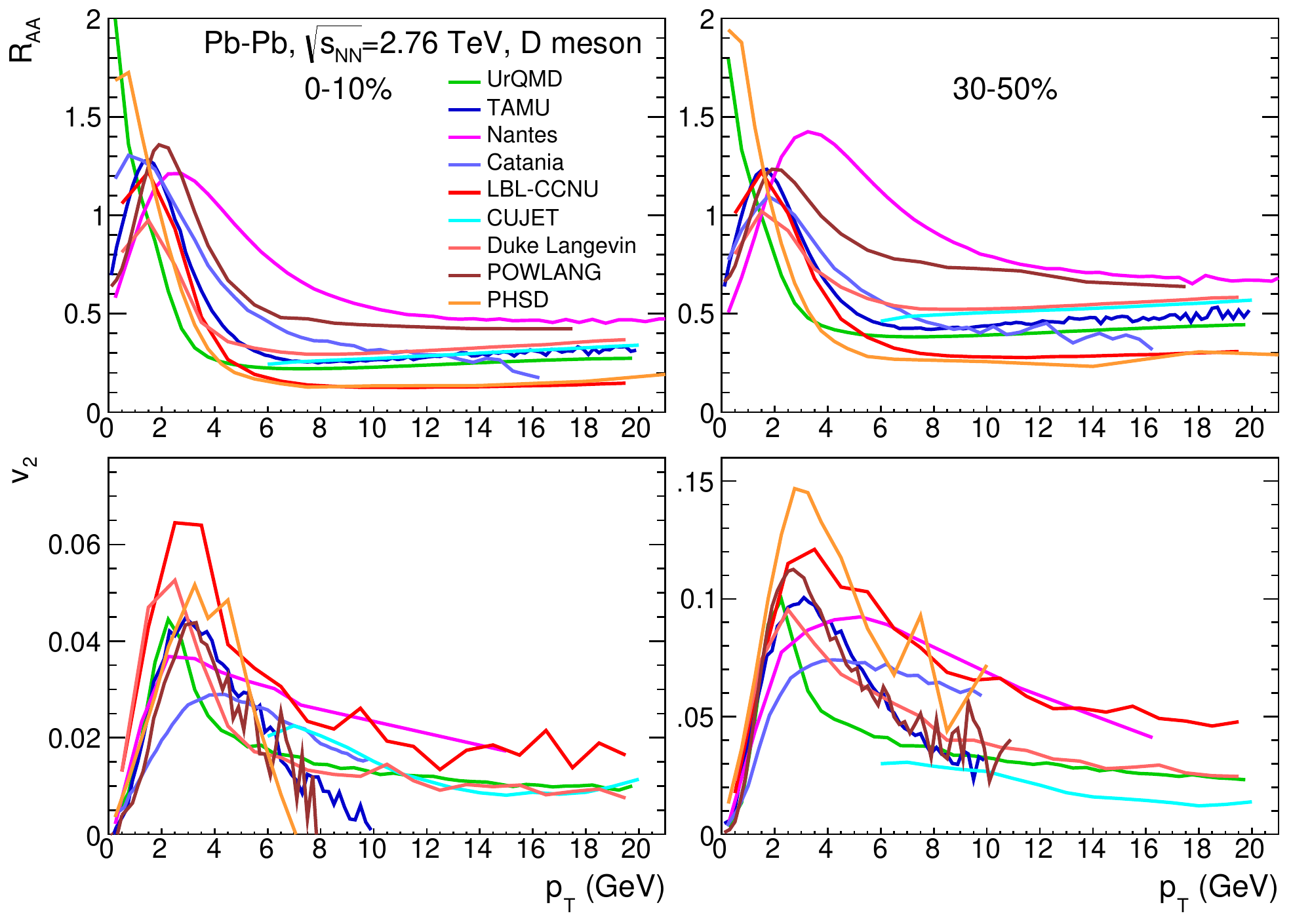}}
\caption{Comparison of the nuclear modification factor (upper panels) and elliptic flow
(lower panels) of $D$ mesons right after the hadronization transition, as obtained from
the $c$-quark spectra from the different evolution models with the elastic pQCD*5 transport
coefficient (displayed in Fig.~\ref{fig:c-quark})
for 0-10\% (left panels) and 30-50\% (right panels) Pb-Pb(2.76\,TeV) collisions.
}
\label{fig:D-meson}
\end{figure}
In Fig.~\ref{fig:D-meson} we summarize the results of the $D$-meson $R_{\rm AA}$ (upper 
panels) and $v_2$ (lower panels)
as they follow from the hadronization schemes described above applied to their respective 
charm-quark outputs with the pQCD*5 interaction (shown in Fig.~\ref{fig:c-quark}).
For the implementations within UrQMD (ICM Wigner functions with thermal quark masses 
of 370\,MeV or more) and PHSD (ICM Wigner functions with finite times estimated from
momentum diffusion), the $D$-meson $R_{\rm AA}$'s do not develop 
significant maximum structures at low $p_T$ (``flow bumps"). A flow bump does develop
for the Duke/LBL-CCNU coalescence model, for the POWLANG in-medium fragmentation scheme, 
for the RRM in TAMU, and most prominently for the Nantes implementation with comparatively 
small light-quark masses and space-momentum correlations accounted for. In the latter 
scheme the coalescence contribution penetrates  
out to rather high $p_T$$\simeq$12-14\,GeV, notable as an enhancement over the pertinent 
$c$-quark $R_{\rm AA}$. This is significantly further out in $p_T$ than in other 
implementations where coalescence effects cease above $p_T$$\simeq$6\,GeV, and thus 
their ordering in suppression at the $c$-quark level is preserved at the $D$-meson level 
(\ie, it is little affected by independent fragmentation). We also remark that the 
$D$-meson $R_{\rm AA}$'s are not necessarily 
norm-conserving, even though the $c$-quark number is conserved; this is due to the 
``chemistry effect" as included by Duke/LBL-CCNU/PHSD (in principle also for TAMU, but 
for clarity not in the present calculation), where, \eg, an 
increase in the $D_s/D$ or $\Lambda_c/D$ ratio in AA relative to $pp$ collisions requires 
a decrease in other charm-hadron species.  

The strong coalescence effect in the Nantes model also shows up in the $D$-meson $v_2$,
roughly doubling the maximum value of the $c$-quark $v_2$ for 30-50\% centrality while 
preserving a rather gradual decrease with $p_T$. The increase in the maximum $D$-meson 
$v_2$ over the $c$-quark one is comparable for the RRM employed by TAMU, 
but this enhancement fades away 
more rapidity in $p_T$ recovering the $c$-quark values for $p_T\gtrsim6$\,GeV, as is the
case for the corresponding $R_{\rm AA}$. The low-$p_T$ increase of the $D$-meson $v_2$ in the
ICMs of UrQMD, LBL-CCNU and PHSD is up to 3\% in absolute value, 
\ie, 30-40\% in relative magnitude; one also finds a broadening of the rather narrow maximum 
structure for the Duke $c$-quark $v_2$. The impact of coalescence on the $D$-meson $v_2$ in 
0-10\% central Pb-Pb collisions is relatively less pronounced, presumably due to the overall 
much smaller bulk-$v_2$ that can be imprinted on the forming $D$-meson.     

\begin{figure}[!t]
\begin{tabular}{cc} 
\begin{minipage}{.5\textwidth}
\centerline{\includegraphics[width=0.95\textwidth]{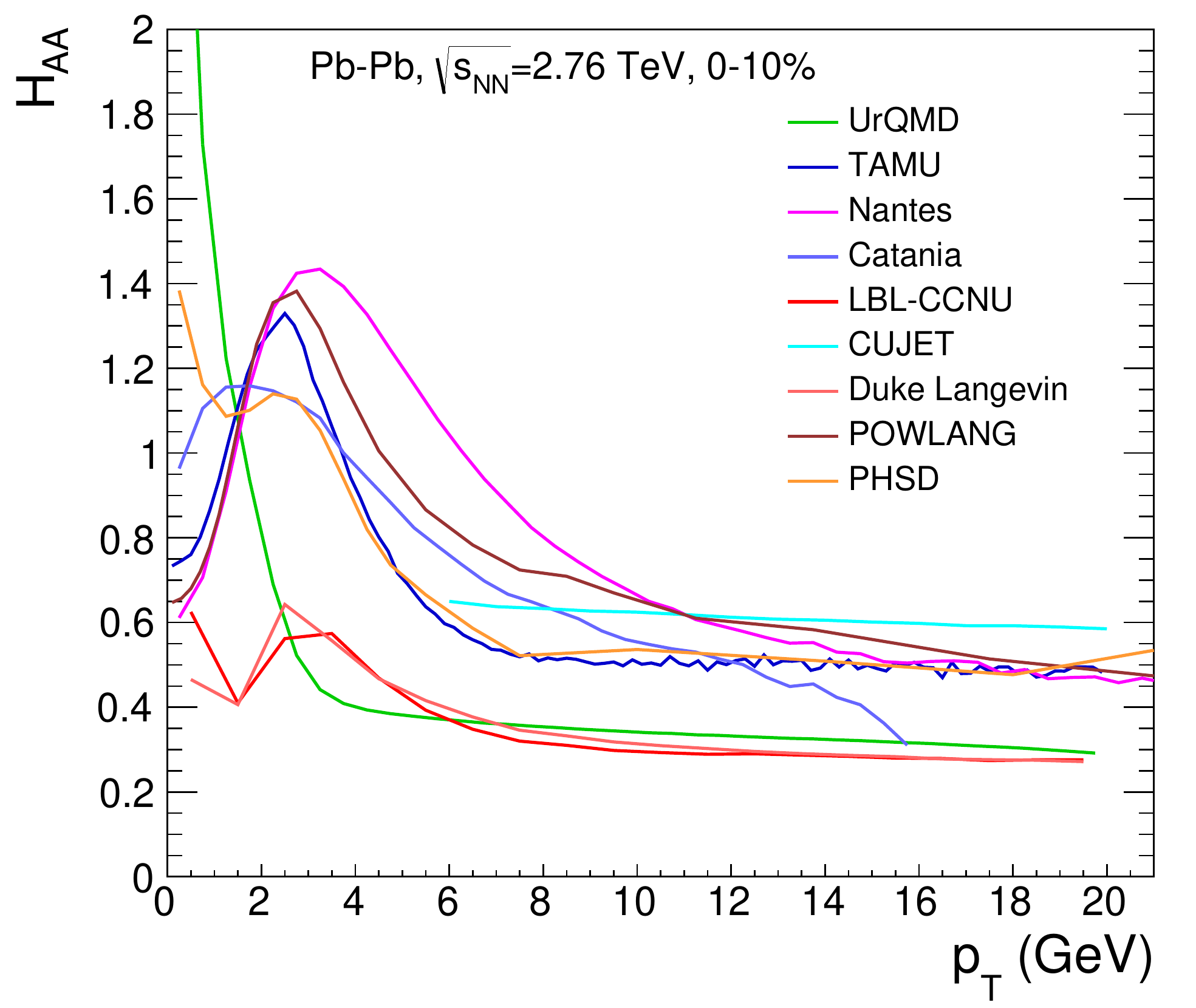}}
\end{minipage} 
\begin{minipage}{.5\textwidth}
\centerline{\includegraphics[width=0.95\textwidth]{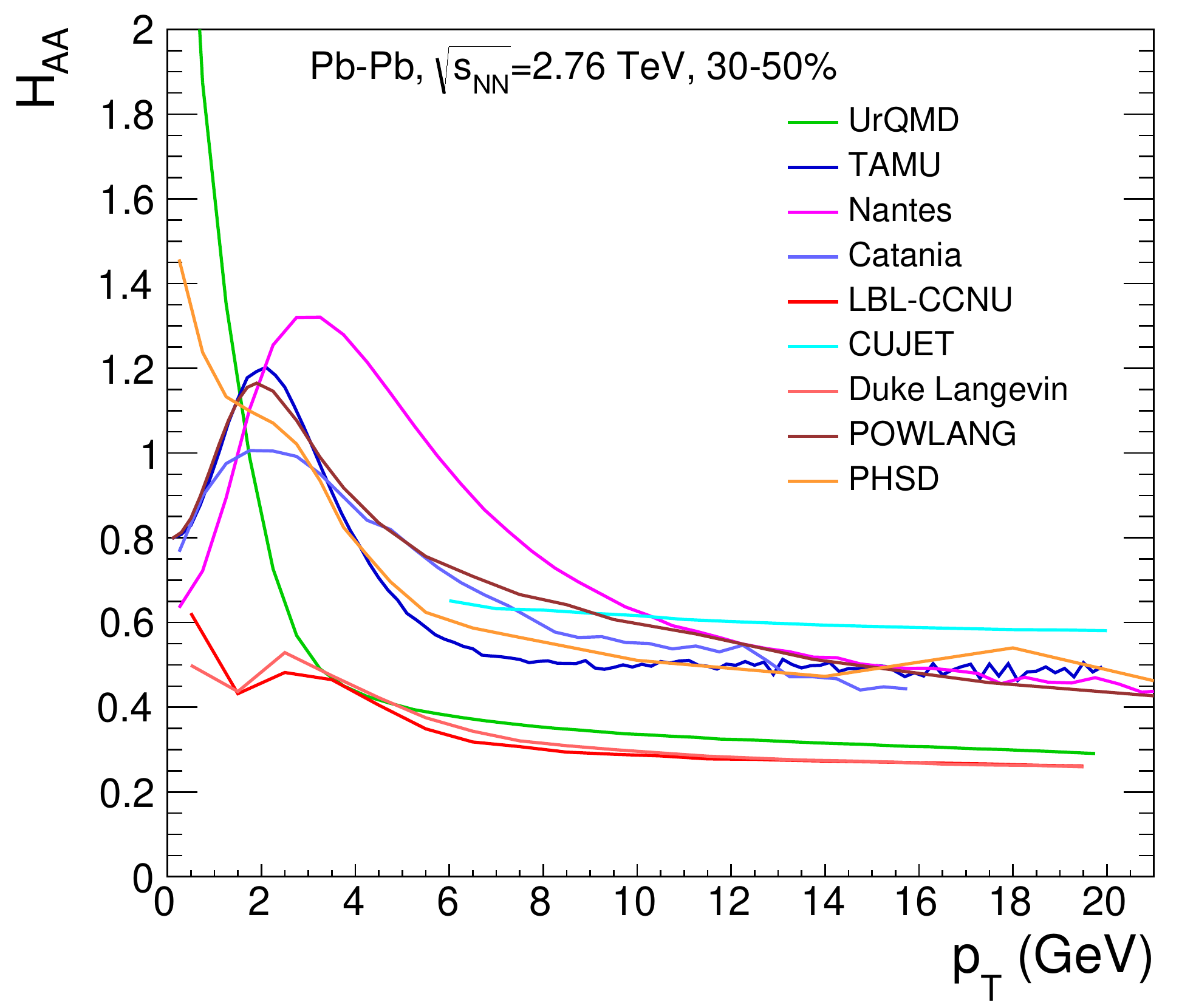}}
\end{minipage}
\end{tabular}
\caption{Comparison of the ratio of $D$-meson to charm-quark $p_T$ spectra, $H_{\rm AA}(p_T=p_t)$,
just after and before hadronization, respectively, in central (left panel) and semi-central (right panel) 
Pb-Pb(2.76\,TeV) collisions for the results from the elastic pQCD*5 QGP transport simulations and 
individual hadronization procedures within the various bulk evolution models.}
\label{fig:haa}
\end{figure}
In an attempt to more directly exhibit the effects of hadronization we introduce the quantity
\begin{equation}
H_{\rm AA} (p_T, p_t=p_T)= \frac{\ud N_D/\ud p_T}{\ud N_c/\ud p_t} =
    \frac{\ud N_D^{\rm coal}/\ud p_T+\ud N_D^{\rm frag}/\ud p_T}{\ud N_c/\ud p_t} \ ,
\end{equation}
evaluated at the same transverse momentum of the $c$-quark ($p_t$) and the $D$-meson
($p_T$); the pertinent ratios from Figs.~\ref{fig:c-quark} and \ref{fig:D-meson} are plotted 
in Fig.~\ref{fig:haa}. In the absence of coalescence effects this ratio simply characterizes
the independent fragmentation function. Correspondingly, at high $p_T$ one finds that the 
different approaches essentially level off in two regimes representing the different 
fragmentation functions, \ie, CUJET (FONLL/BCFY),  Nantes ($p_T=zp_t$), 
Catania (Peterson with $\eps_c$=0.04), PHSD (PYTHIA tuned to FONLL) and 
TAMU (FONLL) vs. LBL-CCNU/Duke (PYTHIA6.4) and UrQMD (Peterson with $\eps_c$=0.05). 
The $H_{\rm AA}$ more clearly exhibits shifts of $c$-quarks to higher $p_T$ in the low- and
intermediate-$p_T$ regime, with marked ``flow bumps" developing for TAMU, Nantes and POWLANG
(as seen before in the $D$-meson $R_{\rm AA}$), and smaller ones for LBL-CCNU/Duke and PHSD. 
All of them are more pronounced for central collisions, as expected.

\begin{figure}[!t]
\centerline{\includegraphics[width=0.97\textwidth]{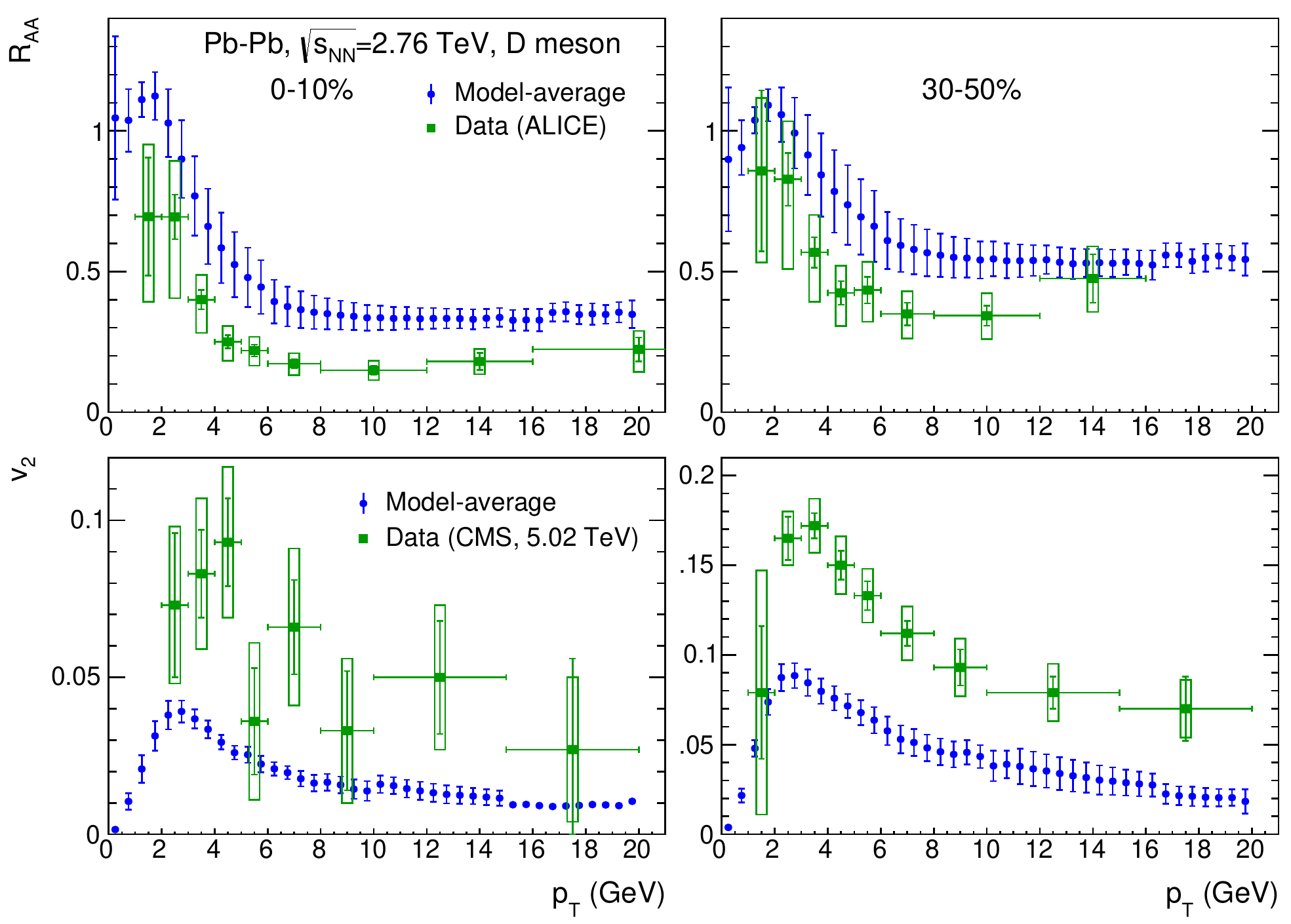}}
\caption{Statistical average of the transport models calculations for $D$-meson nuclear
modification factor (upper panels) and elliptic flow (lower panels) after $c$-quark 
diffusion through the QGP with the elastic pQCD*5 interaction and subsequent hadronization in 
0-10\% (left column) and 30-50\% (right column) Pb-Pb(2.76\,TeV) collisions. The theoretical
``averages" are compared to ALICE data~\cite{Adam:2015sza} for the $R_{\rm AA}$ and CMS 
data\cite{Sirunyan:2017plt} for the elliptic flow (the latter are for Pb-Pb(5.02\,TeV) 
collisions).} 
\label{fig:D-avg}
\end{figure}
Paralleling the charm-quark case, we finish this section by performing a statistical average 
over the $D$-meson $R_{\rm AA}$'s and $v_2$ resulting from the pQCD*5 diffusion calculations 
with different bulk and hadronization models, cf.~Fig.~\ref{fig:D-avg} (as before, LBL-CCNU 
and PHSD results are not included in the averages).
Relative to the $c$-quark case in Fig.~\ref{fig:c-avg}, the most significant increase in the 
percentage uncertainty occurred in the low- and intermediate-$p_T$ region of the $D$-meson 
$R_{\rm AA}$ where the radial flow effect from coalescence processes is most prominent. 
The absolute error also increases in the low- and intermediate-$p_T$ elliptic flow, but since 
the overall signal increases substantially (by $\sim$50\% or more), the relative error did 
not change much. At high $p_T$ where fragmentation prevails, the values and uncertainty in the 
$R_{\rm AA}$ are little affected, while an additional spread is added to the small signal in 
the $v_2$, in part also due to the statistical fluctuations in (some of) the individual 
calculations.  

Let us use the results shown in Fig.~\ref{fig:D-avg} for a preliminary comparison to pertinent 
experimental data~\cite{Adam:2015sza,Sirunyan:2017plt}\footnote{Note that the CMS $v_2$ data 
are for a collision energy of 5.02\,TeV while the calculations are for 2.76\,TeV. This choice
was made based on the higher precision of these data than available ones for 2.76\,TeV, and the 
fact that both calculations and experimental data show little variations in $v_2$ and $R_{\rm AA}$ 
observables when going from 2.76 to 5.02\,TeV. Quantitative comparisons in the future will of course
have to be made at the same energy for theory and data, while the current experimental accuracy 
attained for the $D$-meson $v_2$ at 5.02\,TeV can already give an indication of its constraining 
power.}, in an attempt to assess the heavy-quark transport coefficient. For the nuclear modification
factor, especially for central collisions, the calculations overestimate the experimental 
$R_{\rm AA}$ significantly, \ie, well beyond both error bands, by nearly a factor of 2 for the 
central values (somewhat less for semi-central collisions). At low $p_T$, shadowing is likely to 
play a role in that, but for $p_T\gtrsim5$\,GeV, the studies in Sec.~\ref{sec_initial} imply that the 
suppression is largely due to the hot medium effects where the theoretical averages with the pQCD*5 
interaction remain well above the data. At lower $p_T$, the cleaner observable to gauge the interaction
strength is the $v_2$, which should therefore be a major focus for future precision measurements. 
The calculational average for the pQCD*5 transport coefficient, reaching up to $v_2\simeq$9\% for 30-50\%
centrality, is well below experimental values, again by about a factor of two over most of the $p_T$ 
range out to 20\,GeV. The shape of the theoretical curves is similar to that of the data, but this 
might be a coincidence as different (not mutually exclusive) mechanisms could be responsible for the 
discrepancy (\eg, missing coupling strength to the collective medium at low $p_T$ and elliptic-flow 
fluctuations at high $p_T$). The underestimate of the low-momentum $v_2$ makes the substantial lack 
of interaction strength of the schematic pQCD*5 model especially apparent, as no mechanisms other
than the HQ coupling to the collectively expanding medium are readily conceivable to generate a
large anisotropy at low momentum. Even this preliminary data comparison, with a simple HQ interaction 
in the QGP, demonstrates that the HQ diffusion coefficient, ${\cal D}_s(2\pi T)$, in QCD matter must be 
{\em significantly} smaller than 6 as underlying the calculations in Sec.~\ref{ssec_com-trans}, at 
least for some temperature
region, preferentially where the $v_2$ of the bulk medium is large. The results also indicate that
the theoretical error is controllable and ultimately quantifiable.

The experimental handle on the recombination mechanisms of heavy quarks is likely to
be augmented by the measurement of heavy-strange mesons~\cite{Adam:2015jda,Nasim:2018mdl},
\ie, $D_s=(c\bar s)$ and $B_s=(b\bar s)$ (or even more elusive multi-HQ hadrons, \eg,
$B_c$~\cite{Schroedter:2000ek} or $\Xi^{++}$~\cite{Zhao:2016ccp}). The main idea
is~\cite{He:2012df,Kuznetsova:2006bh,Andronic:2006ky} that the well-established
enhancement of strange quarks in URHICs,
relative to pp collisions, will increase the yields and quantitatively affect the
$p_T$ spectra and elliptic flow of the charm-strange mesons in the presence of
recombination of charm quarks within a heat bath of strange quarks. To render this
a quantiative probe, good control over the recombination mechanism, including the
equilibrium limit and deviations from it, is required. In the following two section,
we scrutinize some of recombination models which have been used in the HF sector to
date.

\subsection{Recombination in Thermal Medium}
\label{ssec_hadro-spec}
In this section we carry out two comparisons to provide more explicit insights into (some 
of) the recombination schemes that are being employed in the approaches discussed above. 
Specifically, we compare the ``standard" implementation of the 
ICM with the RRM using the $c$-quark spectra from the  pQCD*5 Langevin simulation in the
TAMU hydro background, where, for simplicity, we neglect space-momentum correlations
inherent in the RRM. 

Early applications of quark coalescence processes in heavy-ion collisions have been carried 
out with a spatially uniform (``global") distribution functions in 3-momentum space, amounting 
to an instantaneous approximation (see Ref.~\cite{Fries:2008hs} for a review). This allowed 
for a successful description of the hadron-$v_2$ and baryon-over-meson ratios in the light- 
and strange-quark sector in the intermediate-$p_T$ region at RHIC. However, this approximation 
does not conserve energy in the 2$\to$1 hadron formation process and thus cannot recover 
the equilibrium limit of the hadron distributions. 
In Ref.~\cite{Ravagli:2007xx} a resonance recombination model (RRM) has been developed, 
where resonant quark-anti-quark scattering amplitudes are used within a Boltzmann equation, 
which remedies both energy conservation and the equilibrium limit. It has been implemented 
in the heavy-quark context on a hydrodynamic hypersurface in Ref.~\cite{He:2011qa}.

\begin{figure}[!t]
\begin{minipage}{.5\textwidth}
\centerline{\includegraphics[width=1.2\textwidth]{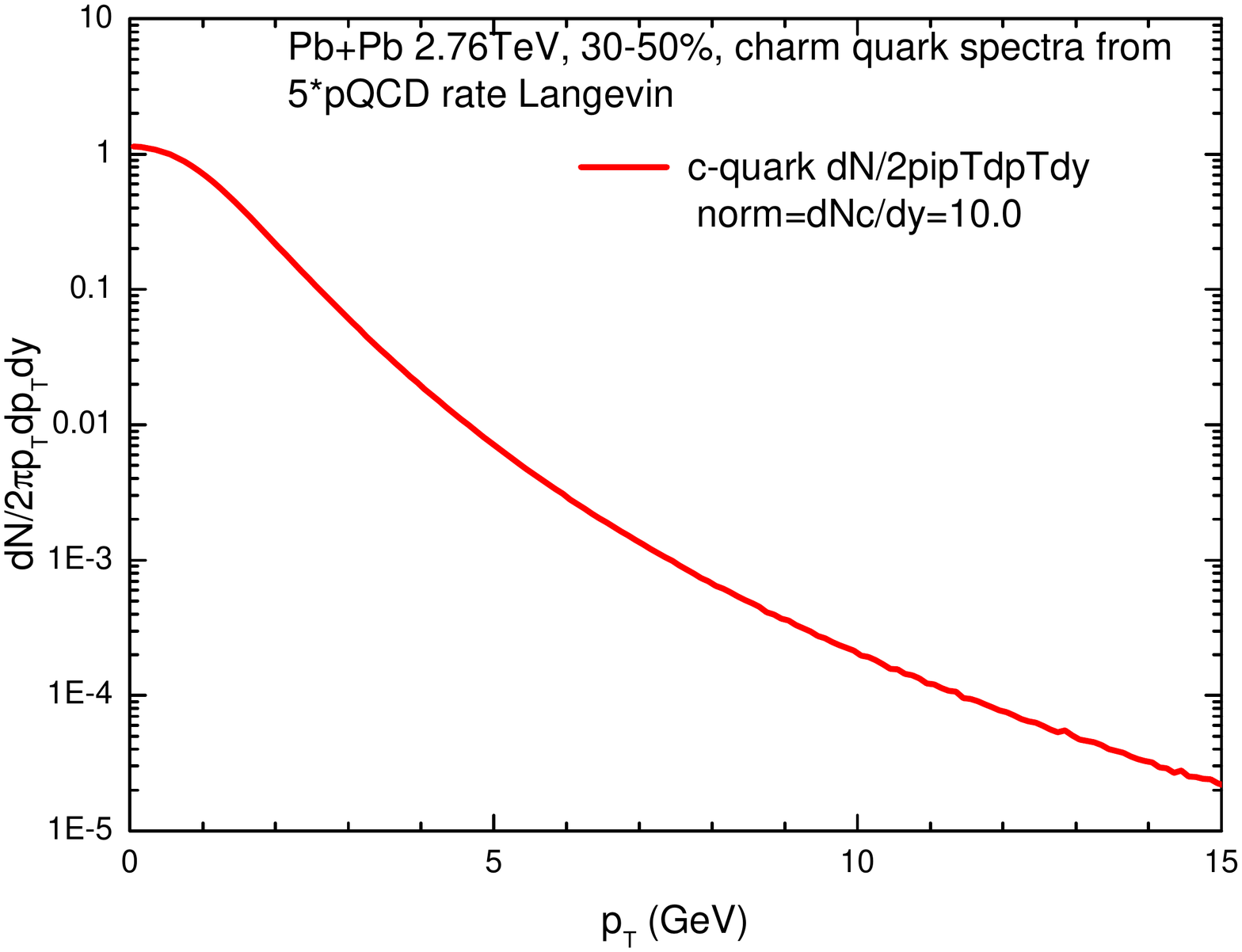}}
\end{minipage}
\begin{minipage}{.5\textwidth}
\centerline{\includegraphics[width=1.2\textwidth]{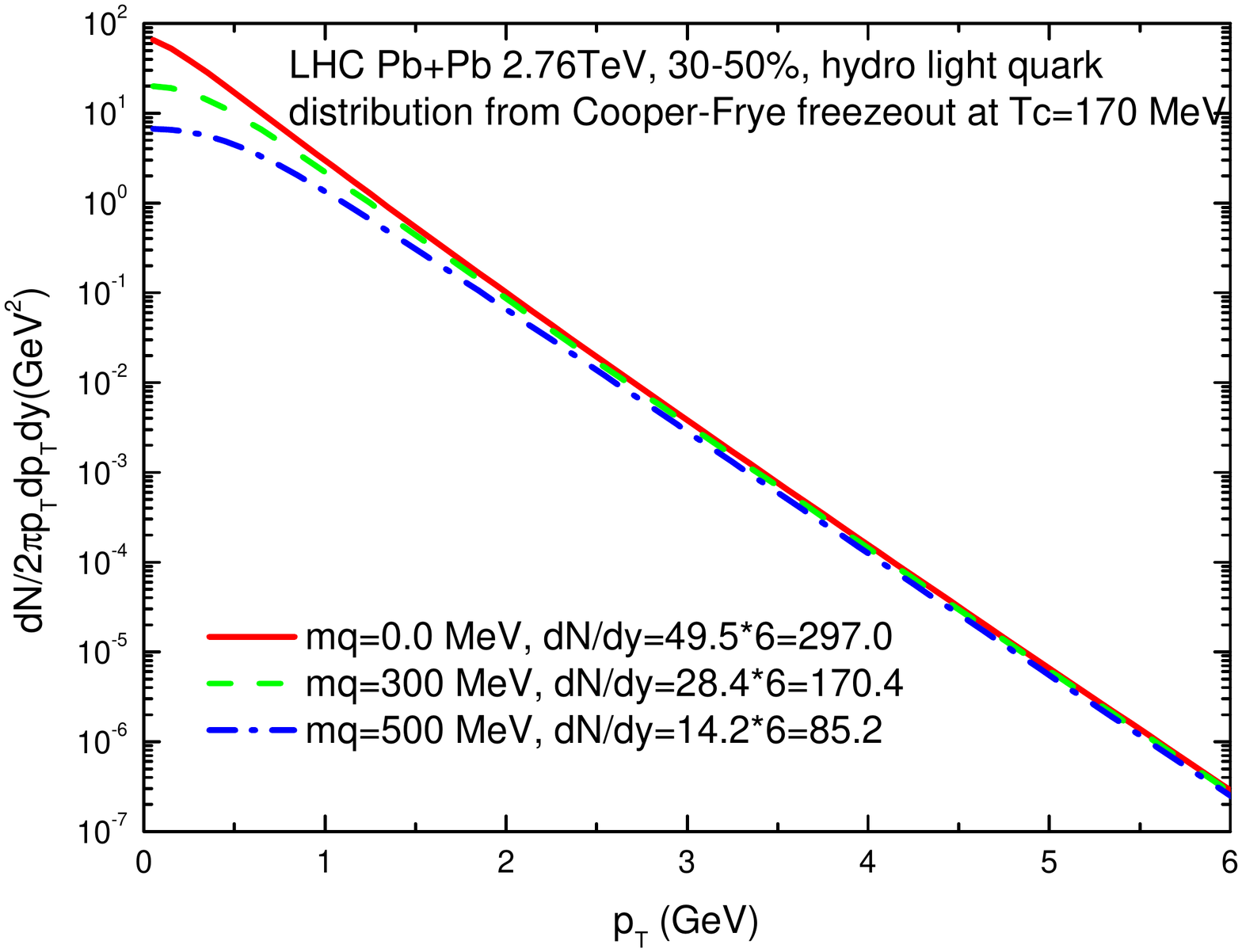}}
\end{minipage}

\vspace{-1.0cm}

\begin{minipage}{.5\textwidth}
\centerline{\includegraphics[width=1.2\textwidth]{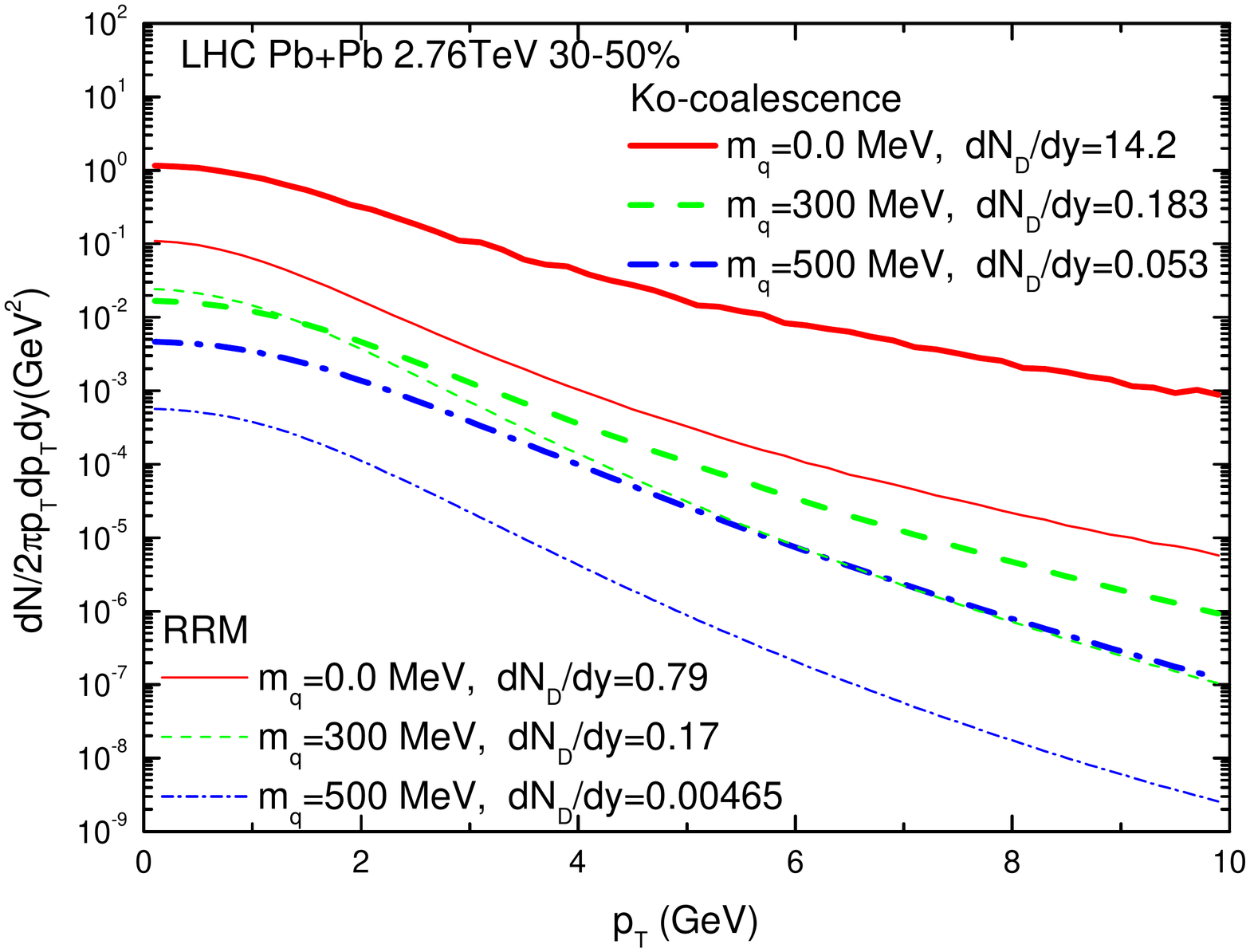}}
\end{minipage}
\begin{minipage}{.5\textwidth}
\centerline{\includegraphics[width=1.2\textwidth]{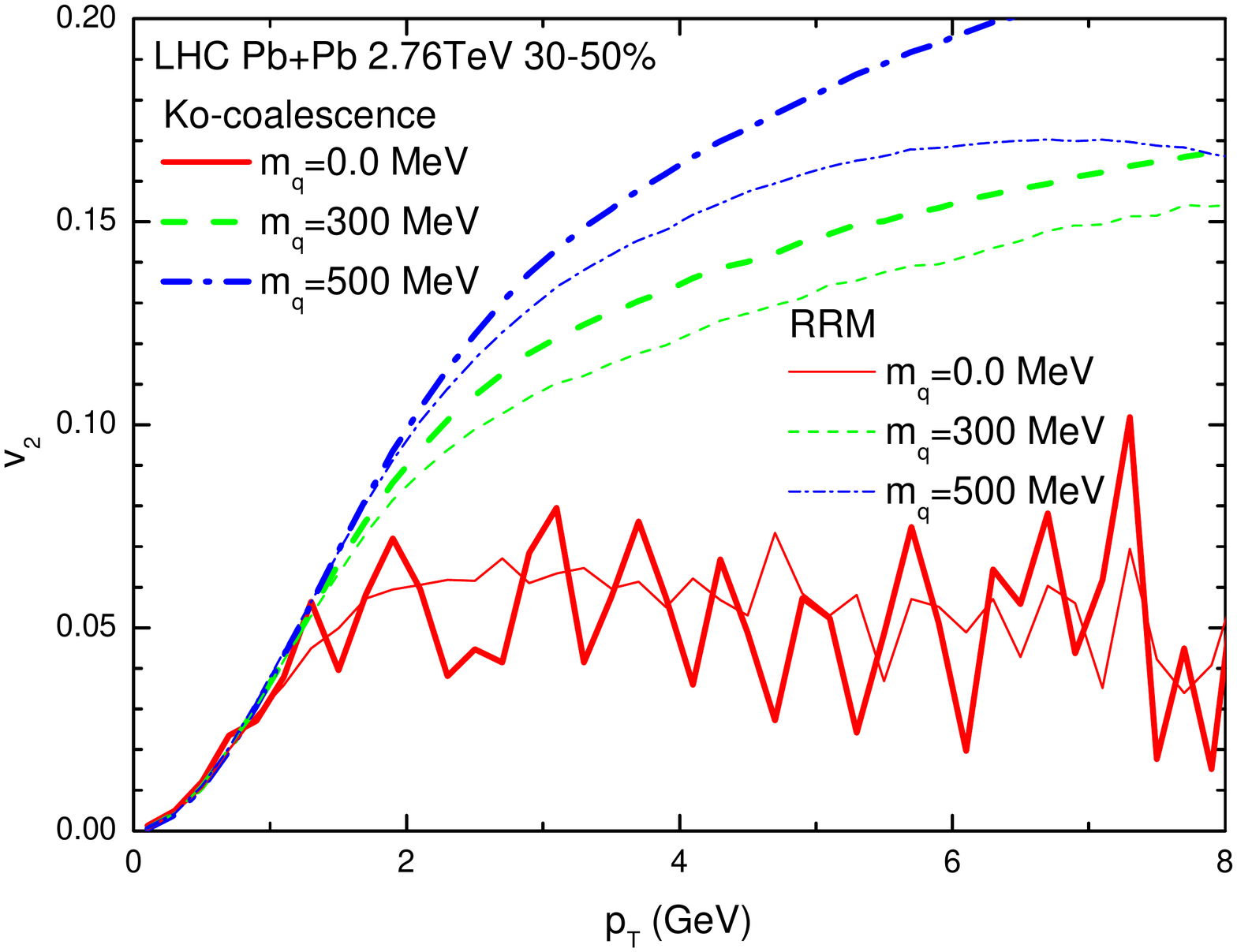}}
\end{minipage}
\caption{Upper panels: charm- (left)  and light-quark (right) $p_t$ spectra from Langevin
simulations and hydro-freezeout at $T_{\rm pc}=170$\,MeV, respectively.
Lower panels: comparison of the $D$-meson $p_T$ spectra (left panel) and elliptic flow 
(right panel) produced through recombination processes at $T_{\rm pc}$=170\,MeV when
using the resonance recombination model (RRM, dashed lines)~\cite{Ravagli:2007xx,He:2011qa} 
and an ICM (solid lines)~\cite{Greco:2003mm,Oh:2009zj} in 30-50\%
Pb-Pb(2.76\,TeV) collisions. The input charm-quark spectra are taken from the Langevin
simulations with the pQCD*5 interactions (within the TAMU hydro evolution) discussed in
Sec.~\ref{ssec_com-trans}.
}
\label{fig:coal-comp}
\end{figure}
Here, we will compare results for $D$-meson spectra from the ICM with the RRM
using the same input $c$-quark spectra and thermal light-quark distributions, which are shown
in the two upper panels of Fig.~\ref{fig:coal-comp}. For definiteness, we employ the $c$-quark
spectrum (normalized to $dN/dy=10.0$) obtained from the Langevin simulations with the
pQCD*5 interaction at the end of the QGP phase within the TAMU hydro model (as described in
Sec.~\ref{ssec_com-trans}), and the light-quark spectra are obtained from Cooper-Frye
freezeout within the same hydro, but for three different parton masses, with correspondingly
different total (integrated) yields as indicated in the figure.

Using these quark momentum distributions, the coalescence calculations are performed via 
a multi-dimensional integration in momentum space without explicit account of the 
space-momentum correlations. The resulting $D$-meson $p_T$ spectra and $v_2$ from the 
ICM~\cite{Greco:2003mm,Oh:2009zj} (thick curves) and from RRM (thin curves) are compared 
in the two lower panels of Fig.~\ref{fig:coal-comp} for the different light-quark masses. 
The absolute yields of $D$ mesons per unit rapidity are also indicated in each case (the 
volume parameter in the coalescence model affects the yields; here it is consistently 
determined from the hydrodynamic hypersurface; \eg, for the present 30-50\% Pb-Pb 2.76\,TeV 
collisions, the fireball volume per unit rapidity is $673.5~{\rm fm^3}$).\footnote{Note that
we do not include any contributions from resonance feeddown nor chemistry effects such as
a charm-quark fugacity factor which would be required to properly normalize the inclusive 
$D$-meson spectra; as such, the ratio ($H_{\rm AA}$) of the direct $D$-meson spectra to 
the input charm-quark spectra is not a meaningful quantity here.} 
The yields of $D$ mesons produced in RRM are comparable to the those produced in the 
ICM for $m_q=300$\,MeV, but substantially smaller for massless and $m_q=500$\,MeV light 
quarks. The $D$-meson $p_T$ spectra from the former are significantly softer than those 
in the latter. This is likely  a consequence of satisfying the equilibrium limit within 
RRM, which was shown to {\em soften} the high-$p_T$ spectra as to approach the 
thermal limit, while in the ICM the collinearity of the coalescing quarks tends to 
strictly {\em add} 3-momentum in the conversion from $c$ quarks to $D$ mesons. 
The $D$-meson $v_2$ obtained from RRM is a bit smaller than that obtained from the 
ICM. This may be due to the fact that the latter tends to recombine charm quarks 
with light quarks for essentially comoving kinematics (parallel momenta), while the RRM 
allows for significant momentum smearing via the isotropic Breit-Wigner cross section, 
thereby reducing the $D$-meson $v_2$ to some degree. 

\subsection{In-Medium Fragmentation}
\label{ssec_in-med-frag}
In this section we discuss the in-medium hadronization scheme implemented in the most recent version of the POWLANG model.
The procedure adopted in Refs.~\cite{Beraudo:2014boa,Beraudo:2015wsd} to model the hadronization 
of heavy quarks in the medium at the end of their propagation in the QGP is the following. 
Once a heavy quark $Q$, during its stochastic propagation in the fireball, has reached a fluid 
cell with a temperature below the decoupling temperature, $T_{\rm dec}$, it is forced to 
hadronize. One then extracts a light antiquark $\overline{q}$ (up, down or strange, with 
relative thermal abundancies dictated by the ratio $m/T_{\rm dec}$) from a thermal momentum 
distribution corresponding to the temperature $T_{\rm dec}$ in the local rest frame (LRF) of 
the fluid; information on the local fluid four-velocity $u^\mu_{\rm fluid}$ provided by 
hydrodynamics allows one to boost the momentum of $\overline{q}_{\rm light}$ from the LRF 
to the laboratory frame. 
A string is then constructed joining the endpoints given by $Q$ and $\overline{q}$ and passed 
to PYTHIA 6.4~\cite{Sjostrand:2006za} to simulate its fragmentation into hadrons (and their 
final decays). This is done as follows: the particle type, energy, polar and azimuthal angle 
of each endpoint are provided to PYTHIA through the PY1ENT subroutine; the PYJOIN subroutine 
allows one to construct the corresponding string; finally a PYEXEC call starts the simulation 
of its fragmentation and the final decays of unstable particles. In case the invariant mass 
of the string is not large enough to allow its decay into at least a pair of hadrons the event 
is resampled, extracting a new thermal parton to associate to the heavy quark. 
In agreement with PYTHIA, in evaluating their momentum distribution, light quarks are taken 
as ``dressed'' particles with the effective masses $m_{u/d}\!=\!0.33$ GeV and $m_s\!=\!0.5$ GeV.
Notice that, while the model allows one to take properly into account the momentum boost given 
to the final hadron by the light quark flowing with the medium, we do not get sizable 
modifications of the heavy-flavour hadrochemistry, \eg, an enhancement of $D_s$ or $\Lambda_c$ 
yields which might occur in the collisions and could be instead accommodated by a direct 
$2\to1$ or $3\to 1$  production mechanism in a coalescence model. Within our framework, 
a string, once formed, is hadronized as in the vacuum, through the excitation -- while 
stretching -- of $q\overline{q}$ pairs (or diquark-antidiquark pairs for the production of 
a baryon-antibaryon pair) from the vacuum: having a strange quark as an endpoint does not 
necessarily imply the production of a $D_s$ meson at hadronization. 

\begin{figure}[!t]
\begin{center}
\includegraphics[clip,width=0.48\textwidth]{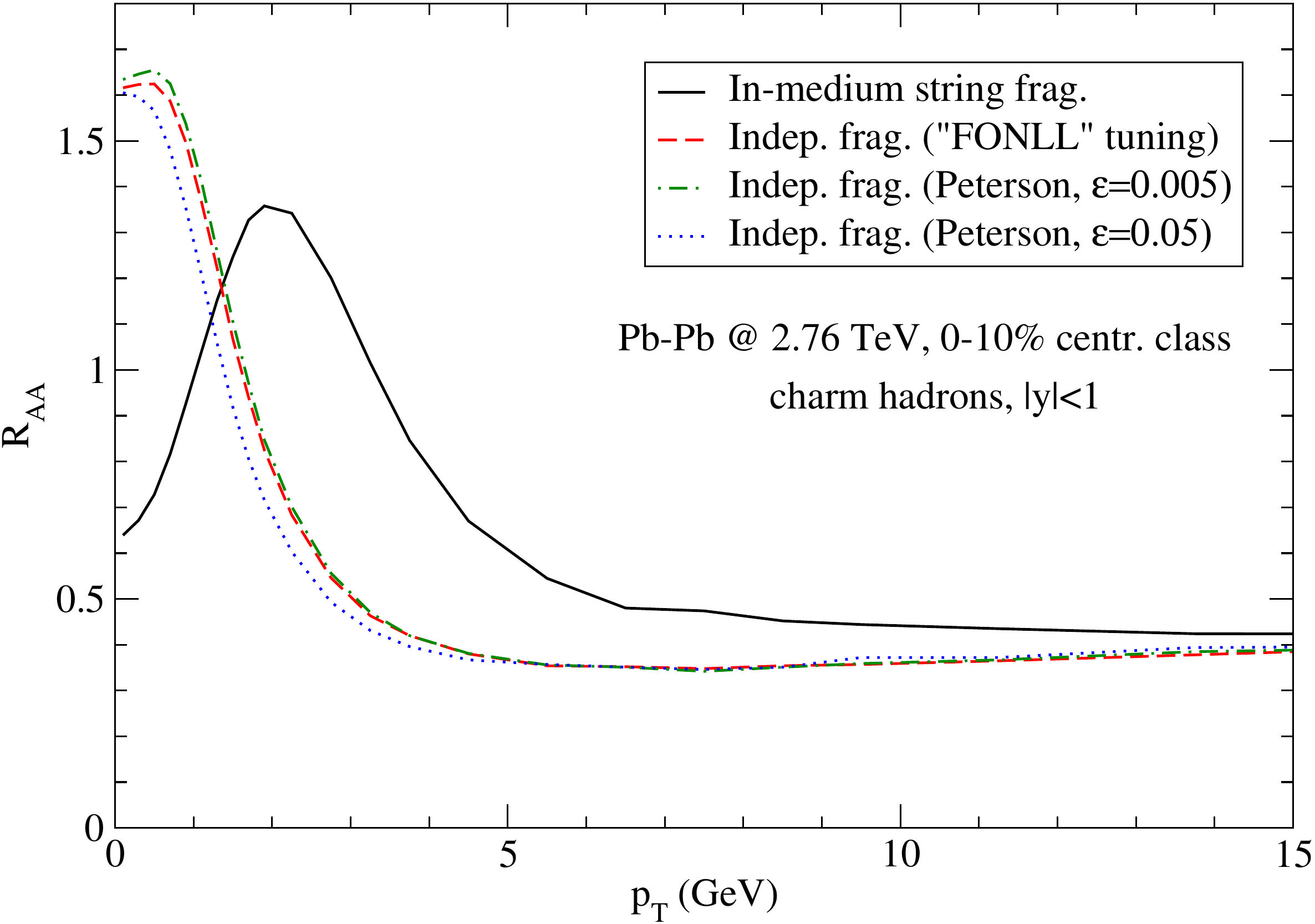}
\includegraphics[clip,width=0.48\textwidth]{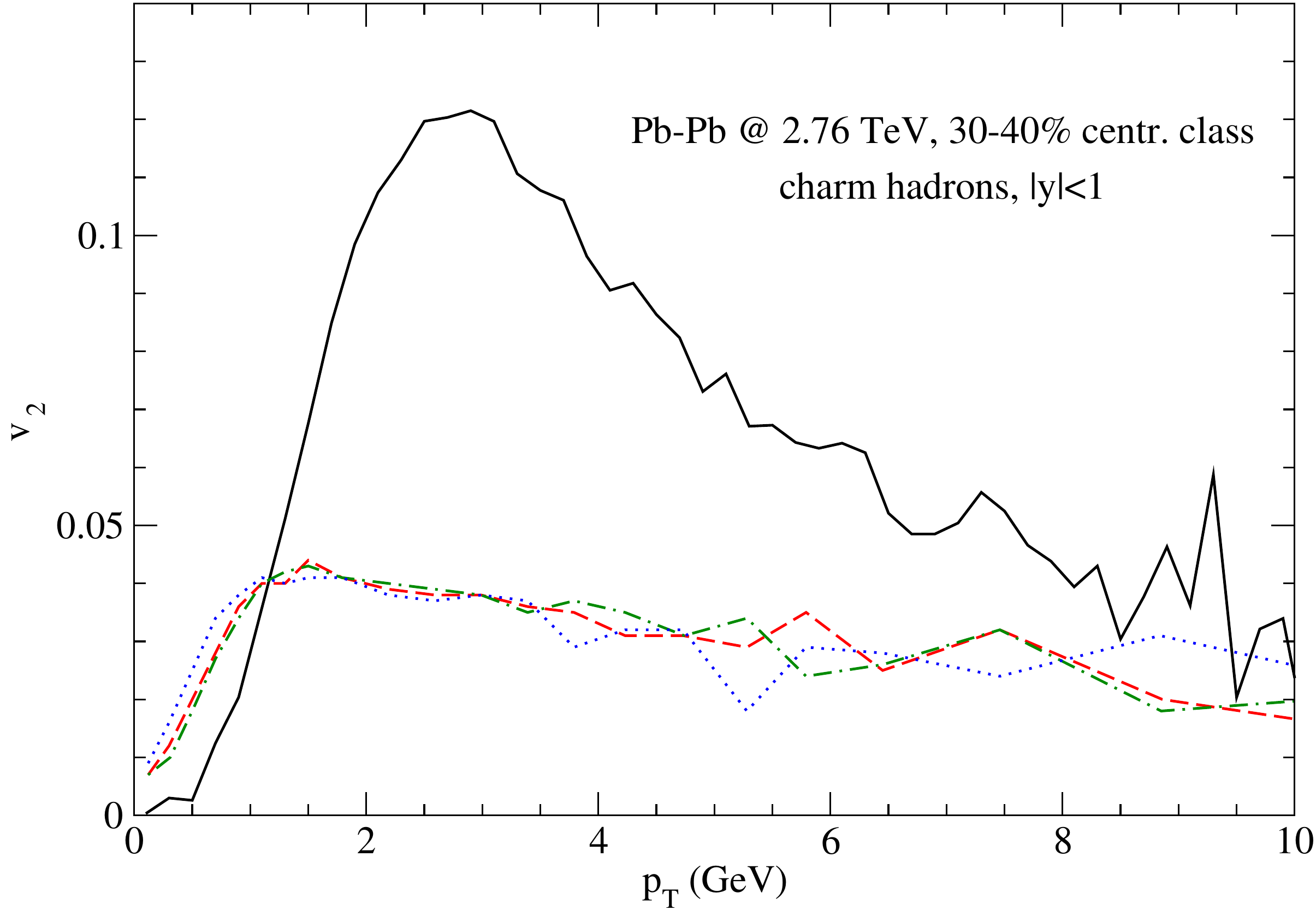}
\caption{Left panel: the nuclear modification factor of charmed hadrons in 0-10\% Pb-Pb collisions. 
Right panel: the elliptic flow of charmed hadrons in 30-40\% Pb-Pb collisions. Results of 
different hadronization schemes are compared. The fragmentation of strings formed via 
recombination with light thermal quarks from the medium leads to a bump in the $R_{\rm AA}$ at 
moderate $p_T$ and to an enhanced $v_2$ arising from the additional radial and elliptic flow 
acquired at hadronization. Hadronization via independent vacuum fragmentation functions leads 
to an $R_{\rm AA}$ reflecting a simple quenching pattern and to a smaller $v_2$.
}
\label{fig:in-medium_frag}
\end{center}
\end{figure}
In order to assess the effect of the in-medium string fragmentation model as described above
we also check the results obtained with the standard independent vacuum fragmentation functions 
(FF's), starting from the same heavy-quark spectrum at the end of the  Langevin evolution. 
We employ the Heavy-Quark Effective Theory (HQET) FF's~\cite{Braaten:1994bz}, with parameters 
referring to the $m_c\!=\!1.5$ GeV case tuned by the authors of FONLL~\cite{Cacciari:2003zu}. 
As a further comparison, we repeat the calculation with Peterson FF's, with parameter 
$\epsilon=0.005$ (representing quite a hard FF, very similar to the one of HQET) and 
$\epsilon=0.05$ (the default value in PYTHIA, corresponding to a softer FF).

We display the results of our study in Fig.~\ref{fig:in-medium_frag}. The curves obtained 
with vacuum FF's are characterized by a rather modest elliptic flow, with maximum value 
around 0.04, simply reflecting the one of the parent $c$ quarks. Furthermore, the nuclear 
modification factor of $D$ mesons simply reflects the quenching of the $p_T$ spectrum due 
to parton energy-loss, the increase at low $p_T$ being due to the conservation of the total 
number of charm quarks during their evolution in the medium. Notice that the results display 
a negligible dependence on the particular FF employed.
On the other hand, the curves obtained with the in-medium string fragmentation display a shift 
of the spectrum from low to moderate values of $p_T$ (``flow bump") and a strong enhancement 
of the elliptic flow. As in the case of coalescence, these features can be qualitatively 
explained as due to the additional radial and elliptic flow inherited by each charmed hadron 
from the light thermal partons (carrying the collective velocity of the medium) picked up  
through hadronization.

\section{Transport Coefficients and Implementation} 
\label{sec_trans}
After having scrutinized the impact of bulk evolution and hadronization models on $D$-meson
observables in nuclear collisions, we now turn to a discussion of HF interactions in QCD 
matter and their manifestation in the transport through the expanding fireball.
In principle, this includes both hadronic and partonic matter, although our focus will
mostly be on temperatures above the pseudo-critical one. Since the QCD transition at vanishing
chemical potential is a continuous crossover, one should also expect a continuous transition 
in the heavy-flavor (HF) degrees of freedom as the temperature is lowered. The hadronization 
mechanisms above are essentially representing the current realization of this transition 
in phenomenological models, and this issue will also be reiterated in the present section.

As was already mentioned in the introduction, HF phenomenology in heavy-ion collisions 
provides a unique opportunity to extract a transport coefficient of the QCD medium, \ie, 
the HF diffusion coefficient, ${\cal D}_s$. However, a mere extraction of this
number, even including its temperature dependence, remains unsatisfactory from the
fundamental point of view of studying the structure of QCD matter. Thus, the goal must 
be to firmly root HF interactions used in heavy-ion phenomenology in the in-medium QCD 
dynamics. For a soft quantity like a transport coefficient, this is a tall order; however,
this is where the benefits of a large quark mass comes in, by providing opportunities for
controlled approximations within suitable theoretical frameworks.
On the one hand, this leads to an effective theory known as heavy-quark effective theory 
(HQET),  which, roughly speaking, reduces the 4-component Dirac spinors to 2-component
Pauli spinors and utilizes a power counting in $\Lambda_{\rm QCD}/m_Q$ and $T/m_Q$. 
On the other hand, ample information on in-medium HF properties is 
available from lattice-QCD (lQCD) computations, including HQ free energies,
which, while not directly identifiable with an in-medium potential, can provide strong
constraints on the latter. Ultimately, one should relate back the insights gained 
from the HF sector to the bulk properties of QCD matter.  

We start this section by a brief comparison of results of several approaches for
calculating HQ transport coefficients in the QGP (Sec.~\ref{ssec_coeffs}), and then 
reverse the strategy of Sec.~\ref{sec_bulk} by using a
common hydrodynamic medium to perform charm-quark Langevin simulations for a few of
these interaction models (Sec.~\ref{ssec_com-hydro}). We assess in-medium HQ interactions
from perturbative-QCD (pQCD),  functional-renormalization-group (FRG) and lattice-QCD (lQCD) 
perspectives 
(Secs.~\ref{ssec_pqcd}, \ref{ssec_frg} and \ref{ssec_lqcd}, respectively),
followed by a study of Boltzmann and Langevin approaches for carrying out HF transport
in heavy-ion collisions (Sec.~\ref{ssec_boltz-lang}).

\subsection{Comparison of Existing Coefficients}
\label{ssec_coeffs}
A rather wide variety of microscopic approaches to compute HQ diffusion coefficients
has been adopted in the literature with applications to heavy-ion phenomenology.
These include perturbatively inspired approaches, which usually include  amendments to the Born
diagrams for HQ scattering off light quarks from the medium to augment the the coupling
strength in the medium. For example, in the SUBATECH approach a running coupling constant in
momentum transfer is implemented reaching close to one at soft momentum 
transfers~\cite{Gossiaux:2008jv,Peshier:2008bg}, while in 
the quasi-particle model (QPM) of the CATANIA group~\cite{Scardina:2017ipo} the ``running" 
essentially occurs in temperature reaching large values near $T_{\rm c}$ to reproduce the QGP 
EoS with large masses $gT$ for the bulk-medium partons~\cite{Plumari:2011mk} (similarly also 
in the PHSD approach). Nonperturbative  
approaches, on the other hand, usually  involve a ladder resummation of the interaction 
kernel including not only Color-Coulomb but also nonperturbative forces. 
For example, within the $T$-matrix framework~\cite{vanHees:2007me,Riek:2010fk,Liu:2016ysz}
employed by the TAMU group remnants of the confining force above $T_{\rm c}$ are 
implemented through potential kernels
constrained by lQCD results for the free energy. In connection with the ladder resummation
the nonperturbative forces lead to the formation of resonance correlations (``pre-hadrons")
which in turn induce a marked increase in the HQ interaction strength when approaching 
$T_{\rm c}$ from above. In the same spirit, the resonance model~\cite{vanHees:2004gq}
used by the UrQMD group is based on resummed $D$-meson $s$- and $u$-channel polegraph
interactions resummed to all orders.    

\begin{figure}[!t]
\begin{center}
\hspace{-0.5cm} \includegraphics[width=0.95\textwidth]{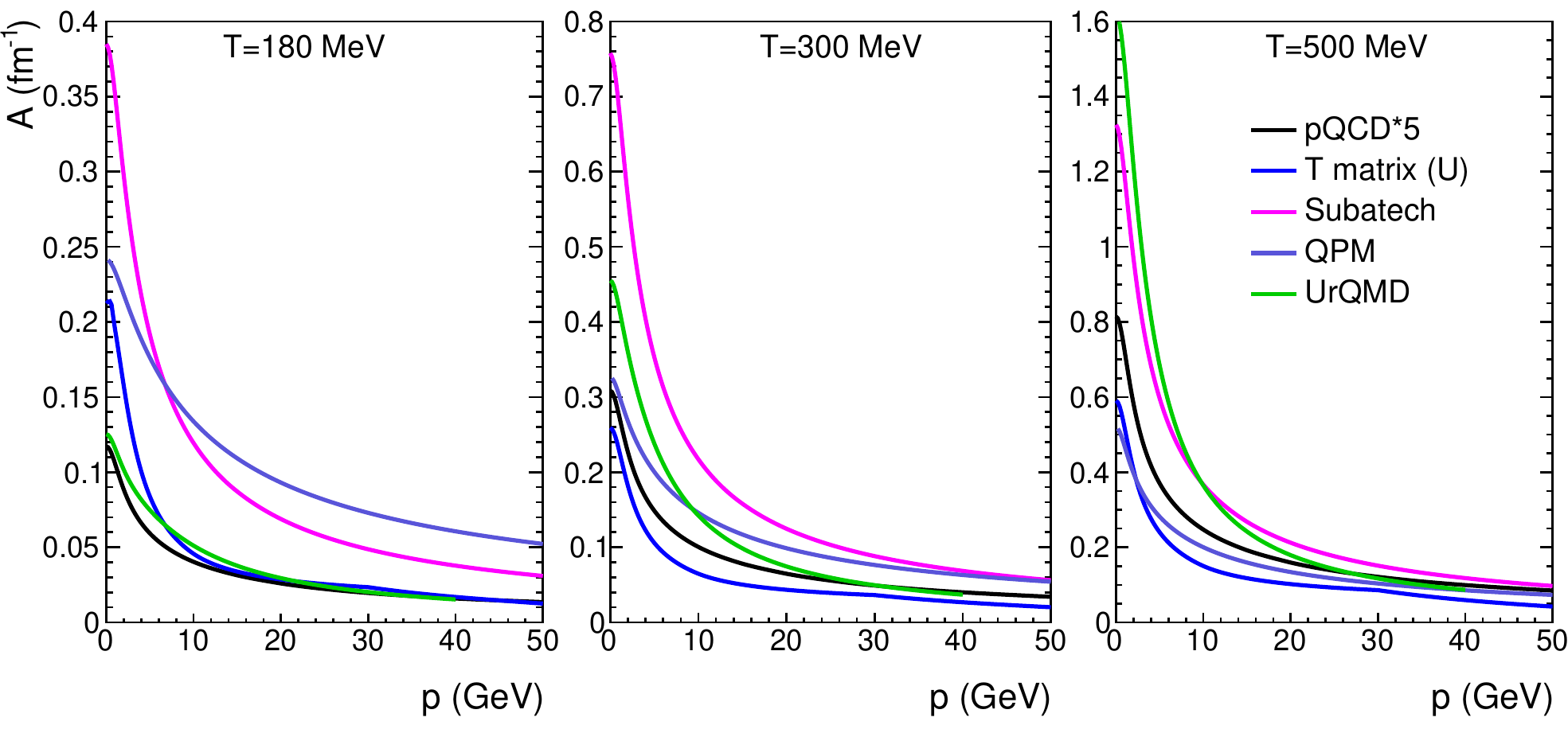}
\end{center}
\caption{Comparison of charm-quark friction coefficients within different models used
in phenomenological applications to heavy-ion data: pQCD*5 interaction as used in 
Sec.~\ref{sec_bulk} (black lines), $T$-matrix model with internal-energy potential as used
by the TAMU group (light-blue lines), $Q^2$-running coupling model of the SUBATECH group
as used in the Nantes transport approach (pink lines), quasi-particle model as used by
the CATANIA group (dark-blue lines) and $D$-meson resonance model as used by the UrQMD
group (green lines). The inverse of the friction coefficient essentially corresponds to 
the thermal relaxation time, $\tau_c=1/A$.
}
\label{fig:coeff}
\end{figure}

In Fig.~\ref{fig:coeff} we compare the charm-quark friction coefficient, $A(p,T)$ 
(a repository of $A$, $B_0$ and $B_1$ as function of $T$=160-600\,MeV in steps of 20\,MeV and
$p$=0-40\,GeV/$c$ in steps 0.2 GeV/$c$ can be found at the website \cite{rrtf:2016})
for several of the above scenarios, as a function of 3-momentum for 3 temperatures spaced
by a factor of 5/3. 
All models show a marked fall-off with 3-momentum. The fall-off is most pronounced in the
SUBATECH model, mostly due to the $Q^2$ running of $\alpha_s$, and in the resonance model
(especially at higher temperatures) where the interaction strength is also concentrated at
low relative momenta of the charm quark and the medium partons, required to excite a $D$-meson 
resonance. Even in the pQCD*5 interaction (underlying the transport calculations carried out in 
Sec.~\ref{sec_bulk}), which does not include a running coupling constant, a significant
fall-off with 3-momentum is found, comparable also to the quasiparticle model (QPM). 
The fall-off is somewhat stronger in the $T$-matrix calculations at low temperatures, where
the force nature changes from a long-range linear potential at low momenta to a color-Coulomb
potential at high momenta; consequently, at higher temperatures, where the remnant confining 
force is essentially screened, the momentum dependence and magnitude of the transport coefficient 
becomes comparable to pQCD*5 and QPM interactions. The strongest increase in temperature is 
found in the pQCD*5 and resonance model, both of which in essence do not include a reduction
in interaction strength with temperature and thus fully pick up on the increase in parton
densities, resulting in an approximately $T^2$ dependence of the low-momentum friction 
coefficient (somewhat weaker in the pQCD*5 case due to the 
increasing screening mass). In the SUBATECH running-coupling model the increase is roughly 
linear in $T$ (the screening of the relatively small Debye mass has a stronger effect),
while in the QPM (with a rather pronounced decrease of $\alpha_s$ with $T$) and the
$T$-matrix approach (with a rather pronounced screening of the confining force) there
is little temperature variation from $T$=180\,MeV to 300\,MeV, while it increases 
appreciably thereafter once color-Coulomb interactions with little temperature dependence 
in the coupling take over.

\subsection{Different Transport Coefficients in a Common Hydrodynamic Medium}
\label{ssec_com-hydro}
%

In this section, we implement HQ transport coefficients from different microscopic 
interactions into a common hydrodynamic medium within the same Langevin scheme. 
Three sets of model calculations of transport coefficients are used and compared: 
(1) a leading-order pQCD calculation multiplied by a $K$ factor of 5 (pQCD*5), as utilized 
in Sec.~\ref{ssec_com-trans} (cf.~black curves in Fig.~\ref{fig:coeff}); 
(2) a pQCD-motivated one-gluon exchange model developed by the Nantes 
group~\cite{Gossiaux:2008jv,Gossiaux:2009qf} 
which includes the effects of a running coupling constant and reduced Debye mass, with a fixed
$c$-quark mass of $m_c$=1.5\,GeV (cf.~pink curves in Fig.~\ref{fig:coeff}); 
and 
(3) an in-medium $T$-matrix formalism developed by the TAMU group~\cite{vanHees:2007me,Riek:2010fk},
with $c$-quark mass varying as $m_c$$\simeq$1.4$\to$1.8\,GeV for $T$$\simeq$500$\to$170\,MeV 
(cf.~blue curves in Fig.~\ref{fig:coeff}). 
The latter two models for the HQ interactions in the QCD medium have had some success
in describing open HF data from RHIC and the LHC using different bulk matter evolution
models. Therefore, applying them within a common hydrodynamic evolution here not only 
provides a direct comparison between the different HQ interactions, but, in turn, also 
helps understand possible differences in the bulk matter and hadronization models that the 
different groups use.

The common hydrodynamic medium we apply in this section is the (2+1)-dimensional viscous 
hydrodynamic model VISHNU developed in Refs.~\cite{Song:2007fn,Song:2007ux,Qiu:2011hf}, 
labelled as ``OSU hydro" in Figs.~\ref{fig:pions} and \ref{fig:protons}. 
We employ the code version and parameter tuning provided by Ref.~\cite{Qiu:2011hf} in the
present study. The QGP fireballs are initialized using the Monte-Carlo Glauber model 
for the initial-entropy density distribution. The starting time of the QGP evolution is 
set at $\tau_0=0.6$~fm and a constant shear-viscosity-to-entropy-density ratio of 
$\eta/s$=0.08 is determined to describe the spectra of soft hadrons emitted from the 
fireballs at both RHIC and the LHC. For the HQ transport through this medium evolution
a Langevin process with the pre-point scheme of Refs.~\cite{Cao:2011et,Cao:2012jt} is 
adopted, together with a leading-order pQCD calculation for the initial HQ $p_t$ spectrum
(without CNM effects).  
The transport coefficients of the three microscopic models employed in this study 
(see above) are suitably converted for use within the pre-point discretization scheme
(and with the Einstein relation enforced for the momentum diffusion coefficient 
starting from the friction coefficient). 

\begin{figure}[!t]
\begin{center}
{\includegraphics[width=0.48\textwidth]{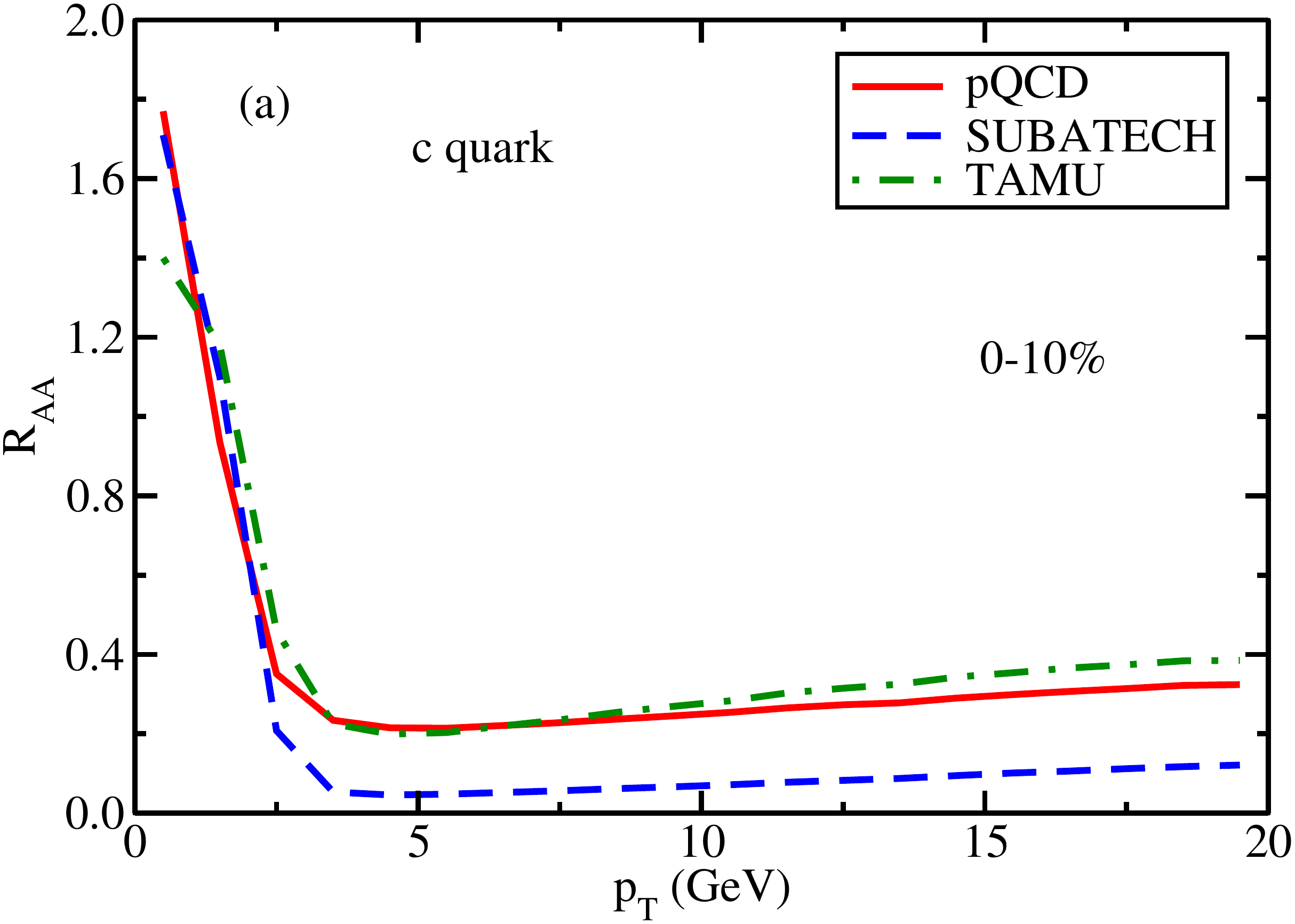}}
{\includegraphics[width=0.48\textwidth]{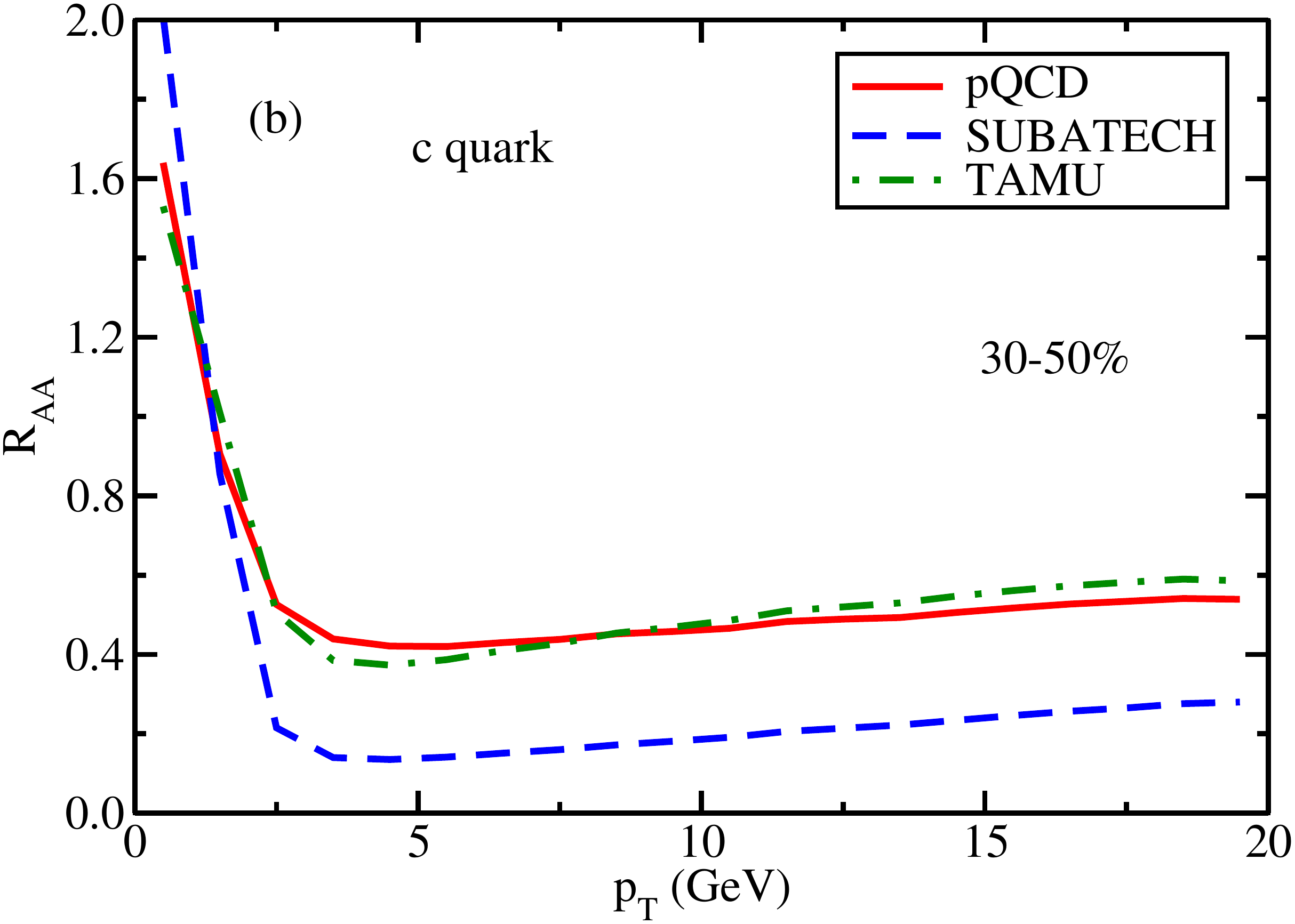}}
{\includegraphics[width=0.48\textwidth]{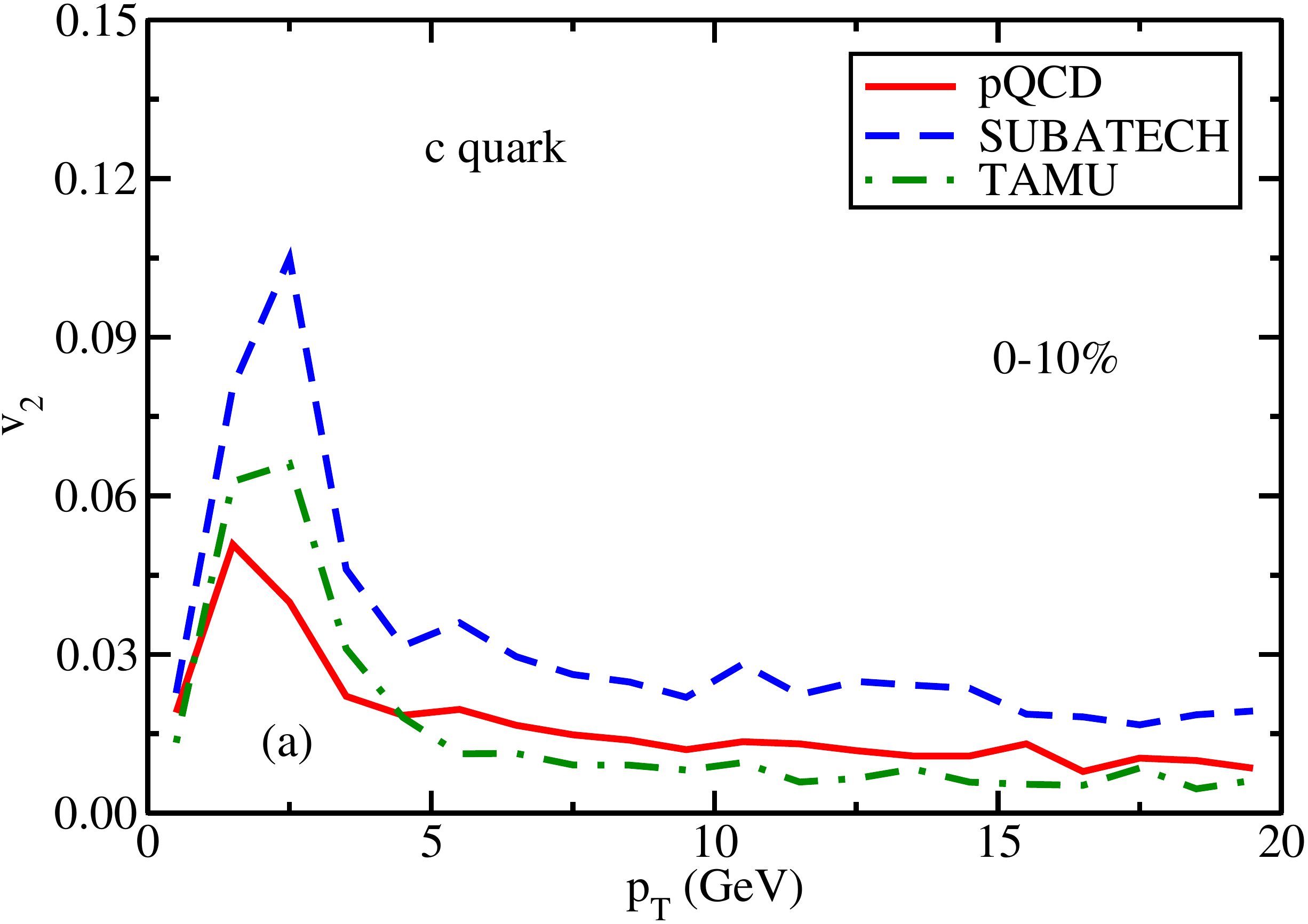}}
{\includegraphics[width=0.48\textwidth]{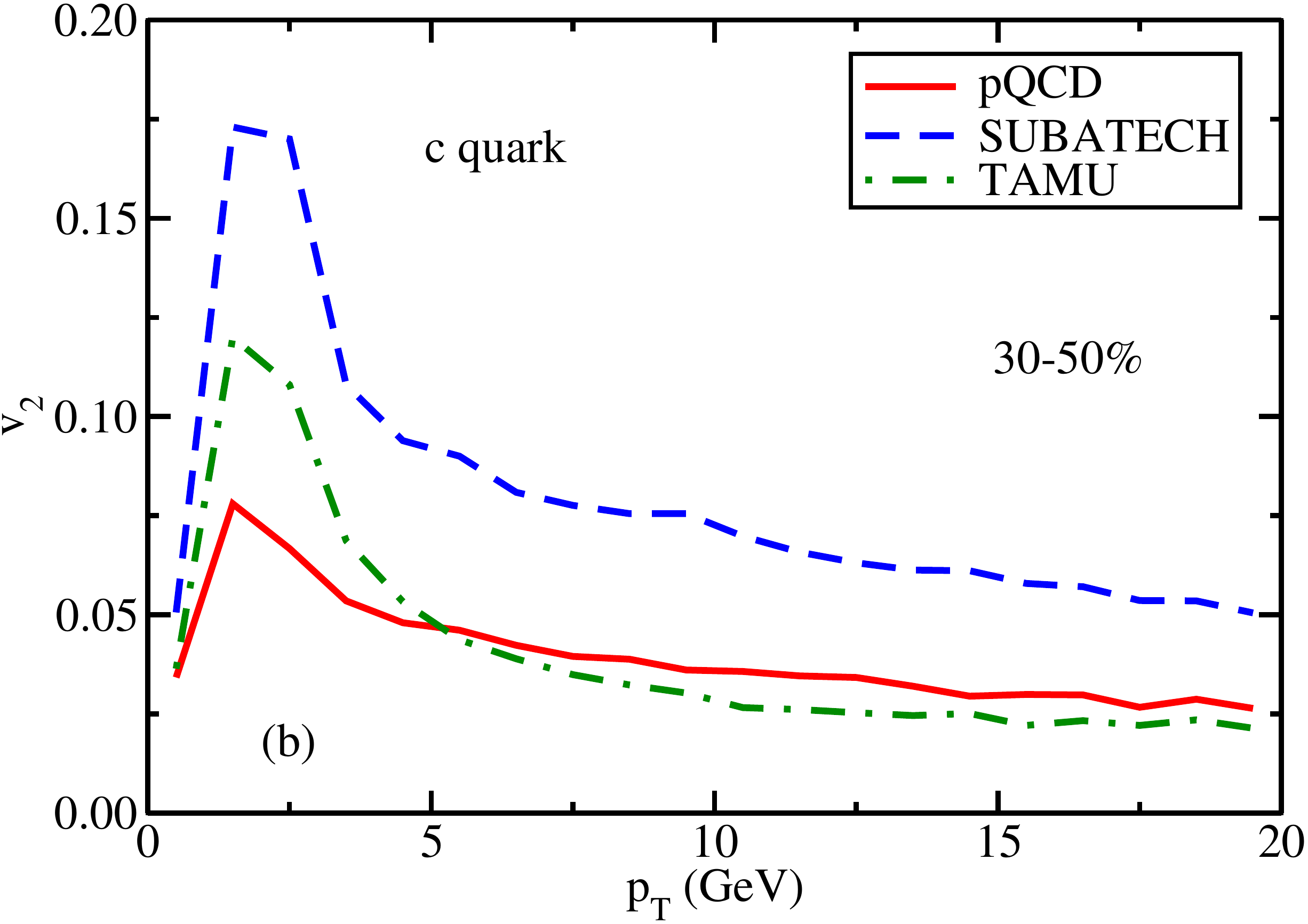}}
\end{center}
  \caption{(Color online) Results from Langevin simulation for the charm-quark 
$R_\mathrm{AA}$ (upper panels) and $v_2$ (lower panels) central (left column) 
and semi-peripheral (right column) 
2.76~TeV Pb-Pb collisions using a common viscous hydrodynamic evolution with 3 different transport
coefficient (solid lines: pQCD*5; dash-dotted lines: nonperturbative $T$-matrix approach, 
dashed lines: pQCD with running coupling).}
\label{fig:c-hydro}
\end{figure}

In Fig.~\ref{fig:c-hydro} we summarize the results for the charm-quark 
$R_\mathrm{AA}$ and $v_2$
in two different centrality bins of Pb-Pb collisions at $\sqrt{s_\mathrm{NN}}$=2.76~TeV,
at the end of the QGP evolution at $T$=165\,MeV. 
The pQCD*5 and the $T$-matrix interactions result in a rather similar charm-quark 
$R_\mathrm{AA}$ over a large range in transverse momentum; a slight difference from the
stronger 3-momentum dependence in the latter, implying a smaller friction coefficient at
large 3-momentum, is still apparent in form of slightly weaker suppression at high $p_t$
(cf.~the two upper panels of Fig.~\ref{fig:c-hydro}). On the other hand, the stronger coupling
of the $T$-matrix interaction at low momentum generates a more pronounced collective 
behavior of the $c$-quarks as signalled by the 
elliptic flow coefficient (cf.~the two lower panels of Fig.~\ref{fig:c-hydro}): the peak value 
of the $v_2$ is up to $\sim$60\% larger for the $T$-matrix interactions than for the
pQCD*5 model; at the same time, the high-$p_t$ $v_2$ from the $T$-matrix is slightly
lower, consistent with the behavior in the $R_{\rm AA}$.  
The much larger transport coefficient in the Nantes-pQCD calculation, relative to the other 
two interactions, leads to a much smaller $R_\mathrm{AA}$ of charm quarks at $p_t$'s 
down to about 2.5\,GeV, accompanied by a larger enhancement below
(dictated by charm number conservation). At the same time, the $v_2$ is also much larger
than in the pQCD*5 and $T$-matrix approach, by about a factor of 2 across all $p_t$. This
factor approximately reflects the difference at the level of the friction coefficient,
$A(p)$. 

As mentioned earlier, both TAMU and Nantes groups are able to describe experimental 
observations of HF $R_\mathrm{AA}$ and $v_2$ with some success using the pertinent
transport coefficients within their respective bulk evolution models (although the 
$T$-matrix model tends to underestimate the observed high-$p_T$ suppression), within 
their respective bulk evolution and hadronization models. The difference in their 
results for the same bulk evolution thus implies significant differences in other 
modeling components,
most notably in the coalescence part (as discussed in Secs.~\ref{ssec_hadro-com} and
\ref{ssec_hadro-spec}), and, to a lesser extent, in the evolution profiles 
from the different hydrodynamic models used by the two groups (as discussed in
Sec.~\ref{ssec_com-trans}).

\subsection{Perturbative Analysis}
\label{ssec_pqcd}

In this section we discuss perturbative treatments of the momentum diffusion of heavy quarks
in a thermal medium.  In doing so, we assume that $T \gg \Lambda_{QCD}$ so that the QCD 
coupling $\alpha_\mathrm{s}$ is small.  It is not clear that this approximation
is useful at physically achievable temperatures.  This issue can be resolved by working 
beyond leading order (LO), to see the size of next-to-leading order (NLO) corrections.

A perturbative calculation of heavy quark transport is not as simple
as a low-order diagrammatic evaluation.  Since the diffusion constant
is defined in terms of low frequency, long-distance behavior,
the evaluation requires diagrammatic resummations, similar to the
evaluation of shear viscosity~\cite{Jeon:1994if}.  The situation simplifies
if one uses the large quark mass $m^2 \gg T^2$, which ensures that the
typical momentum carried in equilibrium is also large,
$p^2 \sim mT \gg T^2$.  Therefore one can instead compute the momentum
diffusion coefficient, and convert it to a spatial momentum diffusion
coefficient using Einstein relations.

Benjamin Svetitsky provided the first complete leading-order
perturbative treatment of heavy-quark diffusion~\cite{Svetitsky:1987gq},
finding~\cite{Svetitsky:1987gq,Braaten:1991jj,Braaten:1991we,Moore:2004tg}
\begin{equation}
  {\cal D}_s = \frac{27}{16\pi \alpha^2_{\mathrm{s}} T}
	\left[ 3 \left( \ln\frac{2T}{\mD}+\frac{1}{2} - \gamma_{\rm
	E} + \frac{\zeta'(2)}{\zeta(2)} \right)
	+ \frac{N_\mathrm{f}}{2}\left( \ln\frac{4T}{\mD}+\frac{1}{2} -
	\gamma_{\rm E} + \frac{\zeta'(2)}{\zeta(2)} \right) \right]^{-1}
	,
\label{eq:D_heavy}
\end{equation}
where $\mD = T\sqrt{6\pi \alpha_{\mathrm{s}}}$ and $N_{\mathrm{f}}$ is
the number of light-quark flavors (3 in most applications).
This was extended to finite quark velocity by Moore and Teaney~\cite{Moore:2004tg}.

\begin{figure}[t]
\begin{center}
\includegraphics[width=9cm,height=7cm]{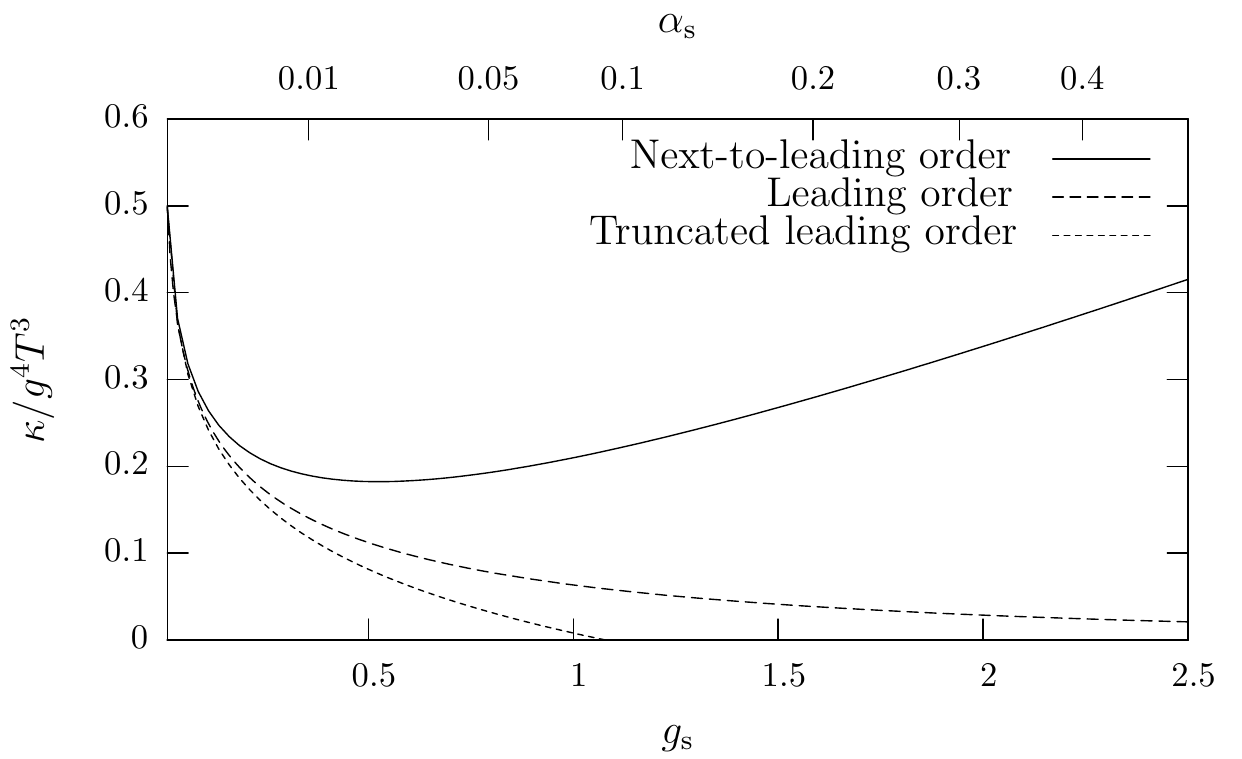}
\end{center}
  \caption{Comparison of leading and next-to-leading order inverse
    heavy-quark diffusion coefficient, $\kappa/T^3 = 2 /({\cal D}_s T)$, 
scaled by the leading-order coupling constant dependence.
 The subleading corrections are large even at coupling values usually 
considered to be very
    small.}
\label{fig_NLO}
\end{figure}

In a pioneering work, Caron-Huot and Moore extended the calculation of
the heavy quark diffusion coefficient to NLO in the
strong coupling~\cite{CaronHuot:2008uh}.  The first corrections arise from the
soft sector ($p\sim \mD$), at order $\mathcal{O}(\sqrt{\alpha_{\mathrm{s}}})$, 
that is, they are non-analytic in the strong coupling $\alpha_{\mathrm{s}}$.  Rather
than discuss the very technical calculation, we will skip to providing
the result, which is
\begin{equation}
  {{\cal D}_s^{\mathrm{NLO}}} = \frac{27}{16\pi \alpha^2_{\mathrm{s}} T}
  \left( \left[ 3 + \frac{N_{\mathrm{f}}}{2} \right]
  \left[ \ln \frac{2T}{\mD}-0.64718 \right]
  + \frac{N_{\mathrm{f}} \ln(2)}{2} + 2.3302 \frac{3 \mD}{T}
  \right)^{-1} \,,
\end{equation}
where the last expression, with numerical coefficient $2.3302$,
represents the NLO corrections.

We illustrate this result in Fig.~\ref{fig_NLO}.  As the figure shows,
for realistic couplings $\alpha_{\mathrm{s}} \simeq 0.3$, the
``subleading'' corrections are several times larger than the leading
order behavior. It appears that the perturbative expansion for this
quantity is especially poorly behaved.   
A key reason for this is that the LO of the HQ thermalization rate is
already ${\cal O}(\alpha_s^2)$, implying that higher orders come in with a larger 
uncertainty as compared to quantities whose LO is ${\cal O}(1)$ (such as the QGP 
pressure or dilepton rate). This makes it difficult to make much progress in 
computing the medium's effects on heavy quarks purely with perturbative tools.

\subsection{Functional Renormalization Group}
\label{ssec_frg}

A continuum approach for non-perturbative computations of transport
coefficients on the basis of single-particle spectral functions of
quarks and gluons, $\rho_A$ and $\rho_q$, has been put forward in
Ref.~\cite{Haas:2013hpa}. It has been used for the shear viscosity in
in quenched QCD~\cite{Haas:2013hpa,Christiansen:2014ypa}, with in-medium 
gluon propagators obtained with functional renormalization group (FRG) techniques. 
For high temperatures the results compare well with hard-thermal-loop
(HTL) results. This agreement extends to surprisingly low temperatures, 
$T\gtrsim 2 T_c$, and is supported by diagrammatic similarities of the 
(resummed) perturbative approach to the fully non-perturbative setting. At
temperatures below $T_c$ the result is compatible with the viscosity in a 
glueball resonance gas. Within this approach the diagrammatic similarities 
of standard perturbative resummation schemes with the fully non-perturbative 
diagrammatics is apparent. In the case of two-point correlation functions 
of the energy momentum tensor it leads to a seven-loop exact formula that 
can be reduced to three-loop resummed expressions in terms of full 
vertices and propagators \cite{Haas:2013hpa,Christiansen:2014ypa}.

For the computation of the heavy-quark diffusion coefficient, an analogous
starting point is the relation of the momentum diffusion coefficient,
$\kappa$, to the force-force correlator,
\begin{equation}\label{eq:kappa}
\kappa = \int dt \langle F(t) F(0)\rangle\,,\qquad {\rm with}  \qquad  
F[q,A](t) =\int d^3  \bar q(t,\vec x) t^a E^a_y(t,\vec x) q(t,\vec x) \ ,
\end{equation}
see, \eg~\cite{Svetitsky:1987gq,CasalderreySolana:2006rq}. This correlation 
can be approximated by that of two chromo-electric fields connected 
by Wilson lines. In the present approach it is more convenient to directly 
compute the correlation function, eq.~(\ref{eq:kappa}). The latter has a 
seven-loop exact representation given by  
\begin{equation}\label{eq:kappaop}
\kappa = \int dt F[\hat q, \hat A](t)\, F[\hat q, \hat A](0)\,,
\qquad {\rm with} \qquad 
\hat A_\mu =\langle A_\mu \phi\rangle_c \frac{\delta}{\delta 
\phi } + \phi\,,\qquad \phi=(A_\mu, q , \bar q)\,, 
\end{equation}
where the subscript ${}_c$ indicates the connected part. The real-time 
two-point functions, $\langle \phi_1 \phi_2\rangle_c$, can be
expressed through the respective single-particle spectral functions of
quarks and gluons in QCD. These can be either computed directly or with
the help of MEM-type methods from Euclidean correlation functions. In 
the latter approach the largest systematic error arises from the 
low-frequency tail of the reconstructed real-time correlation functions. 
In the present diagrammatic approach this systematic error is averaged 
over via the frequency loop-integrals~\cite{Haas:2013hpa,Christiansen:2014ypa}. 
Both methods, the direct computation and the reconstruction, have been applied 
in QCD and low-energy effective 
models~\cite{Haas:2013hpa,Christiansen:2014ypa,Pawlowski:2015mia,Strauss:2012dg,Kamikado:2013sia,Tripolt:2013jra,Yokota:2016tip},
and the respective results can be used for the computation of the diffusion 
coefficient. 

As mentioned above, this approach bears diagrammatic similarities to
the standard perturbative approach described in Sec.~\ref{ssec_pqcd}. 
For example, the lowest order contribution in eq.~(\ref{eq:kappaop}) 
arises from diagrams with two quark and one gluon spectral function 
(lines) between the $F$'s. In particular, this
similarity can be used to discuss the convergence of the perturbative
approach as well as its regime of validity. A particularly simple example 
for this structure is the Debye mass which
agrees in next-to-leading order with the full non-perturbative result
for temperatures $T\gtrsim 2T_c$~\cite{Cyrol:2017qkl}. The
failure for temperatures $T\lesssim 2 T_c$ can be readily explained by
the influence of the non-perturbative confinement physics and scale
at these temperatures.

In future the combination of these approaches will also allow for
functionally assisted analytic computations: the perturbative setting
allows for analytic computations while the non-perturbative approach
is used to access and determine the validity regime of the (resummed)
perturbative approach.

\subsection{Information and Constraints from Lattice QCD}
\label{ssec_lqcd}

QCD calculations can contribute to understanding of heavy-flavor
production in hot medium in several different ways. Lattice-QCD can provide
some information  on the heavy quark diffusion coefficient. These calculations
can be compared to the calculations based on a weak-coupling expansion, which
are valid are sufficiently high temperature. 
Diagonal and off-diagonal charm susceptibilities can provide
information on the charm degrees of freedom across the QCD transition.
Finally, spatial and temporal correlators provide information on
in-medium properties of charm hadrons and/or about their dissolution in
hot medium. Below we will discuss the status of these calculations in
more detail. 

\subsubsection{Heavy-quark diffusion coefficient}
The spatial HQ diffusion coefficient can be defined in
terms of spectral functions corresponding to current-current
correlators of heavy quarks
\begin{equation}
\sigma(\omega,\vec{p})=\frac{1}{\pi}\int dt e^{i \omega t} \int d^3 x 
e^{i \vec{x} \cdot \vec{p}} 
\langle \left[ J_{i}(t,\vec{x}),J_i(0,0) \right] \rangle, 
\end{equation}
where $J_i=\bar \psi_h \gamma_i \psi$ with $\psi_h$ being the
heavy quark field.
The spatial diffusion coefficient is defined as 
\begin{equation}
{\cal D}_s=\lim_{\omega \rightarrow 0} \sigma(\omega,\vec{p}=0)/(\omega \chi_q \pi).
\end{equation}
Here $\chi_q$ is the quark number susceptibility for heavy quarks.
In the case of a large quark mass, $M \gg T$, the structure of the
spectral function has a simple form for $\vec{p}=0$:
\begin{equation}
\sigma(\omega,0)=\frac{1}{\pi} \chi_q \frac{\omega \eta}{\omega^2+\eta^2}
\frac{T}{M},
\label{spf_low}
\end{equation}
where $\eta=T/(M {\cal D}_s)$ is drag coefficient entering the Langevin 
equation~\cite{Petreczky:2005nh} ($\tau_Q=1/\eta$ is the thermal HQ relaxation time). 
In other words, for zero spatial momentum
the spectral function has a transport peak at $\omega \simeq 0$.
For $p \ll T$ the structure of the spectral function can be worked out
and it is determined by the same constant $\eta$ \cite{Petreczky:2005nh},
\ie, for small momenta there is no dependence of the drag 
coefficient on the momentum.
As one can see from Eq.~(\ref{spf_low}) 
the width of the transport peak is very small for large quark mass.
This makes the lattice determination of the HQ diffusion
coefficient very challenging~\cite{Petreczky:2005nh,Petreczky:2008px}.
However, the difficulty associated with the large quark mass can
be turned into an advantage. Namely, one can integrate out the HQ 
degrees of freedom in the spirit of HQ effective theory
and reduce the current-current correlator to the correlator of the 
chromo-electric field strength~\cite{CaronHuot:2009uh}. 
The corresponding spectral function in the $\omega \rightarrow 0$ limit
gives the momentum diffusion coefficient $\kappa=2 M T \eta$~\cite{CaronHuot:2009uh}.
Furthermore, this spectral function does not have a peak around $\omega=0$, instead the
high-$\omega$ and the low-$\omega$ regions are smoothly connected 
\cite{CaronHuot:2009uh,Burnier:2010rp}. From the point of view of reconstructing
the spectral function from the lattice data this has a clear advantage since for
determination of $\kappa$ one has to determine the intercept rather than the width 
of the transport peak, and therefore the lattice determination of $\kappa$ may be 
more easily feasible.
Lattice determinations of $\kappa$ in quenched QCD have been reported in 
Refs.~\cite{Banerjee:2011ra,Francis:2015daa}. A prerequisite for the determination
of the transport coefficient $\kappa$ is sufficiently accurate data for the electric
field strength correlator. Due to gluonic nature of the correlation functions
the lattice data are very noisy and the use of noise reduction techniques is mandatory
\cite{Banerjee:2011ra,Francis:2015daa}. In addition one has to perform calculations
at several lattice spacings and perform a continuum extrapolation. This step has so far
been performed only in Ref.~\cite{Francis:2015daa}. Given the lattice data one relies
on a fit ansatz that smoothly connects the known high-$\omega$ asymptotics of the spectral
function with the form $\kappa \omega$ for small $\omega$. This ansatz is not unique and
the use of different ans\"atze translates into systematic errors in the determination of 
$\kappa$.  The detailed analysis of Ref.~\cite{Francis:2015daa} results in a value of 
\begin{equation}
\kappa/T^3 =1.8 - 3.4.
\label{kappa_res}
\end{equation}
for $T=1.5\,T_c$ (where $T_c\simeq 270$\,MeV for quenched QCD).
This corresponds to a range of values for $2\pi T {\cal D}_s$ of 3.7-7.0. This result agrees 
with findings presented in Ref.~\cite{Banerjee:2011ra} at fixed lattice spacing within errors.
It is also comparable to the value of $\sim$6 from the ``pQCD*5" interaction employed in the 
various bulk and hadronization models discussed in Secs.~\ref{sec_bulk} and \ref{sec_hadro}. 

Attempts to determine the spatial HQ diffusion coefficient from current-current
correlators have been presented in Ref.~\cite{Ding:2012sp} in quenched QCD:
\begin{equation}
2 \pi T {\cal D}_s =1.8 \pm 0.5 (stat.)^{+1.3}_{-0.5} (syst.), T=1.46 T_c.
\end{equation}
This is significantly smaller than the value of ${\cal D}_s$ reported above. Note, however, 
that not all systematic effects have been taken into account in this analysis. As discussed
before it is difficult to determine reliably the width of the transport peak.

For phenomenological applications it would be important to perform calculations
in full QCD. With the current technology this is not possible since the noise
reduction techniques are only available for quenched QCD. One possible way
to deal with noise in full QCD would be to use a gradient-flow 
method~\cite{Kitazawa:2017qab,Kitazawa:2016dsl,Datta:2015bzm,Petreczky:2015yta}.

The formulation of the HQ diffusion in terms of electric field strength
correlators, or equivalently in terms of force-force correlators acting on
the heavy quark, turned out to be very useful when calculating the momentum
diffusion coefficient in the weak coupling expansion~\cite{CaronHuot:2008uh} 
or in AdS/CFT~\cite{CasalderreySolana:2006rq}. The value of $\kappa$ from
the lattice calculation given by Eq.~(\ref{kappa_res}) is in the range of    
the NLO weak coupling result of Ref.~\cite{CaronHuot:2008uh} shown in 
Fig.~\ref{fig_NLO} if the value of $\alpha_s \simeq 0.26$ is used; however, as
emphasized in Sec.~\ref{ssec_pqcd}, the perturbative series is badly convergent 
at even smaller values of the coupling.
Weak-coupling techniques could still be useful to better constrain the shape
of the spectral function of the chromo-electric field strength at intermediate 
frequencies. Since these do not involve the $\omega \rightarrow 0$ limit they 
could be useful to guide analyses of lattice calculations.

\subsubsection{Charm fluctuations and correlations and charm degrees of freedom in hot matter}
\begin{figure}[htb]
\centerline{\includegraphics[width=0.55\textwidth]{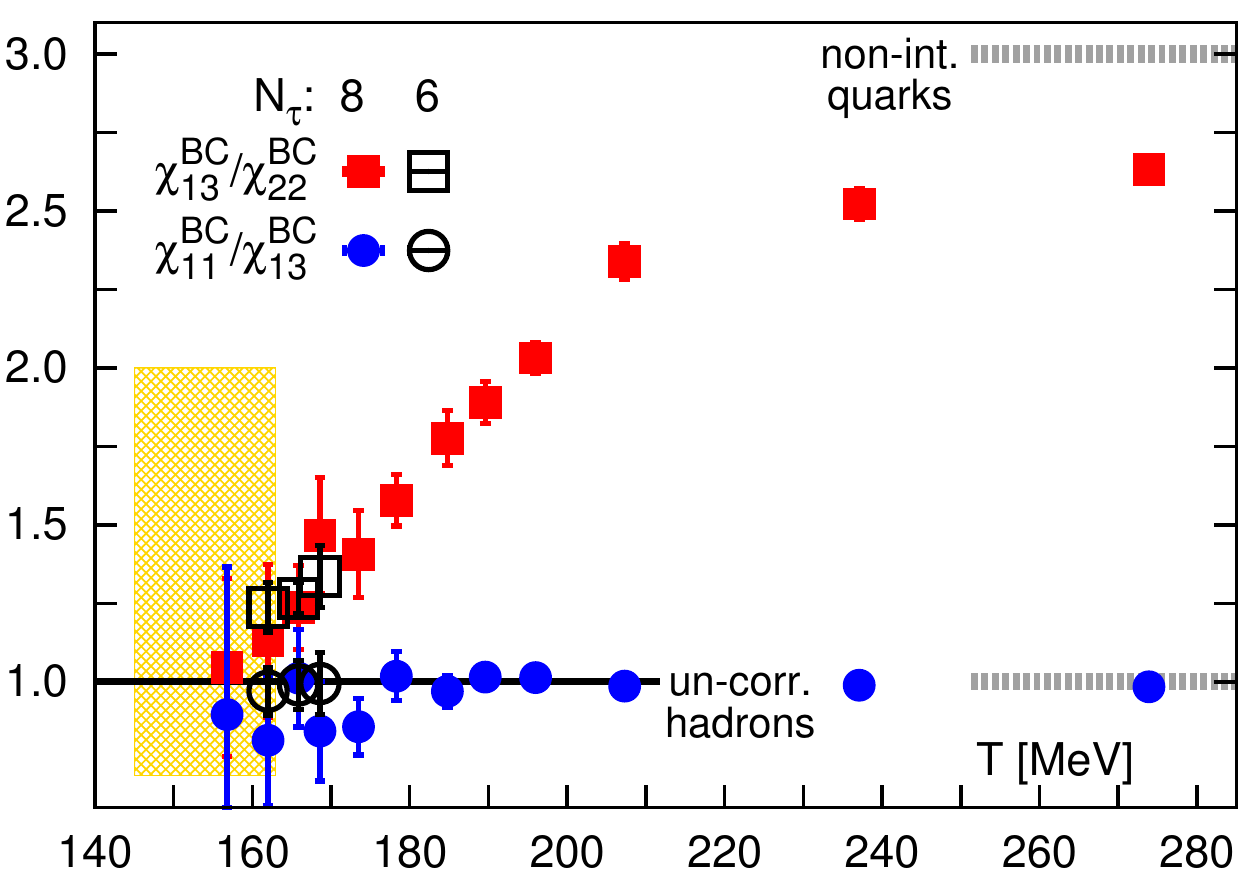}}
\caption{Ratio of baryon number charm correlations as functions of temperatures. The horizontal lines
correspond to HRG and to quark gas. The ratio of correlations involving the same number of derivatives
in baryon chemical potential but same number of derivatives with respect to charm chemical potential
are always one because sectors with $|C|=2,3$ do not contribute because of the large charm-quark 
mass~\cite{Bazavov:2014yba}.}
\label{fig:BC}
\end{figure}

\begin{figure}[th]
\centerline{\includegraphics[width=0.57\textwidth]{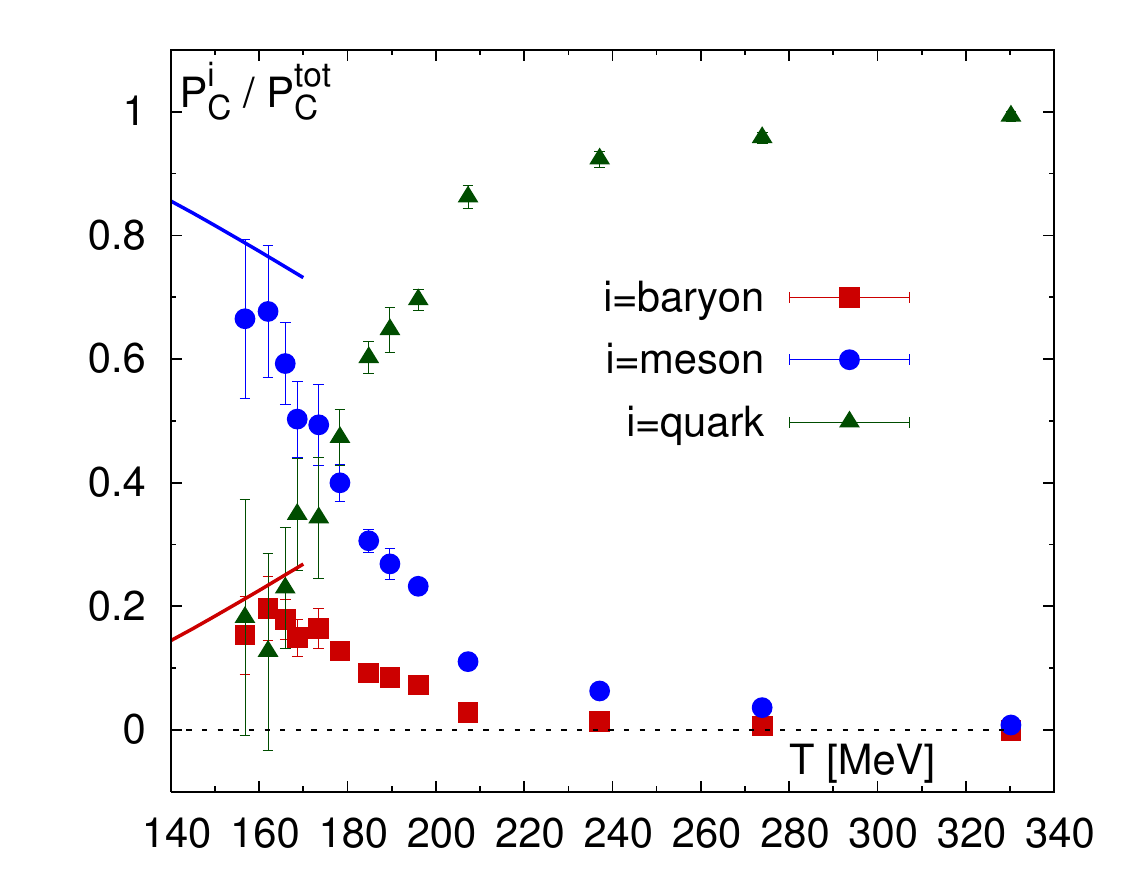}}
\caption{The partial pressure of charm quarks, charm mesons and charm baryons, normalized by
the total charm pressure, as function of the temperature~\cite{Mukherjee:2015mxc}.}
\label{fig:partial}
\end{figure}

Derivatives of the QCD pressure with respect to the chemical potential, 
\begin{eqnarray}
\chi_n^X &=&T^{n} \frac{\partial^n (p(T,\mu_X,\mu_Y)/T^4)}{\partial \mu_X^n} 
\\
\chi_{nm}^{XY}&=&T^{n+m}\frac{\partial^{n+m}(p(T,\mu_X,\mu_Y)/T^4)}{\partial \mu_X^n \partial \mu_Y^m} 
\ , 
\end{eqnarray}
define fluctuations of a conserved charge $X$ or correlations between conserved charge $X$ and 
conserved charge $Y$.  These have been calculated on the lattice including the case of charm 
$X=C$~\cite{Bazavov:2014yba}. Fluctuations and correlation of
conserved charges are sensitive to deconfinement and provide information on the relevant degrees of
freedom. At low temperature the fluctuations and correlations can be understood in terms of hadron 
resonance gas (HRG) model~\cite{Bazavov:2012jq,Borsanyi:2011sw,Bazavov:2014yba}, while at high 
temperatures they can be understood in terms of quark degrees of 
freedom~\cite{Bellwied:2015lba,Ding:2015fca,Bazavov:2014yba,Bazavov:2013uja}.
This is demonstrated in Fig. \ref{fig:BC} in terms of baryon number charm correlations. In fact these correlations
together with charm fluctuations $\chi_2^C$ can clarify the nature of charm degrees of freedom. Below $T_c$ charm fluctuations
and correlations can be described in terms of HRG (cf.~Fig.~\ref{fig:BC}). Above $T_c$ the partial pressure
of the charm degrees of freedom can be written as sum of partial pressures of charm mesons, charm baryons 
and charm quarks~\cite{Mukherjee:2015mxc}.
Using lattice data on $\chi_2^C$, $\chi_{22}^{BC}$ and $\chi_{13}^{BC}$ one can obtain the partial 
pressures of charm quarks, $p_q(T)$, charm mesons, $p_M(T)$, and charm baryons, $p_B(T)$,
which are shown in Fig.~\ref{fig:partial}. At $T_c$ the partial baryon and meson pressures agree with HRG
prediction, while the partial charm-quark pressure is consistent with zero within errors.
As the temperature increases the partial meson pressure and baryon pressure decrease and become very
small for $T>200$ MeV. This can be interpreted as gradual melting of charm hadrons above $T_c$.
The important point here, however, is that hadron like excitations in the open charm sector may exist
above $T_c$. Quarks dominate the charm pressure only for $T>200$ MeV. At these temperatures charm
quark properties, like in-medium mass and width can be extracted from charm fluctuations, $\chi_2^C$, 
see Ref.~\cite{Petreczky:2009cr}.
As shown there the quasi-particle model with $T$-dependent effective charm-quark mass works 
well~\cite{Petreczky:2009cr}.

\subsubsection{Charm meson correlators}
Properties of charm hadrons are encoded in the spectral functions.
Temporal and spatial correlators that can be calculated in lattice QCD
are related to the spectral functions. The temporal correlators are
simple periodic Laplace transformations of the spectral functions. Therefore,
many attempts to reconstruct the spectral functions by using a Bayesian 
approach have been presented in the literature, mostly focusing on hidden 
heavy-flavor mesons (see, \eg,, Ref.~\cite{Mocsy:2013syh}).
Due to the fact the the temporal meson correlators are defined only for
Euclidean time separation $\tau<1/(2 T)$ there is a limited sensitivity
to the in-medium modification of the spectral 
functions~\cite{Petreczky:2008px,Mocsy:2007yj,Riek:2010fk}.

Alternatively, one can consider spatial meson correlation functions, which seem to
be much more sensitive to the in-medium modifications of the spectral 
functions~\cite{Bazavov:2014cta}.
However, the relation of the spatial meson correlators to the spectral functions is
more complicated. It is given by a double integral transformation~\cite{Bazavov:2014cta}.
Nevertheless, some qualitative information on the in-medium modifications of the open-charm 
mesons can be obtained. It turns out that open-charm meson spectral functions
are modified already below $T_c$ \cite{Bazavov:2014cta}. The in-medium modifications
are large above $T_c$, and for $T>250$ MeV the spatial meson correlators are compatible
with the propagation of an uncorrelated quark anti-quark pair, \ie,  with the dissolution 
of $D$-meson states. This is consistent with the findings of the previous section based 
on baryon charm correlations. 

First attempts to study $D$-meson spectral functions have been presented
in Ref.~\cite{Kelly:2017opi} and the findings are in agreement with
the study of spatial correlators.

\subsection{Boltzmann vs. Langevin}
\label{ssec_boltz-lang}

In this section we compare the two transport implementations for HF propagation through
QCD matter that have been most widely employed at low and intermediate momenta, \ie, 
Boltzmann ($BM$) and Langevin ($LV$) approaches. In particular, we will elaborate on both 
benefits and drawbacks of both schemes.   

\subsubsection{Fokker-Planck and Boltzmann transport equations}
The Boltzmann equation for the HQ distribution function can be written in a compact form as:
\begin{equation}
 p^{\mu} \partial_{\mu}f_{Q}(x,p)= {\cal C}[f_q,f_g,f_{Q}](x,p)
\label{B_E}
\end{equation}
where ${\cal{C}}[f_q,f_g,f_{Q}](x,p)$ is the relativistic Boltzmann-like collision integral 
where the phase-space
distribution function of the bulk medium $f_q,f_g$ can be evaluated solving the Boltzmann-equation
also for quarks and gluons~\cite{Ferini:2008he,Ruggieri:2013ova}.

It is well known that the relativistic collision integral for two-body collisions can be written 
in a simplified form~\cite{Prino:2016cni,Svetitsky:1987gq} in the following way:
\begin{eqnarray}
{\cal C}[f_{Q}] = \int d^3q \left[ \,w(\textbf{p+q},\textbf{q})f_{Q}(x,p+q)  
-w(\textbf{p},\textbf{q})f_{Q}(x,p)\right] 
\label{C22}
\end{eqnarray}
where $w(\textbf{p},\textbf{q})$ is the rate of collisions of a heavy quark per unit of 
momentum phase space which changes the its momentum from $\textbf{p}$ to $\textbf{p-q}$. 
It is directly related to the scattering matrix
${\cal M}_{(q,g)+Q \rightarrow (q,g)+Q }$:
\begin{eqnarray}
w(\textbf{p},\textbf{q})=
\frac{1}{128 \, \pi^2} \int \frac{d^3k}{(2\pi)^3}\,f_{q,g}(x,p)\,
\, \frac{|{\cal M}_{(q,g)Q}|^2}{ E_p\,E_k\,E_{p-q}\,E_{k+q}}\, \delta^0 (E_p+E_k-E_{p-q}-E_{k+q})
\end{eqnarray}
we recall that the scattering matrix is the real kernel of the dynamical evolution for both
the Boltzmann approach and the Fokker-Planck one.
Of course all the calculations discussed in the following
will originate from the same scattering matrix for both cases.

The non-linear integer-differential Boltzmann equation can be significantly simplified
employing the Landau approximation whose physical relevance can be associated
to the dominance of soft scatterings with momentum transfers, $q=|\bf q|$, which are 
small compared to the particle momentum, $p$. Namely, one expands 
$w(\textbf{p+q},\textbf{q})f(x,p+q)$ around $q$,
\begin{eqnarray}
w({\bf p+q},\textbf{q})f_{Q}(\textbf{x},\textbf{p+q}) \approx 
w(\textbf{p},\textbf{q})f(\textbf{x},\textbf{p})
+q_i \frac{\partial}{\partial p_i} (\omega f) 
+\frac{1}{2}q_i q_j \frac{\partial^2}{\partial p_i \partial p_j} (\omega f ) 
\label{expeq_000}
\end{eqnarray}
and inserts this into the Boltzmann collision integral, Eq.(\ref{C22}), 
to obtain the Fokker-Planck Equation:
\begin{eqnarray}
\frac{\partial f}{\partial t}=\frac{\partial}{\partial p_i} \left[ A_i({\bf p}) f + \frac{\partial}{\partial p_j}[B_{ij}(\bf p)]\right] \ . 
\label{eq:FP}
\end{eqnarray}
The transport coefficients defined by $A_i=\int d^3q \,w({\bf p}, {\bf q}) q_i=A({\bf p})p_i$ 
and $B_{ij}=\int d^3q \,w({\bf p}, {\bf q}) q_i q_j$
are directly related to the so-called drag ($\gamma$) and momentum diffusion coefficient ($D_p$)
that are determined by the underlying scattering matrix figuring in the transition
probabilities, $w({\bf p}, {\bf q})$.

In a locally isotropic medium, the diffusion tensor $B_{ij}$ can be reduced to two independent
components that determine the diffusion in the directions transverse and longitudinal
relative to the HQ momentum, $B_0$ and $B_1$, respectively:
\begin{eqnarray}
B_0=(\delta_{ij}-\frac{p_i p_j}{p^2}) \, B_{ij} \, \,\,\,\, , \,\,\,\, \,B_1=\frac{p_i p_j}{p^2} \, B_{ij}
\end{eqnarray}
In principle, the 3 transport coefficients are related to each other through the
dissipation fluctuation theorem (DFT), leaving two independent ones. In practice, especially
at high momenta, this is not readily satisfied which limits the applicability of
the Fokker-Planck approximation. Therefore, to ensure the HQ distribution to converge to
the correct the equilibrium distribution, $f_{eq}(p)=e^{-E/T}$, most of the groups have 
enforced the Einstein relation by expressing $B_1$ through $A(p)$. The deviations from
using the explicitly calculated coefficients can serve as a quality check of the 
approximation~\cite{vanHees:2004gq}.
For the following study, we adopt the same implementation as in the calculations reported
in Secs.~\ref{sec_bulk} and ~\ref{sec_hadro}, \ie, the drag ($A$) and transverse 
diffusion ($B_0$) coefficients are calculated from the pQCD*5 matrix elements and 
the longitudinal one is adjusted to $B_1 = TEA$ within the 
post-point Ito scheme of realizing the Langevin process.\footnote{The BM vs.~LV comparison 
can, in fact, help to assess different implementations of the Einstein relation (or 
DFT) which is usually not automatically satisfied for 
a given model calculation of the different transport coefficients. It turns out that 
employing the friction coefficient, $A$, from the model calculation to enforce the DFT 
for the longitudinal diffusion coefficient (as done here) results in better agreement 
with the BM results than, \eg, using the calculated $B_1$ and readjust $A$ (and 
$B_0$); typically, the calculated $A$ and $B_0$ are better compatible with 
the DFT than $B_1$ in relation to $A$ or $B_0$.}
The Boltzmann equation is solved numerically by dividing coordinate space into a
three-dimensional lattice and using the test particle method to sample
the distributions functions.  The collision integral is solved by mean of a stochastic 
implementation of the collision probability~\cite{Ferini:2008he,Ruggieri:2013ova,Xu:2004mz}.

Before turning to the numerical results in the following section, we recall that the 
semi-classical nature of the Boltzmann equation implies that the surrounding medium consists 
of well-defined quasi-particles, \ie, quantum effects inducing a finite energy resolution  
for a given momentum state are neglected. However, the Fokker-Planck equation, realized via 
a more general Langevin process, does not rely on this assumption; the underlying transport 
coefficients can be evaluated with the full off-shell effects included in the medium's spectral 
function~\cite{KadanoffBaym:1962,Danielewicz:1982kk,Berrehrah:2014kba,Liu:2016ysz}. 
This situation may be relevant for a strongly coupled QGP where strong quantum effects due to 
intense rescattering among the medium particles could render the Langevin approach preferrable 
over the Boltzmann one, provided the HQ mass is large enough to warrant a soft-interaction 
approximation while maintaining the heavy quark a good quasi-particle.    

\subsubsection{Numerical Results Comparing Boltzmann and Langevin Simulations}
\label{sssec_bm-lv-num}
To investigate the differences that arise from the two different transport implementations,
we report test calculations performed with simple LO pQCD HQ scattering matrix elements in 
two versions
(a) with a constant coupling $\alpha_s=0.4$ and $K$ factor, similar to what was done for the 
comparisons in Sec.~\ref{sec_bulk}, as
well as for a quasiparticle model (QPM)~\cite{Plumari:2011mk}. 
We will consider four values for the HQ mass, corresponding to values for 
charm ($M_Q$=1.3\,GeV) and bottom quarks ($M_Q$=5\,GeV), and two intermediate values
($M_Q$=2 and 3\,GeV). 
We expect that the differences between the Boltzmann and the Langevin dynamics are regulated 
by the ratio $M_Q/T$, but a key role is also played also by the differential cross section 
that determines the momentum transfer per collision~\cite{Das:2013kea} as well as the
masses of the medium particles (assumed to be massless for the pQCD case (which differs
from the studies in Sec.~\ref{sec_bulk}) and massive for the QPM).

\begin{figure}[t]
\begin{center}
\includegraphics[width=30pc,clip=true]{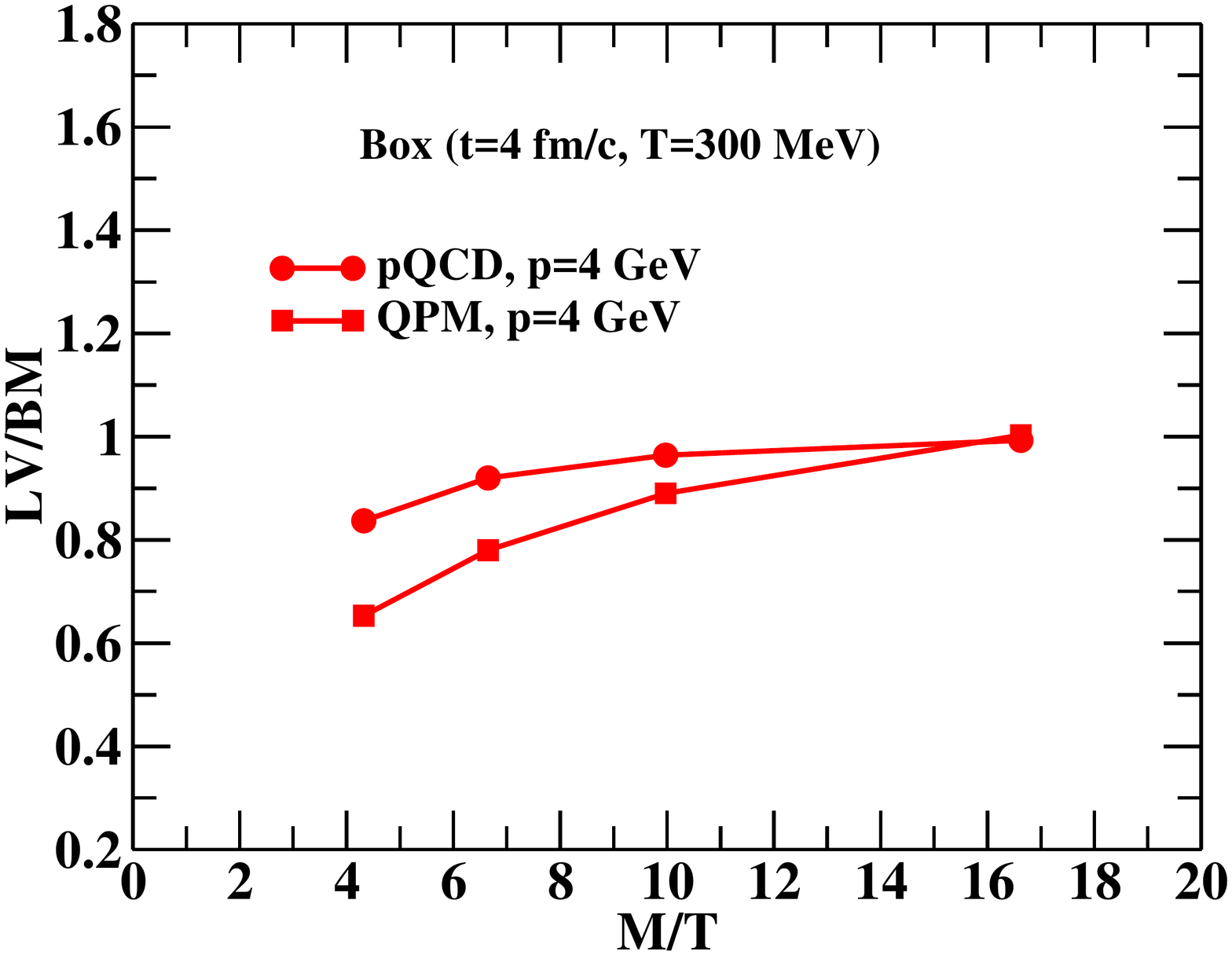}
\caption{Ratio of the HQ $R_{\rm AA}$ in a Langevin simulation over that in a Boltzmann 
simulation for HQ transport in a massless QGP in a box at fixed temperature, $T$=300\,MeV, 
recorded after an evolution time of $t$=4\,fm  for a value of $p$=4\,GeV of the HQ 
momentum. The results are plotted as a function of the HQ mass scaled by temperature, 
and the different symbols correspond to different underlying HQ interactions with 
medium partons.
}
\label{fig:bm-lv}
\end{center}
\end{figure}
During the expansion of the QGP matter in AA collisions, $M_Q/T$ increases by about a factor 
of three due to the wide range of temperatures explored by the expanding QGP matter
(\eg, $T$$\simeq$160-500\,MeV at the LHC), and an additional amount due to a variation of the 
in-medium HQ masses, usually leading to an increase of the mass as $T_{\rm c}$ is approached 
from above (\eg, from $m_c$=1.3\,GeV to 1.8\,GeV~\cite{Riek:2010fk}), which augments the 
range of the $M_Q/T$ values to $\sim$3-11). A study as a function of $M_Q/T$ is 
therefore most transparent if carried out in a box of bulk matter at fixed $T$ with 
periodic boundary conditions. Toward this end we have performed a calculation in a box 
at $T$=300\,MeV for different $M_Q/T$ values for the case of an underlying scattering 
matrix in the HQ scattering off the medium partons from 
(a) a ``pQCD*3" scenario (similar to Sec.~\ref{ssec_com-trans} with $\alpha_s=0.4$ and 
$m_D=gT$ (corresponding to $m_D\simeq0.67$\,GeV), but with massless medium partons and 
a $K$-factor of 3 to reproduce the same charm-quark diffusion coefficient as in the pQCD*5 
scenario (which uses massive partons), and (b) a QPM with $\alpha_s\simeq0.62$ and 
$m_D$=0.85\,GeV in a medium of massive partons with $m_g$=0.69\,GeV and $m_q$=0.46\,GeV. 
In Fig.~\ref{fig:bm-lv} we show the ratio of the momentum distribution, 
$LV/BM=f_{LV}(p,t_f)/f_{BM}(p,t_f)$, at a time $t_f$=4\,fm, determined such that 
$R_{AA}\approx \, 0.4$ for the $BM$ case, in the range of what has been found in 
Sec.~\ref{ssec_com-trans}, recall.~Fig.~\ref{fig:c-avg}.
This condition is chosen with the aim of 
comparing $BM$ and $LV$ dynamics under conditions that mimic the one observed experimentally,
even if we are considering a bulk matter at fixed $T$. 

For both HQ interactions, the $b$-quark case leads to negligible deviations in the
HQ $R_{AA}$ between Boltzmann and Langevin simulations. For the  pQCD*3 case (filled
circles in Fig.~\ref{fig:bm-lv}), also the lower $M_Q/T$ values, in the range of possible 
$c$-quark masses near $T_c$, only lead to  small to moderate deviations, in the 5-20\% range.
The filled squares, corresponding to the $LV/BM$ ratio in the QPM model are larger, 
leading to a deviation of up to 35\% for the smallest $M_Q/T$ ratio of 4 (or $\sim$25\% for
$M_Q/T$=6.7 applicable for $c$ quarks at lower temperatures, $T$$\le$250\,MeV).
Here, the larger Debye mass, $m_D$, and the heavier medium scattering centers in the QPM,
relative to the (massless) pQCD*3 scenario, 
lead to a differential scattering cross section which is more isotropic which affects 
the small-momentum transfer approximation in the Langevin process.
We also note that the differences are expected to become smaller again at lower momenta
and larger times as the HQ distributions get closer to the universal equilibrium limit
which is of course realized in both $BM$ and $LV$ approaches.

\begin{figure}[t]
\begin{center}
\includegraphics[width=0.49\textwidth]{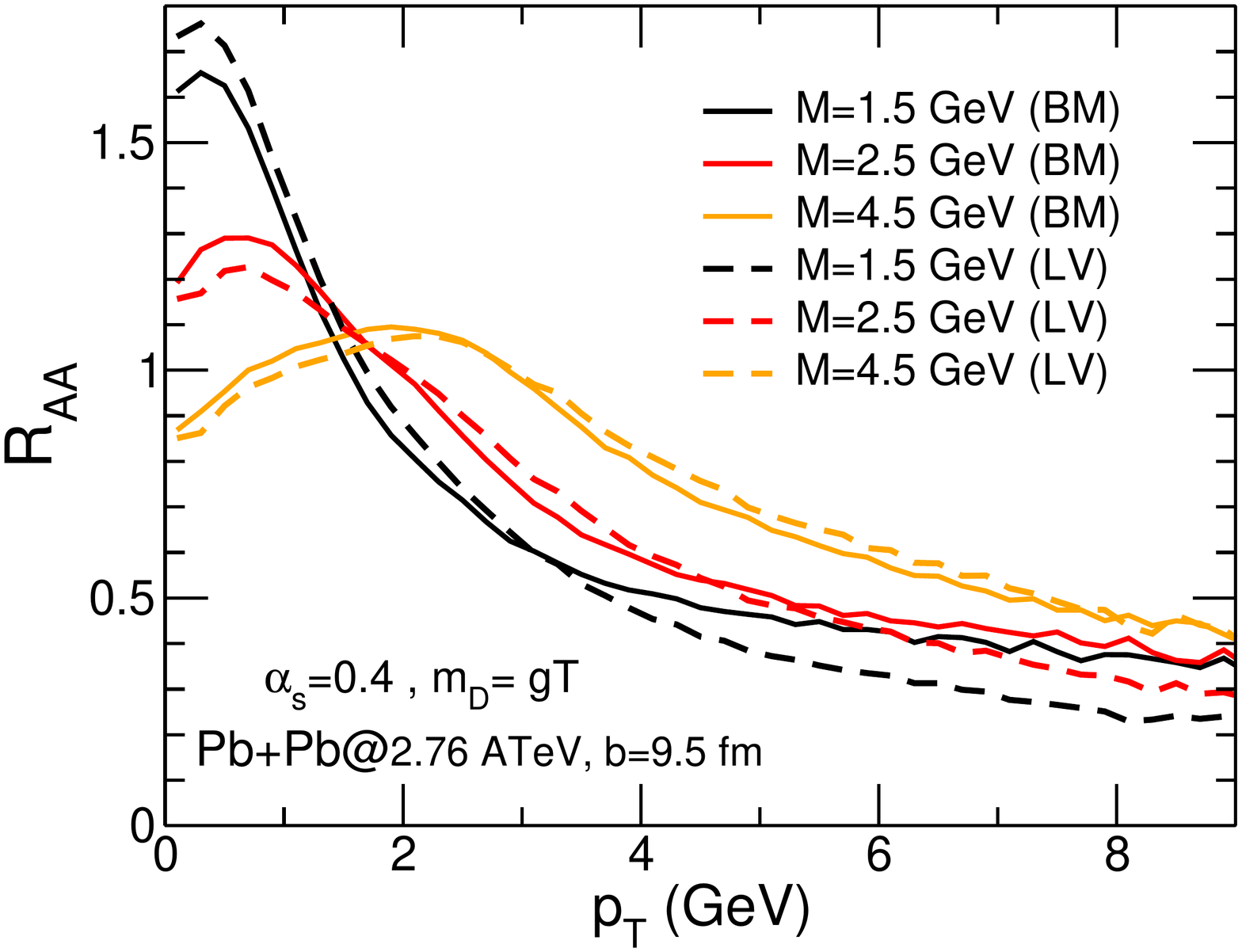}
\includegraphics[width=0.5\textwidth]{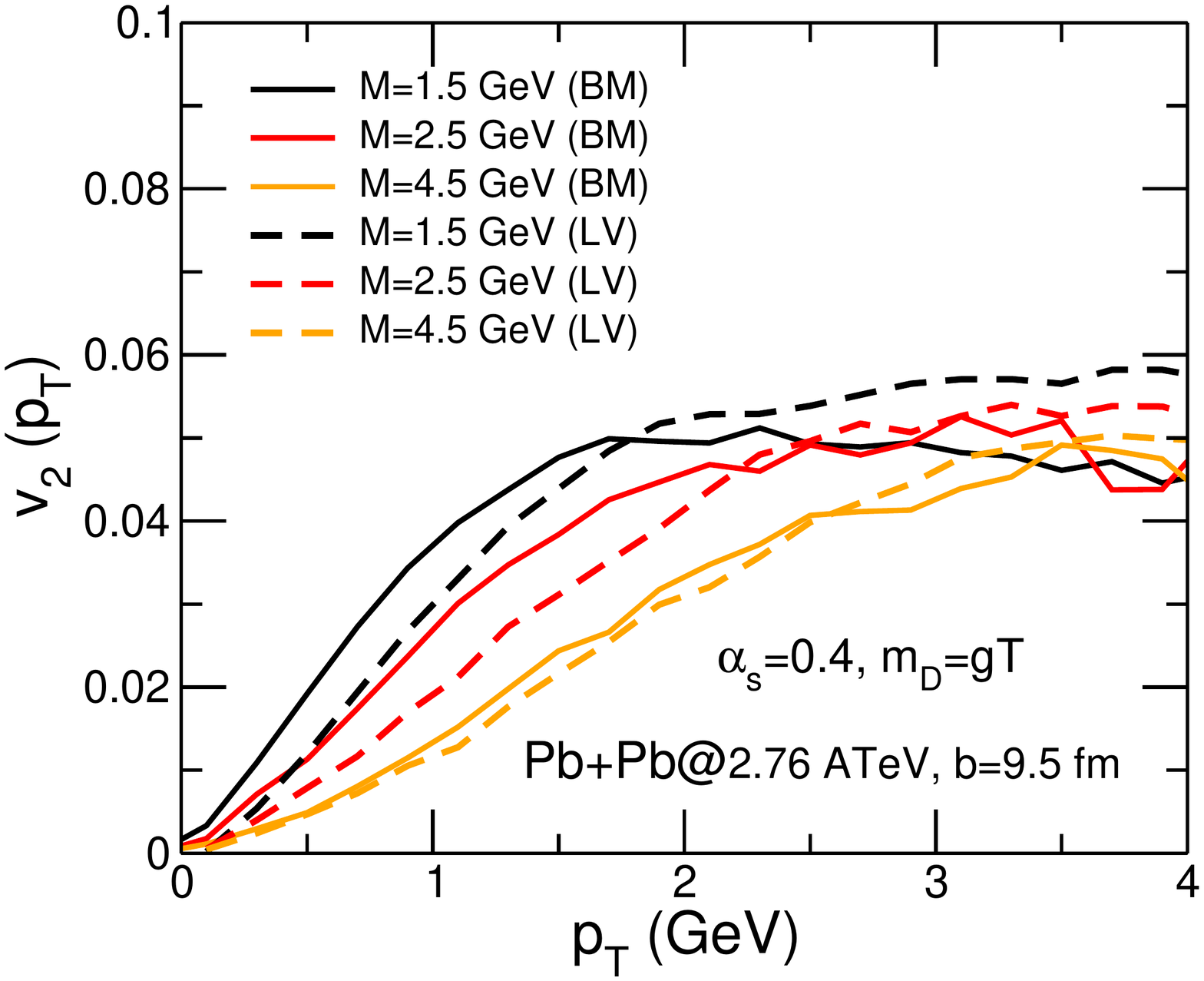}
\caption{Nuclear modification factor (left panel) and elliptic flow (right panel) for 
heavy quarks in semi-central Pb+Pb($\sqrt{s_{NN}}=2.76 \, \rm TeV$) collisions (at $b=9.5$\,fm) 
for different values of the HQ mass, $M_Q$ (indicated by the different line colors), in a 
Boltzmann (solid lines) and in a Langevin approach (dashed lines).}
\label{fig-raa-v2}
\end{center}
\end{figure}
Finally, a comparison has been made under conditions of a more realistic simulation for 
Pb+Pb($\sqrt{s_{NN}}$=2.76\,TeV) collisions at impact parameter $b$=9.5\,fm, 
as were considered in Sec.~\ref{sec_bulk} for the 30-50\% centrality class, using the 
same underlying scattering matrix as employed there, \ie, 
LO pQCD with $\alpha_s=0.4$. In Fig.~\ref{fig-raa-v2} the pertinent nuclear modification
factor and elliptic flow are plotted for heavy quarks of varying mass at the end of the QGP 
phase. Boltzmann transport generally leads to a larger $R_{AA}$ relative to Langevin dynamics.
For a mass of $M_Q=1.5$\,GeV corresponding to charm quarks, the $R_{AA}$ from $BM$ is about 
25-30\% larger in the intermediate $p_t$ region where it tends to saturate.
On the other hand, the elliptic flow is slightly larger for $BM$ dynamics at low $p_t$
but smaller at high $p_t$. For bottom quarks, both $R_{AA}(p_t)$ and $v_2(p_t)$ are 
nearly identical in the two approaches.
In general, the deviations become larger with increasing $p_t$ as the Gaussian distribution
in the energy loss underlying the Langevin approximation becomes less accurate while the
BM approach captures the full differential distribution following from the microscopic
scattering matrix element. In addition, one expects radiative contributions to become
relevant, whose interferences effects are not easily captured in either BM or LV 
descriptions.
The differences increase slightly for more central collisions, as can be expected for a longer 
duration of the QGP medium. 
While there are no significant differences of $R_{AA}$ and $v_2$ for bottom quarks, 
some difference can arise for more exclusive observables, \eg, angular correlations
between $B$ and $\bar B$~\cite{Scardina:2017}.

\section{High-$p_T$ Energy Loss and $\hat{q}$}
\label{sec_eloss}

In this section we discuss various aspects pertaining to the description of high-$p_T$
heavy quarks propagating through the QGP. While the previous sections focused on elastic
interactions which are parametrically dominant at low $p_T$, radiative processes are 
expected to become dominant at high $p_T$. The understanding of the transition between the 
two regimes is an important ingredient for quantifying the temperature and momentum
dependence of HQ transport coefficients. In Secs.~\ref{ssec:qhat}  and \ref{ssec:jet-energy} 
we discuss basic ingredients to, and definitions of, high-$p_T$ transport coefficients, most
commonly quantified via the average transverse-momentum transfer per mean-free-path, $\hat{q}$.  
In Secs.~\ref{ssec_dglv} and \ref{ssec_scet} we discuss two different state-of-the-art 
approaches to high-energy HQ energy loss (which include radiative contributions and their 
coherence), and in Sec.~\ref{ssec_delE} we compare their results for the path length 
dependence of the fractional energy loss to other implementations used in HF phenomenology.

\subsection{Transverse momentum broadening and QGP properties}
\label{ssec:qhat}

Different approaches to calculating the non-abelian parton energy  loss are available in the
literature~\cite{Zakharov:2000iz,Baier:1996sk,Gyulassy:2000er,Wang:2001ifa,Arnold:2002ja}. 
The medium-induced radiative spectrum  generally depends  on the multiple scattering of the 
propagating parton in the medium and the transverse-momentum transfer distribution in the 
scatterings. These can be schematically expressed as 
$ \int {d\Delta z}  \; {1}/{\lambda_g(z)} \cdots $  and  
$\int  d^2{\vc q}_\perp \; {1}/{\sigma_{el}} {d\sigma_{el}^{\; {\mathrm{med}}}}/{d^2 {\vc q}_\perp}\cdots$,
respectively. Note that for soft gluon emission in the eikonal limit only the gluon scattering
length enters the expression for the medium-induced radiative spectrum. It is obvious that without 
further approximations this spectrum depends on $\lambda_g(z)$ and the typical inverse range of the 
interaction, $m_D \sim 1/r_D$.

There are several possibilities for relations between the interaction length and the momentum transfer
from the QCD medium.
\begin{itemize}
\item The interaction length and momentum transfer are largely independent, providing a 2D 
parameter space. Such a scenario  would require rather involved multi-parameter fits to data and 
has not been explored so far in the literature.
\item  Assuming local thermal equilibrium, density and temperature can be related at any space 
time point. The range of the interaction and parton scattering cross section can be estimated  
and depend on the typical coupling between the jet and the medium gluon. The interaction length 
is then obtained form the QGP density and the scattering cross section~\cite{Gyulassy:2000er}.
\item  One can use thermal field theory to relate the relevant medium parameters to the temperature 
$T$. This is similar to the situation described in the previous item, but without explicitly 
evaluating the scattering cross sections and densities~\cite{Arnold:2002ja}.
\item  An approach to energy loss in the limit of infinite energies and infinite number 
of scatterings assumes that at {\em any scale} the  transverse-momentum broadening of any size 
is given by a 2D Gaussian random walk~\cite{Zakharov:2000iz,Baier:1996sk}. By discarding the 
detailed kinematic information that pertains to  parton scattering one can relate the radiative 
intensity spectra to the transport parameter as $\hat{q} \sim m_D^2 /\lambda_g$.
\item In deep inelastic scattering the radiative spectrum can be related to higher-twist 
matrix elements of field operators~\cite{Wang:2001ifa}. The interaction length can be thought 
of as the inter-nucleon distance. The application to the QGP case is by analogy.
\end{itemize}

The momentum broadening of a parton that propagates in dense QCD mater is often discussed in 
relation to parton energy loss. An impact parameter space resummation is used together with 
a small-impact parameter approximation of the Fourier transform of the differential scattering 
cross section. Although such an approach is  analytically appealing, it gives jet distributions 
that may differ substantially from the exact formula~\cite{Gyulassy:2002yv}.
Let us elaborate on this in more detail. 
The normalized elastic cross section (triggering a gluon emission) in Fourier space can be 
written as
\begin{equation}
\frac{d \tilde{\sigma}_{el} }{d^2 {\bf q}}({\bf b}) =
\int \frac{d^2 {\bf q}}{(2 \pi)^2} \; e^{-i{\bf q} \cdot {\bf b}}
\frac{1}{\pi} \frac{m_D^2}{({\bf q}^2+m_D^2)^2} =
\frac{m_D \, b}{4 \pi^2} K_1(m_D\, b)   \approx
\frac{1}{4 \pi^2} \left(1- \frac{\xi \, m_D^2\, b^2}{2}
+ {\cal O}(b^3)  \right) \ ,
\label{ftdc}
\end{equation}
where $b=|{\bf b}|$ is the magnitude of the impact parameter vector; in the quadratic term 
in Eq.~(\ref{ftdc}) a factor $\log [2/(1.08\, m_D\, b)]$ has been absorbed into a $b$-independent 
constant, $\xi$, which is the source of the leading logarithmic energy dependence in momentum 
space. Starting from a jet propagating in the ``$\hat{z}$'' direction, 
$dN^{(0)}/d^2 {\bf p} = \delta^2({\bf p})$, the approximation of large-number, small-impact parameter 
scatterings reduces the momentum space distribution to a classic Moliere form, 
\begin{equation}
dN({\bf p}) = \int d^2{\bf b} \; e^{i{\bf p}\cdot {\bf b} }
\frac{1}{(2\pi)^2} \frac{e^{-\frac{\chi\, m_D^2 \, \xi \, b^2}{2}}  }
{\chi\, m_D^2 \, \xi }
= \frac{1}{2\pi} \frac{e^{-\frac{p^2}{2\, \chi  \, m_D^2 \, \xi} }}
{\chi\, m_D^2\, \xi }
 \ .
\label{gauss}
\end{equation}
The resulting distribution is of Gaussian form with a width of $\chi\, m_D^2\, \xi$. Within
this Gaussian approximation the average broadening is 
$$\left\langle {\bf p}^2  \right\rangle = 2\chi\, m_D^2\, \xi \ ,   \qquad \xi \sim O(1)\;.$$
The factor of 2 arises from the two-dimensional random walk. The opacity $\chi$ is the number
of scatterings. With these caveats, for a non-expanding homogeneous medium a transport coefficient
$ \hat{q}  = 2  {m_D^2}/{\lambda} $ can characterize the typical {\em soft} momentum transfer
between the jet and the medium.

In heavy-ion collisions it is not possible to define a model-independent length, as
the medium has a varying density as a function of position and time. The Gaussian broadening 
result can be rewritten as
$$\left\langle {\bf p}^2  \right\rangle = \int \hat{q}(z)  \, d \Delta z  \ , \qquad    
\hat{q}(z)  = 2 \frac{m_D^2(z)}{\lambda_g (z)}.   $$
Even though the transport parameter $\hat{q}$ alone does not even describe the simpler problem of 
transverse momentum broadening, the above definition captures properties of the medium 
without mixing in large logarithms of the jet energy. It should be noted that presently there exists 
no derivation of the strong-coupling constant renormalization in the presence of a medium. 
Therefore, the coupling constant in the above formulas is kept fixed.

One should not overrate the meaning of the quantitative values for $\hat{q}$ since they 
are rather model dependent. These values will differ at different space-time points 
(${x_\perp, \tau})$, as the density, temperature and transport
properties depend on both the geometry and evolution of the medium. If  averages are performed, 
the way in which different space-time points are weighted must be explicitly specified.
The same applies to the temperature, Debye screening scale, interaction length and 
combinations thereof. Even in this case there will be residual dependences on the type of 
thermal QCD medium that is assumed, \eg,  a gluon-dominated plasma vs.~a quark-gluon plasma, or
the number of active quark flavors.

\subsection{Energy dependence of the transport coefficient}
\label{ssec:jet-energy}
In this subsection, we concentrate on a dynamical QCD medium, \ie, the thermal motion and
recoil of the constituents is accounted for.
However, we note that the derivation and subsequent discussion
is similar in the static medium case, so this section (with
straightforward corrections) is applicable to static QCD medium as well.

In a dynamical QCD medium, the perturbative interactions between a high-energy parton
and the QGP can be characterized by the HTL resummed elastic collision rate~\cite{Burke:2013yra}
\begin{equation}
\frac{d\Gamma_{\rm el}}{d^2 \mathbf{q}}
= C T^3 \frac{\alpha_s^2 } {\mathbf{q}^2(\mathbf{q}^2 + m_D^2)} \ , 
\label{CollRate}
\end{equation}
where $C=4 C_A (1+N_f/6)$ is a constant with $C_A$ the value of the Casimir operator for the 
propagating parton and $N_f$ the number of active light-quark flavors in the QGP. 

The transport coefficient, \ie, the average transverse-momentum transfer
squared per mean-free-path, is then defined as
\begin{equation}
\hat{q} = \int^{q_{\rm max}} d^2 \mathbf{q}\, \mathbf{q}^2
\frac{d\Gamma_{\rm el}}{d^2 \mathbf{q}}
\label{qhat}
\end{equation}
where $q_{\rm max} \approx 6ET$ is an ultraviolet (UV) cut-off. If coupling constant is assumed 
to be constant, $\hat{q}$ reduces to~\cite{Burke:2013yra}
\begin{equation}
\hat{q} \approx C T^3 {\alpha_s^2 \ln(6 E T /m_D^2)}.
\label{qhatEdep}
\end{equation}

As the transport coefficient is a {\it medium property}~\cite{Burke:2013yra}
that controls the parton energy loss, this parameter should not depend on the
energy of the jet. However, from Eq.~\ref{qhatEdep}, we see that $\hat{q}$ has
a logarithmic jet energy dependence.

A practical prescription to remove the energy dependence from the transport coefficient is to 
include the running of the strong coupling constant as in vacuum in the jet-medium interaction 
vertices. That is, if the running coupling is defined as in Ref.~\cite{Field} (where $\Lambda_{QCD}$ 
is the perturbative QCD scale),
\begin{equation}
\alpha_s (Q^2)=\frac{4 \pi}{(11-2/3 N_f) \ln (Q^2/\Lambda_{QCD}^2)},
\label{alpha}
\end{equation}
then Eq.~(\ref{qhat}) can be solved similarly to the procedure in Ref.~\cite{Peigne:2008nd}, yielding
\begin{equation}
\hat{q} \approx C T^3 \, \alpha_s(m_D^2) \, \alpha_s(6 E T) \,  \ln(6 E T /m_D^2) \ ,
\end{equation}
which straightforwardly leads to
\begin{equation}
\hat{q} \approx C T^3 \, \frac{4 \pi}{(11-2/3 N_f)} \, \alpha_s^2(m_D^2) \ .
\label{qhatNoEdep}
\end{equation}
Therefore, to leading logarithmic accuracy, this  leads to a cancellation of the
logarithmic terms that arise from the power-law tails of Moliere multiple
scattering and consequently to a transport coefficient that does not depend on the jet energy.

\subsection{Dynamical energy loss formalism}
\label{ssec_dglv}
The dynamical energy loss formalism is an approach based on finite-temperature
field theory with a hard-thermal loop (HTL) resummation that incorporates 
that the scattering partners in the QGP are dynamical (\ie, moving) partons. 
Furthermore, it takes into account finite-size effects of the medium as relevant
for QGP droplets created in URHICs. The main ingredients of this model are: 
i) radiative~\cite{Djordjevic:2009cr,Djordjevic:2008iz} and collisional~\cite{Djordjevic:2006tw} 
energy loss computed in the same theoretical framework (finite-size dynamical QCD medium), 
ii)  magnetic-mass (non-perturbative) effects~\cite{Djordjevic:2011dd} in radiative energy loss, 
consistently included into the energy loss through a sum rule procedure, 
iv) a running coupling~\cite{Djordjevic:2013xoa}. 
The model is implemented within a numerical procedure that takes into account parton 
production and fragmentation functions, as well as path length and multi-gluon 
fluctuations~\cite{Djordjevic:2013xoa}. 
Additionally, the dynamical energy loss uses no fit parameters and is able to treat both 
light and heavy partons, so that it can provide predictions for an extensive set of 
observables. The details of the framework are briefly outlined below.
\\

{\bf General framework}\\
To calculate the quenched spectra of final hadrons, the formalism uses the generic pQCD convolution
\begin{eqnarray}
\frac{E_f d^3\sigma}{dp_f^3} = \frac{E_i d^3\sigma(q)}{dp^3_i}
 \otimes
{P(E_i \rightarrow E_f )}
\otimes D(q \to H_q) \otimes f(H_q \to e, J/\psi) \ ;  
\label{schem} 
\end{eqnarray}
the subscripts "$i$"  and "$f$" correspond, respectively, to ``initial" and ``final", $q$ denotes 
both quarks and gluons, while the different factors on the {\it rhs}  mean the following: 
\begin{itemize}
\item[{\it (i)}] $E_i d^3\sigma(Q)/dp_i^3$ denotes the initial parton spectrum. 
The spectrum is extracted from Ref.~~\cite{Kang:2012kc} for gluons and light quarks, 
and from Ref.~\cite{Cacciari:2012ny,Cacciari:1998it} for charm and bottom quarks. 
\item[{\it (ii)}]  $P(E_i \rightarrow E_f )$ is the energy loss probability, generalized to 
include both collisional and radiative energy loss in a realistic finite-size dynamical QCD 
medium (a short review on the energy loss mechanism is given in the following 
subsection), as well as multi-gluon~\cite{Gyulassy:2002yv} and path length 
fluctuations~\cite{Wicks:2005gt}. 
\item[{\it (iii)}] $D(q\to H_q)$ is the fragmentation function of quark/gluon $q$ to hadron $H_q$.  
We use DSS~\cite{deFlorian:2007aj}, BCFY~\cite{Braaten:1994bz} and KLP~\cite{Kartvelishvili:1977pi} 
fragmentation functions for light hadrons, $D$ mesons and $B$ mesons, respectively.
\item[{\it (iv)}] In the case of heavy quarks, there can also be a decay of hadron $H_q$ into 
single electrons or $J/\psi$. This is represented by the functions $f(H_q \to e, J/\psi)$. The 
decays of $D$ and $B$ mesons to non-photonic single electrons, and decays of $B$ mesons to 
non-prompt $J/\psi$ are obtained according to Ref.~\cite{Cacciari:2012ny}.
\end{itemize}
In the dynamical energy loss calculations, the four steps outlined by Eq.~(\ref{schem}) are treated 
separately, in the order defined by the above expression.
\\

{\bf Assumptions}\\
The following assumptions are used in the dynamical energy loss approach:
\begin{itemize}
\item[{\it (i)}] The final quenched energy is sufficiently large so that the Eikonal approximation 
can be employed.
\item[{\it (ii)}] The radiative and collisional energy loss can be treated separately, so that the change 
of the spectrum can be first calculated due to radiative, and then due to collisional energy loss. This 
approximation is reasonable when collisional and radiative energy loss processes can be decoupled from 
each other (which follows from the HTL approach~\cite{Djordjevic:2007hp} that is used in dynamical energy 
loss calculations) and when the radiative/collisional energy losses are sufficiently small (which is
the essence of the soft-gluon, soft-rescattering approximation).
\item[{\it (iii)}] The parton-to-hadron fragmentation functions are the same for Pb+Pb and 
$e^+e^-$ collisions; this is expected to be valid for hadronization outside the QGP.
\item[{\it (iv)}] Multiple gluon emissions can be independently treated in multi-gluon fluctuations. 
This is a reasonable assumption~\cite{Blaizot:2012fh,Fickinger:2013xwa} within 
(the above mentioned) soft-gluon approximation.
\end{itemize}

\vspace{0.5cm}

{\bf Energy loss calculations} \\
The expression for the radiative energy loss in a finite-size dynamical QCD 
medium~\cite{Djordjevic:2009cr,Djordjevic:2008iz}, 
obtained within the HTL approximation, at $1^{st}$ order in opacity is given by:
\begin{eqnarray}
\frac{\Delta E_{rad}}{E} = \frac{C_R {\alpha_s}}{\pi} \frac{L}{\lambda_{\rm dyn}}
\int dx \frac{d^2k}{\pi} \frac{d^2q}{\pi} \,v(\mathbf{q}) \left(1{-}
\frac{\sin \frac{(\mathbf{k}{+}\mathbf{q})^2{+}\chi}{x E^+}L}
{\frac{(\mathbf{k}{+}\mathbf{q})^2{+}\chi}{x E^+} L}\right)
  \frac{2 (\mathbf{k}{+}\mathbf{q})}{(\mathbf{k}{+}\mathbf{q})^2{+}\chi} 
\left(\frac{(\mathbf{k}{+}\mathbf{q})}{(\mathbf{k}{+}\mathbf{q})^2{+}\chi}{-}
\frac{\mathbf{k}}{{\mathbf{k}}^2{+}\chi}\right) \ .
\label{eloss}
\end{eqnarray}
Here, $E$ is initial parton energy, $L$ is the length of the QGP fireball, 
$C_R=\frac{4}{3}$ and  $C_2(G)=3$; $\bk$ and $\bq$ denote 
transverse momenta of radiated and exchanged (virtual) gluons, respectively, 
$\chi\equiv M_q^2 x^2 + m_g^2$ where $M_q$ is the bare quark mass and $x$ is the longitudinal 
momentum fraction of the quark carried away by the emitted gluon; $m_g=m_D/\sqrt 2$ is 
the effective (asymptotic) thermal mass for gluons with hard momenta 
$k\gtrsim T$~\cite{Djordjevic:2003be}; ${\lambda_{\rm dyn}}=(3 {\alpha}_T)^{-1}$ is the 
mean-free-path in the dynamical QCD medium; $v(\textbf{q})$ corresponds to the effective cross 
section, which for the case of finite magnetic mass~\cite{Djordjevic:2011dd}, $m_M$, is given by:
\begin{eqnarray}
v(\mathbf{q})=\frac{{m^2_D}-{m^2_M}}{({\mathbf{q}}^2 + {m^2_D}) ({\mathbf{q}}^2+ {m^2_M})} \ ,
\label{vqMagMass}
\end{eqnarray}
reducing to a well-known HTL effective cross section~\cite{Arnold:2002ja,Djordjevic:2009cr} 
in the case of zero magnetic mass:
\begin{eqnarray}
v(\mathbf{q})=\frac{{m^2_D}}{{\mathbf{q}}^2({\mathbf{q}}^2 + {m^2_D}) } \ .
\label{vqNoMagMass}
\end{eqnarray}

In dynamical energy loss approach, Eq.~(\ref{vqMagMass}) is dominantly used, 
since several non-perturbative approaches~\cite{Maezawa:2010vj,Nakamura:2003pu}, 
suggest that at RHIC and the LHC magnetic mass is different from zero, $0.4 < m_M/m_D < 0.6$.

Collisional energy loss, calculated in the finite-size dynamical QCD medium (\ie, in a framework 
consistent with the radiative energy loss) is discussed in detail in Ref.~\cite{Djordjevic:2006tw}.  
For collisional energy loss, Eq.~(14) from this reference is used (not spelled out here since
it is rather lengthy).
\\

{\bf Running coupling}\\
The running coupling is defined as in Ref.~\cite{Field}:
\begin{equation}
\alpha_s (Q^2)=\frac{4 \pi}{(11-2/3N_f) \ln (Q^2/\Lambda_{QCD}^2)}.
\end{equation}
Here, $N_f$=2.5(3) is used at RHIC (LHC) and $\Lambda_{QCD}=0.2$ GeV as the perturbative QCD scale.

In the case of the running coupling, the Debye mass $m_D$~\cite{Peshier:2006ah} 
is obtained by selfconsistently solving the equation:
\begin{eqnarray}
\frac{m^2_D}{{\Lambda}^2_{QCD}} \ln\left(\frac{m^2_D}{{\Lambda}^2_{QCD}}\right)
=\frac{1+N_f/6}{11-2/3N_f} \left(\frac{4\pi T}{{\Lambda}_{QCD}}\right)^2 \ .
\label{mu}
\end{eqnarray}
For the collisional energy loss, the coupling was introduced according to Ref.~\cite{Peigne:2008nd}, 
while for the radiative energy loss the coupling was introduced according to Ref.~\cite{Djordjevic:2013xoa}.
\\

{\bf Path length and multi-gluon fluctuations} \\
Path length fluctuations take into account that partons can traverse different paths through the QGP
fireball, while multi-gluon fluctuations take into account that the energy loss is a distribution. 

For the radiative energy loss, the numerical method for including multi-gluon fluctuations is 
based on the approach developed in Ref.~\cite{Gyulassy:2002yv}. A generalization of this approach 
is developed for the dynamical energy loss case~\cite{Djordjevic:2013xoa}, which includes energy 
loss in a finite-size dynamical QCD medium, together with magnetic-mass and running-coupling effects.

For collisional energy loss, the full fluctuation spectrum is approximated by a 
Gaussian~\cite{Moore:2004tg,Wicks:2005gt}. The mean of the Gaussian is determined by the average 
energy loss and the variance by
$\sigma_{coll}^2 = 2 T \langle \Delta E^{coll}(E_i,L)\rangle$, where $\Delta E^{coll}(E_i,L)$ is 
given by Eq.~(14) in Ref.~\cite{Djordjevic:2006tw}.

Path length fluctuations are included in the energy loss probability according to~\cite{Wicks:2005gt}:
\begin{eqnarray}
P(E_i\rightarrow E_f=E_i-\Delta_{rad}-\Delta_{coll})=\int dL \, P(L) \, P_{rad} (\Delta_{rad}; L) \otimes
P_{coll}(\Delta_{coll}; L) \ .
\end{eqnarray}
Here, $P(L)$ is the path length distribution extracted from Ref.~\cite{Dainese:2003wq} assumed to
be the same for all parton varieties, as it corresponds to a geometric quantity.

\subsection{Next-to-leading order calculation of heavy-flavor spectra in heavy-ion collisions}
\label{ssec_scet}

In the past several years new theoretical developments in the description of hard probes in 
heavy-ion collisions were enabled by the introduction of an effective theory of jet propagation 
in matter, the so-called Soft Collinear Effective Theory with Glauber Gluons, 
SCET$_\mathrm{G}$~\cite{Idilbi:2008vm,Ovanesyan:2011xy}.
The collinear in-medium splitting functions, the building blocks in  parton shower
formation~\cite{Ovanesyan:2011kn,Ovanesyan:2015dop}, were obtained to first order in 
opacity. This allows for a unified description of vacuum and medium-induced branching. 
Applications so far, beyond the traditional energy loss approach, have been limited to 
light hadrons~\cite{Kang:2014xsa}, jets~\cite{Chien:2015hda} and jet 
substructure~\cite{Chien:2015hda,Chien:2016led}.

An important step toward generalizing such a unified description to heavy flavor is to 
include quark masses into SCET$_\mathrm{G}$. The SCET$_\mathrm{M}$ Lagrangian with quark 
masses in the vacuum was obtained in Ref.~\cite{Leibovich:2003jd}. The introduction of HQ
masses requires a specific power counting, where $M_Q/p^+ \sim \lambda$ is of the order of 
the small power counting parameter in SCET. This is also consistent with the power counting 
for the dominant transverse-momentum component of the Glauber gluon exchange 
$\sim (\lambda^2, \lambda^2, \lambda)$. Hence, to lowest order, the new effective theory 
of HQ propagation in matter~\cite{Kang:2016ofv} is denoted as   
SCET$_\mathrm{M,G}=$SCET$_\mathrm{M}\otimes$SCET$_\mathrm{G}$.

The three splitting processes where the heavy quark mass plays a role are $Q\to Qg$, $Q\to gQ$ 
and $g\to Q\bar Q$.
Going beyond the energy loss limit of soft-gluon emission, a more careful consideration 
of parton splitting and deflection kinematics is necessary which was also achieved 
in Ref.~\cite{Kang:2016ofv}. The full expressions for the splitting functions, 
${\cal P}^{\mathrm{med}}_{i\to jk}(z,\mu)$, are lengthy and not reproduced here. 
However, we emphasize again the main idea of separating the perturbative splitting processes 
induced by Glauber gluon interactions from the medium, which themselves can be non-perturbative. 
As such, the  expressions derived in Ref.~\cite{Kang:2016ofv} are applicable for both the QGP 
and cold nuclear matter but one has to take into account the different transport properties 
of these strongly-interacting systems.

\begin{figure}[t]
\begin{center}
\includegraphics[width=0.45\textwidth]{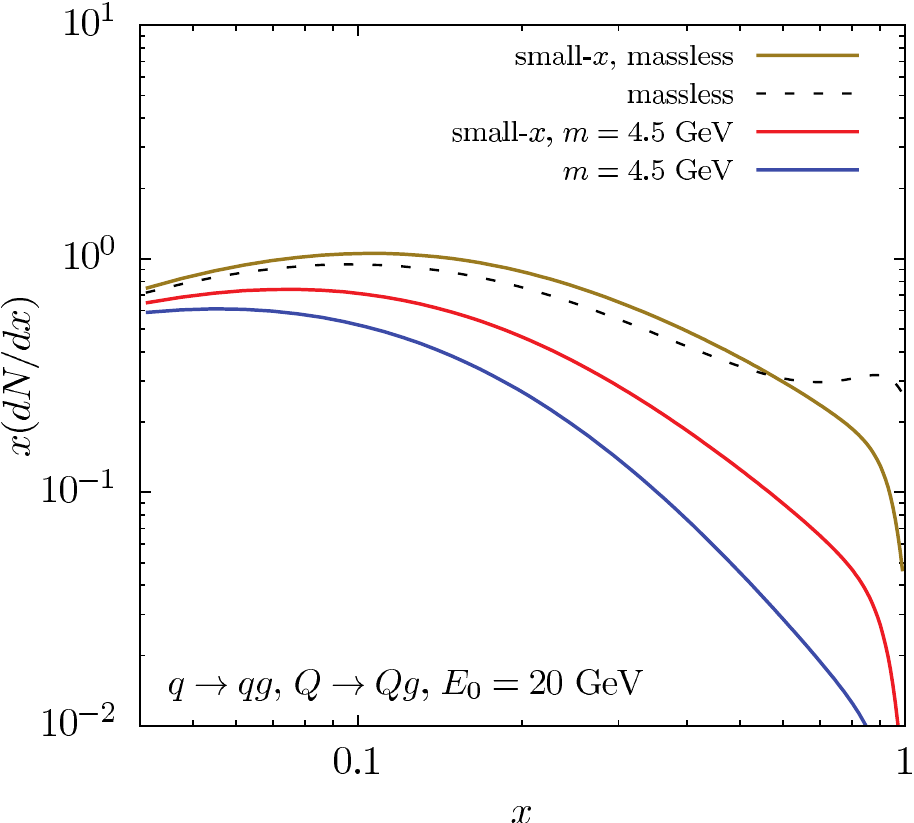} 
\hspace{0.8cm}
\includegraphics[width=0.45\textwidth]{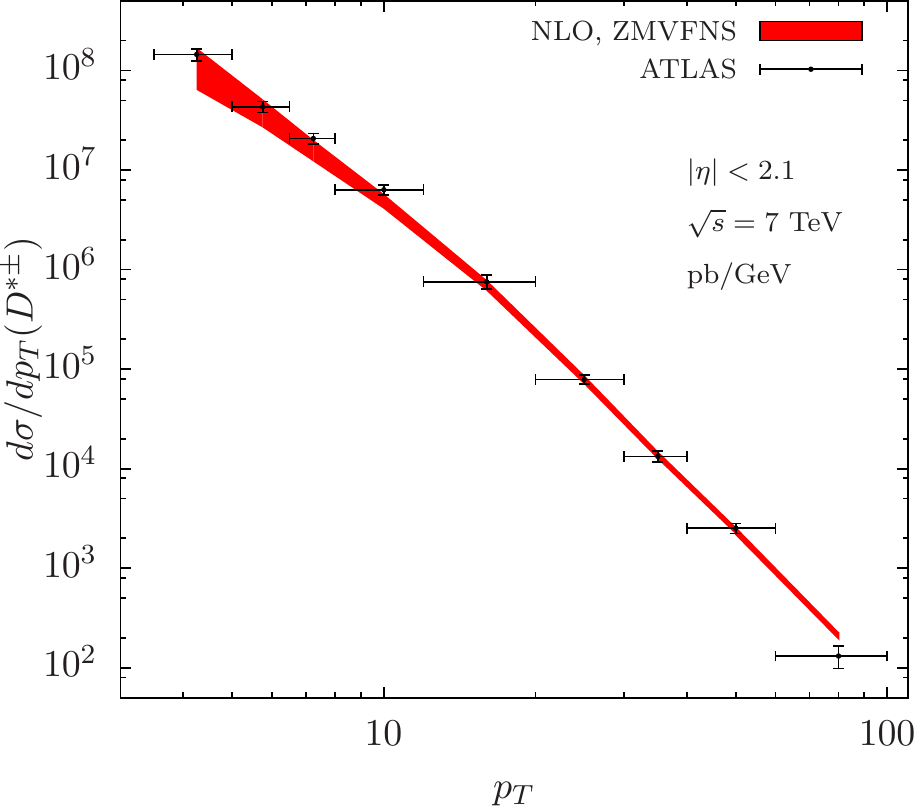} 
\end{center}
\caption{Left panel: comparison of the intensity spectra, $x(dN/dx)$, for the heavy-quark $\to$ 
quark splitting process. The massive results for the full splitting,  $Q\to Qg$, are in blue, the 
corresponding small-$x$ results are in red. We have chosen the mass $m_b=4.5$~GeV. We also plot 
the massless results $q\to qg$ for both the full splitting in dashed black and the small-$x$ 
limit in green. Right  panel: differential cross sections for $pp\to D^{*\pm}X$ at 
$\sqrt{s}$=7\,TeV. Data are from ATLAS~\cite{Aad:2015zix}.}
\label{fig:qq100}
\end{figure}

The soft-gluon emission limit, \ie, the limit when $x = k^+/p^+ \ll 1$, is the only limit where
a radiative energy loss interpretation of the general splitting processes described above can 
be given. It is easy to see that the $Q\to g Q$ and  $g\to Q\bar Q$ splittings are formally 
suppressed. Taking the small-$x$  limit in $Q \to  Q g $ yields 
\begin{eqnarray}
x \left(\frac{dN^{\mathrm{SGA}}}{dxd^2{\vc k}_\perp}\right)_{Q \rightarrow Qg } &=&  
\frac{\alpha_s}{\pi^2} C_F \int {d\Delta z}    \frac{1}{\lambda_g(z)} 
\int  d^2{\vc q}_\perp  \frac{1}{\sigma_{el}} \frac{d\sigma_{el}^{\rm med}}
{d^2 {\vc q}_\perp}   
\nonumber \\
&& \times \frac{2 \vc{k}_{\perp}  \cdot \vc{q}_{\perp}} 
{[{k}_{\perp}^2+x^2m^2][({\vc k}_{\perp}-{\vc q}_{\perp})^2 + x^2M_Q^2]}
 \left [ 1-\cos \frac{({\vc k}_{\perp}-{\vc q}_{\perp})^2+x^2M_Q^2}{xp^+_0} \Delta z \right], 
\nonumber \\
\label{smallx1}
\end{eqnarray} 
a much simpler result, see also Ref.~\cite{Djordjevic:2003zk}. The comparison of the full 
splitting kernels with the soft-gluon limit results and the comparison of massless and 
finite-mass partons is given in Fig.~\ref{fig:qq100}. We show results for splitting functions 
averaged over the binary-collision distributed jet production in central Pb+Pb collisions 
at $\sqrt{s_\mathrm{NN}}=5.02$ TeV at the LHC in a gluon-dominated plasma. Note the 
pronounced differences between the massless and massive cases. It is also important to 
observe that in the energy region where mass effects are most important the  differences 
between the full splitting functions and the soft-gluon approximation are large.  

Traditionally, energy loss calculations have focused on a scenario where only heavy quarks 
fragment into heavy mesons. This leads to simple arguments about mass and color charge 
ordering of light hadron, $D$-meson and $B$-meson suppression. The splitting functions above 
imply that both light partons and heavy quarks can fragment into heavy mesons. This will, of 
course, have implications for their quenching pattern. It is important to identify the regions 
where the uncertainty due to the possibly  different production mechanisms is minimal. 
In $pp$ collisions a good description of heavy-meson production can be achieved using the 
fragmentation functions of Refs.~\cite{Kneesch:2007ey,Kniehl:2009ar,Kniehl:2009mh,Kniehl:2012ti} 
in which light parton fragmentation into heavy mesons is included. The calculation combines 
the zero-mass variable-flavor number scheme (ZMVFNS)~\cite{Collins:1986mp,Stavreva:2009vi} 
and the $pp\to HX$ NLO framework~\cite{Aversa:1988vb,Jager:2002xm}, to obtain
\begin{equation}
\label{eq:sighadX}
\frac{d\sigma^{H}_{pp}}{dp_Td\eta}  =  
\frac{2 p_T}{s}\sum_{a,b,c}\int_{x_a^{\rm min}}^1\f{dx_a}{x_a}f_a(x_a,\mu)\int_{x_b^{\rm min}}^1\f{dx_b}{x_b} f_b(x_b,\mu)   \int^1_{z_c^{\rm min}} \frac{dz_c}{z_c^2}\frac{d\hat\sigma^c_{ab}(\hat s,\hat p_T,\hat \eta,\mu)}{dvdz}D_c^H(z_c,\mu).
\end{equation}
One example of $D$-meson production at the LHC is shown in the right panel of Fig.~\ref{fig:qq100}. 
It is important to note that the contributions of HQ and gluon fragmentation to $D$ mesons 
are approximately equal, and similarly for $B$ mesons. Recent global analysis of  $D^*$ production, 
which includes novel open heavy flavor-in-jet measurements favors even larger gluon fragmentation 
contribution~\cite{Anderle:2017cgl}. 

Going beyond the soft-gluon approximation requires a new treatment of the medium-induced 
parton shower. Incorporating this contribution consistent with NLO calculations can be 
schematically expressed as
\begin{equation}
\label{eq:AA}
d\sigma^H_{\mathrm{PbPb}} = d\sigma^{H, {\rm NLO}}_{pp} + d\sigma^{H, {\rm med}}_{\mathrm{PbPb}},
\end{equation}
where $d\sigma^{H, {\rm NLO}}_{pp}$ is the NLO cross section in the vacuum, and 
$d\sigma^{H,{\rm med}}_{\mathrm{PbPb}} =\hat \sigma^{(0)}_i \otimes D_i^{H,\mathrm{med}} $ 
is the one-loop medium correction. Using the medium-induced splitting functions, 
${\cal P}^{\mathrm{med}}_{i\to jk}(z,\mu)$, we find for the medium-modified quark and gluon 
fragmentation functions, $D_i^{H,\mathrm{med}}$,  
\begin{eqnarray}
\label{eq:Dmed}
D_q^{H,\mathrm{med}}(z,\mu)  = & \int_z^1\f{dz'}{z'}D_q^H\left(\f{z}{z'},\mu \right)
{\cal P}^{\mathrm{med}}_{q\to qg}(z',\mu)-D_q^H(z,\mu)\int_0^1dz' 
{\cal P}^{\mathrm{med}}_{q\to qg}(z',\mu) \nn 
\\
&+\int_z^1\f{dz'}{z'} D_g^H\left(\f{z}{z'},\mu\right) {\cal P}^{\mathrm{med}}_{q\to gq}(z',\mu) \, , 
\\
D_g^{H,\mathrm{med}}(z,\mu) = &  \int_z^1\f{dz'}{z'}D_g^H\left(\f{z}{z'},\mu \right)
{\cal P}^{\mathrm{med}}_{g\to gg}(z',\mu)-\f{D_g^H(z,\mu)}{2}\int_0^1dz' \left[  
{\cal P}^{\mathrm{med}}_{g\to gg}(z',\mu)\right.  
\nn \\
&\left. +2 N_f {\cal P}^{\mathrm{med}}_{g\to q\bar q}(z',\mu)\right]  +\int_z^1\f{dz'}{z'} 
\sum_{i=q,\bar q} D_i^H\left(\f{z}{z'},\mu\right){\cal P}^{\mathrm{med}}_{g\to q\bar q}(z',\mu) \, .
\end{eqnarray}

\begin {figure*}[t]
\begin{center}
\includegraphics[width=0.45\textwidth]{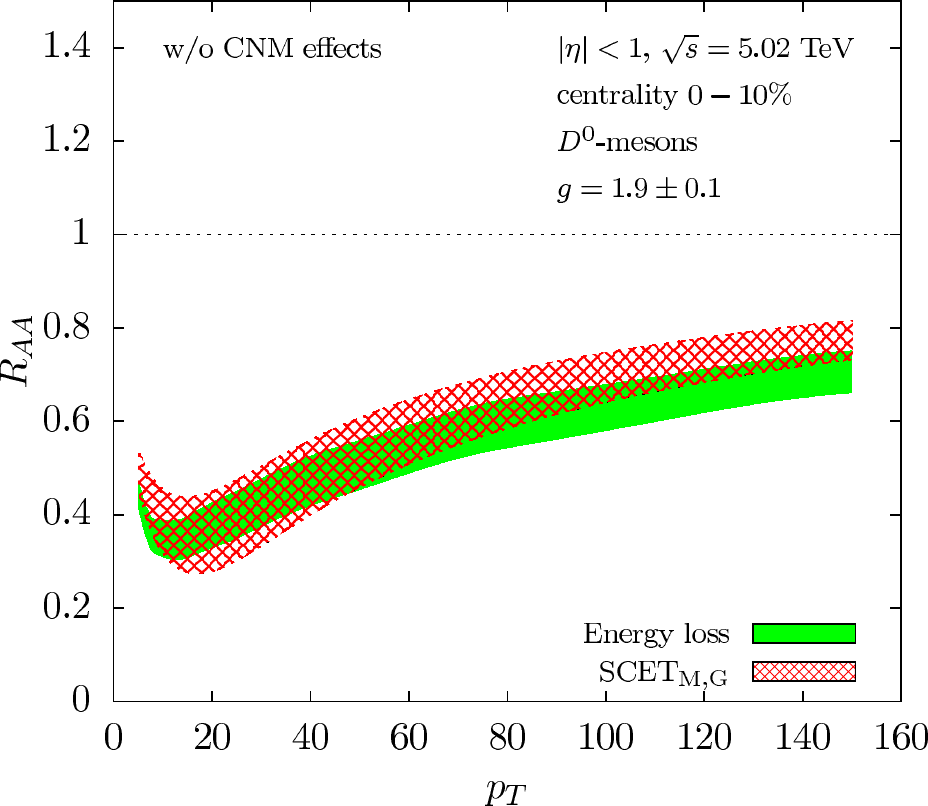} 
\hspace*{1cm}
\includegraphics[width=0.45\textwidth]{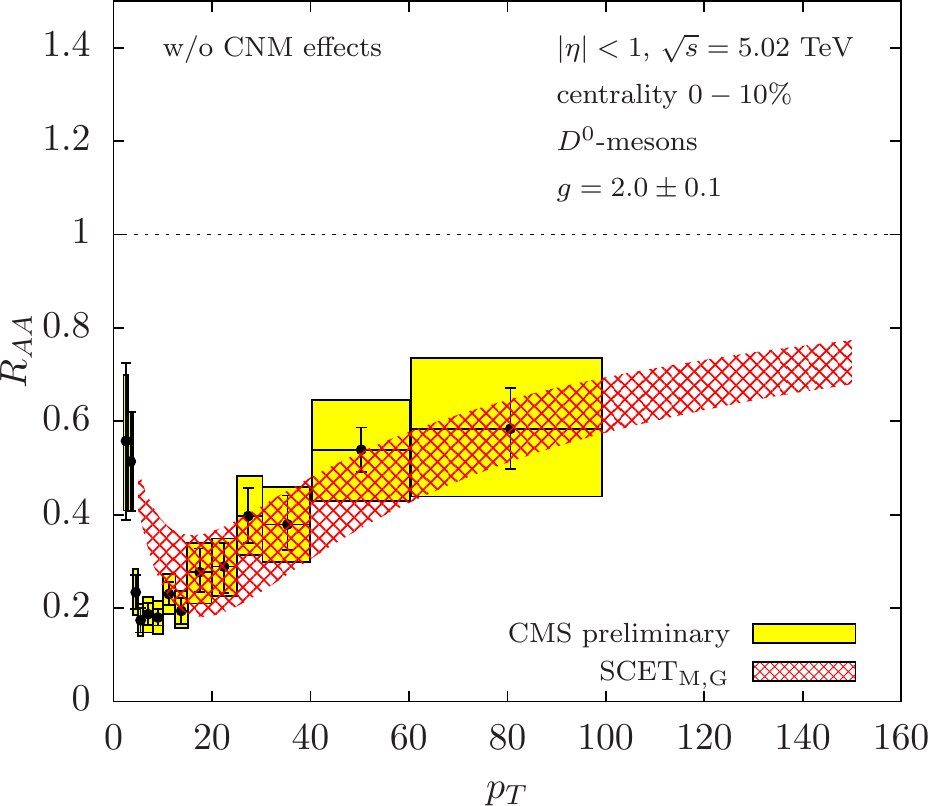} 
\end{center}
\caption{The nuclear modification factor, $R_{\rm AA}$, for $D^0$ mesons as a 
function of transverse momentum in 0-10\% central Pb+Pb(5.02\,TeV) collisions at the LHC. 
Left panel: comparison of the results obtained within the traditional approach to energy loss 
(green band) to those based on SCET$_{\mathrm{M,G}}$ (hatched red band); the bands reflect 
the range of the strong coupling constant, $g$=1.9$\pm$0.1. 
Right panel: a SCET$_{\mathrm{M,G}}$ calculation of $D^0$-meson suppression (with a slightly
readjusted coupling of $g$=2.0$\pm$0.1) compared to preliminary CMS data.}
\label{fig:RAA-D}
\end{figure*}

The suppression of heavy mesons that originate from gluon fragmentation can be considerably 
stronger than the suppression of heavy mesons that originate from heavy-quark fragmentation. 
The nuclear modification factors become equal only at very high $p_T$, where the larger 
``energy loss'' of gluons is offset by its softer fragmentation function. A practical way, 
however, of determining the region where the perturbative calculations can be used to probe 
the properties of the medium is to compare the $R_{\rm AA}(p_T)$ from  the energy loss and 
the full NLO calculation. Results are presented in Fig.~\ref{fig:RAA-D}. For $D$ mesons the 
results are fairly comparable within uncertainties; a comparison to recent CMS data in
0-10\% central Pb+Pb collisions at the LHC is shown in the right panel of  Fig.~\ref{fig:RAA-D}.  
For $B$ mesons there is significant deviation below $p_T \sim 20$~GeV. At those transverse 
momenta collisional energy loss~\cite{Wicks:2005gt}, described elsewhere in this document,
and/or heavy-meson dissociation~\cite{Adil:2006ra,Sharma:2009hn,Sharma:2012dy} is expected
to play a role.
However, it is important to realize that there is uncertainty in the absolute magnitude of the 
suppression based on medium-induced splitting/radiative processes that has not been discussed 
in the literature until very recently~\cite{Kang:2016ofv}. For the purpose of presenting 
results in this report we take the soft-gluon emission, heavy-quark energy loss limit. 
Theoretical model assumptions are listed in a separate section.

\vspace{0.5cm}

{\bf Theoretical model assumptions} \\
We describe below the theoretical model assumptions that go into the SCET$_{\mathrm{M,G}}$  
calculation of open heavy flavor. 
\begin{itemize}
\item  Most of the model dependence comes form the treatment of the background QGP medium
adopted here. 
Jet production, being rare such that $ \sigma(E_T > E_{T\; \min}) T_{AA}(b)  \ll 1$,
follows binary collision  scaling $ \sim d^2N_{\rm bin.}/d^2{\bf x}_\perp $.
In contrast, the medium is assumed to be distributed according to the participant number
density, $ \sim d^2N_{\rm part.}/d^2{\bf x}_\perp $.  We take into account longitudinal
Bjorken expansion. It was shown that transverse expansion does not affect the overall cross 
section suppression much, however it leads to a smaller high-$p_T$ elliptic flow $v_2$. This 
is the reason for which the elliptic flow at high $p_T$ in the numerical results section
turns out to be quite large.

\item  
We assume local thermal equilibrium and a gluon-dominated plasma. The medium formation time 
is taken to be $\tau_0  = 0.3$~fm at the LHC. The local density of the medium then reads  
 \begin{equation}
\rho =  \frac{1}{\tau}  \frac{d^2 ( dN^g/dy)}{d^2 {\bf x}_\perp}
\approx  \frac{1}{ \tau } \frac{3}{2}
\left| \frac{d\eta}{dy} \right| \frac{d^2 (dN_{\rm ch}/d\eta) }{d^2 {\bf x}_\perp}
\;  .
\label{ydep}
\end{equation}
Here,  $dN_{\rm ch}/d\eta = \kappa N_{\rm part}/ 2$ with $\kappa \approx 8.25 $
for Pb+Pb collisions at $\sqrt{s_{NN}}=2.76$~TeV at the LHC. The parameter $\kappa$ can be 
constrained by experimentally measured charged-particle rapidity density. Since a gluon-dominated
plasma has fewer degrees of freedom than a QGP, it is hotter at equal space-time points.
 
\item  
The temperature at any space-time position can be obtained from the density as
\begin{equation}
T(\tau, {\bf x}_{\perp}) = \  ^3\!\sqrt{ \pi^2 \rho(\tau,{\bf
x}_{\perp}) / 16 \zeta(3) }\;, \tau > \tau_0 \ .
\label{tempdet}
\end{equation}
The Debye screening scale is given by $m_D = gT$, recalling that we work
in the approximation of a gluon-dominated plasma, \ie, $N_f = 0$. The relevant
gluon mean free path is easily evaluated:  $\lambda_g = 1/ \sigma^{gg} \rho$
with $\sigma^{gg} = (9/2) \pi \alpha_s^2 / m_D^2$. 

\item 
We assume an effective fixed coupling $g$ between the jet and the medium. At present, 
there are no reliable results for the renormalization of the strong coupling in the presence 
of a medium.  Typical values at the LHC are in the range $g$=1.9-2.2, typical values at the 
RHIC are in the range $g$=2.0-2.3. The value of this coupling is adjusted for comparison 
to other calculations in the numerical section. For parton splitting processes an 
additional $\alpha_s$ associated with the splitting vertex occurs. In fixed-order calculations,
such as NLO~\cite{Kang:2016ofv,Kang:2017frl}, we evaluate $\alpha_s$ at the  hard scale, 
$Q^2$, in the process.
If multiple gluon emission in the energy loss limit~\cite{Vitev:2002pf} or full QCD evolution 
for parton showers~\cite{Kang:2014xsa,Chien:2015vja} are considered,   
$\alpha_s$ runs with the transverse mass of the emitted parton relative to the jet axis.

\end{itemize}

\subsection{Reference Results in an Infinite and Finite QGP}
\label{ssec_delE}
\begin{figure}[!t]
\begin{center}
\includegraphics[width=1.0\textwidth]{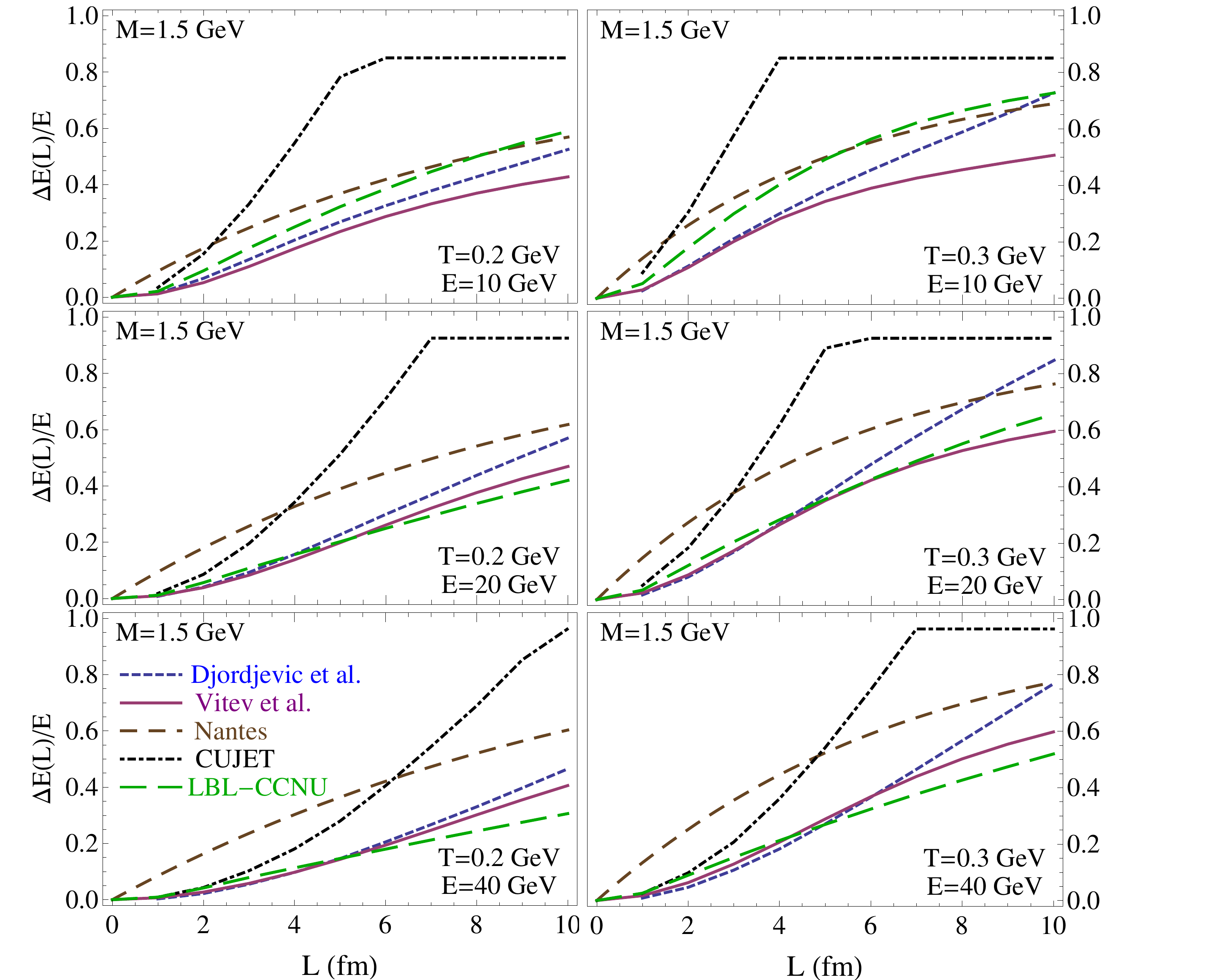}
\end{center}
\vspace{-0.2cm}
\caption{Fractional {\em radiative} energy loss, $\Delta E/E$, of $c$-quarks as a function
of their path length in a fixed-temperature QGP for various model calculations (SCET NLO 
energy loss limit [Vitev et al.], DGLV [Djordjevic et al.], higher-twist [LBL-CCNU], 
running-coupling elastic [Nantes] and internal-energy elastic [CUJET] formalisms), 
for two temperatures ($T$=0.2 and 0,3\,GeV 
in the left and right columns, respectively) and three initial quark energies ($E$=10, 
20, 40\,GeV in the upper, middle and lower panels, respectively).}
\label{fig:delE-c}
\end{figure}

\begin{figure}[!t]
\begin{center}
\includegraphics[width=1.0\textwidth]{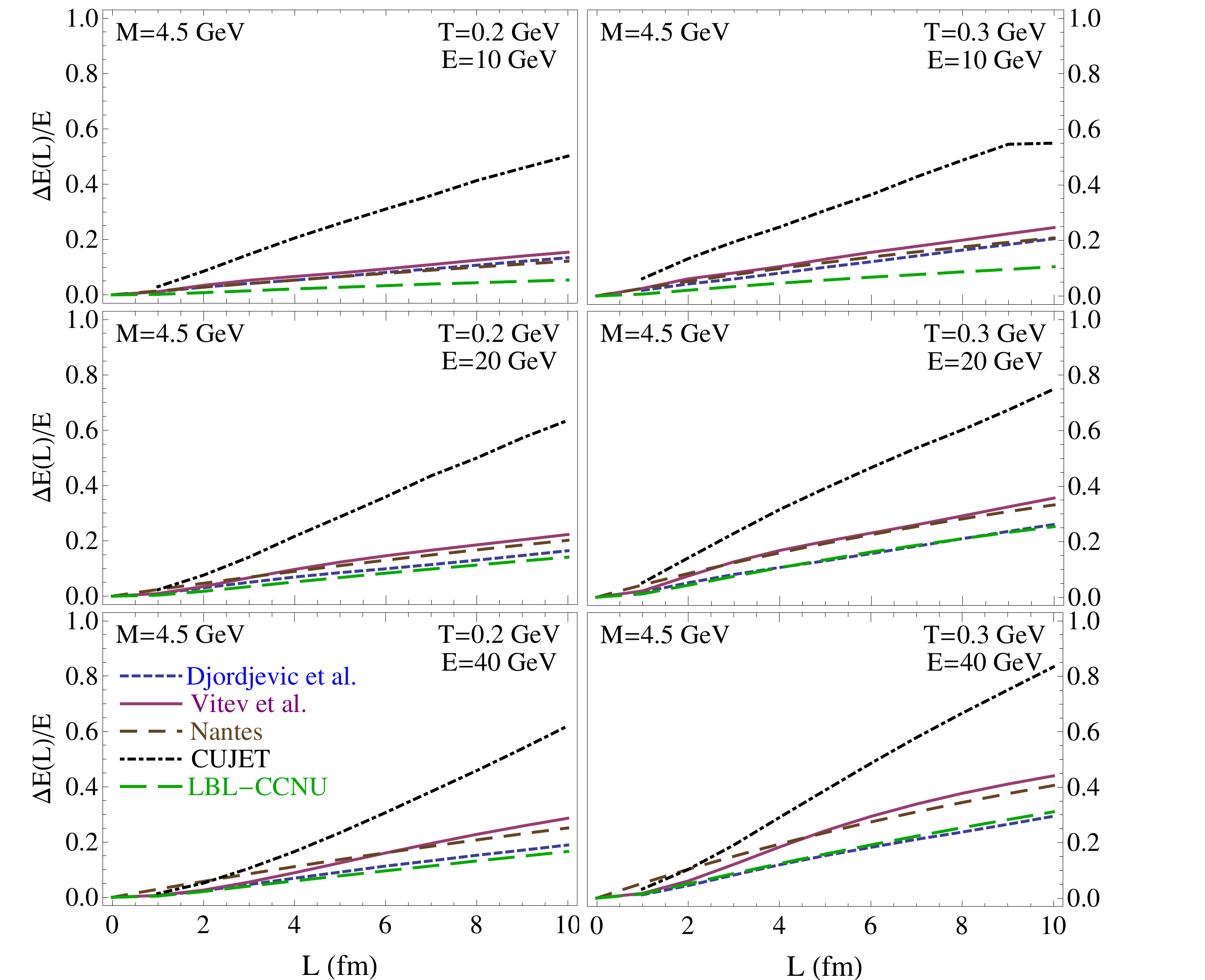}
\end{center}
\vspace{-0.2cm}
\caption{Same as Fig.~\ref{fig:delE-c} but for $b$-quarks.}
\label{fig:delE-b}
\end{figure}

In this section we provide reference results in a QGP at fixed temperature for radiative 
energy loss models for high-$p_t$ heavy quarks, as well as an application to the 
$D$-meson nuclear modification factor and elliptic flow for the models discussed in 
Secs.~\ref{ssec_dglv} (dynamical energy loss) and \ref{ssec_scet} (SCET)
which were not included in the comparisons using the pQCD*5 elastic interaction 
conducted in Secs.~\ref{sec_bulk} and \ref{sec_hadro}.

In Figs.~\ref{fig:delE-c} and \ref{fig:delE-b}, the relative energy loss, 
$\Delta E/E$, for $c$- and $b$-quarks, respectively,
is displayed as a function of the path length $L$ for four model calculations.
In practice, the most relevant range for HF phenomenology in heavy-ion collisions
turns out to be about $L\lesssim5$\,fm.
We recall that in the SCET NLO approach of Vitev et al.~(as described in Sec.~\ref{ssec_scet}), 
multi-gluon radiation is included with $\alpha_s=0.4$ resulting in 
$\hat{q}=2 m_D^2/\lambda_g= 0.36(1.20)$\,GeV$^2$/fm for $T=0.2(0.3)$\,GeV.
The Djordjevic et al.~calculations correspond to the radiative part evaluated within the 
DGLV formalism (as described in Sec.~\ref{ssec_dglv}), which also includes multi-gluon
emission, but differs, \eg, in the choice of the gluon propagators (HTL vs.~Debye-screened 
vacuum in SCET, which, however, in practice are quite similar) and the prescription for the 
cutoff in transverse momentum, and includes the Ter-Mikayelian effect. 
Nevertheless, the two approaches show good agreement for $c$-quark energy loss, with
some deviation developing only for relatively large path lengths, especially at higher 
temperature and lower $c$-quark energy, where the Vitev et al.~calculation tends to level 
off in a more pronounced way (the Djordjevic et al.~calculations for the energy loss, 
$\Delta E$, do not include multi-gluon emission fluctuations). 
For small path lengths, $L\lesssim$3-4\,fm (somewhat decreasing with temperature), we 
observe a hierarchy where the more energetic $c$-quarks lose a smaller fraction of 
their energy, while the effective path length dependence is $\propto L^\gamma$ with 
$\gamma>1$ (also found in the BDMPSZ approach not shown here). At later times, the exponent 
for the effective path length dependence reduces, first becoming linear ($\gamma$) and 
then turning over ($\gamma<1$) well before the expected saturation for values of 
$\Delta E/E$ close to unity is reached. The change in $\gamma$ is more pronounced for 
smaller initial parton energies.
In the LBL-CCNU approach~\cite{Cao:2016gvr,Cao:2017hhk} the number of emitted gluons per 
time step is computed from a radiation spectrum obtained from the higher-twist energy loss 
formalism. The radiation rate is proportional to $\alpha_s \hat{q}$ where $\hat{q}$ is 
based on pQCD elastic scattering~\cite{Cao:2016gvr}, with a running coupling constant 
at scale $Q^2$=$2ET$ for the HQ-gluon vertices and a constant $\alpha_s$=0.15 for the gluon's 
Debye mass and coupling to thermal partons (assumed to be massless). The resulting 
$\hat{q}$ amounts to 0.16(0.41)\,GeV$^2$/fm for $T$=0.2(0.3)\,GeV.  
Effects from both finite formation times and multi-gluon emission are accounted for.
The fractional energy loss of charm quarks obtained from this model turns out to be 
comparable to the other pQCD-based calculations displayed in Fig.~\ref{fig:delE-c},
with, however, different values for $\alpha_s$ (and $\hat{q}$).  
The Nantes calculation~\cite{Aichelin:2013mra,Gossiaux:2012cv} is characterized by
a running $\alpha_s$, reaching rather large values, and a reduced Debye mass in the 
elastic HQ scattering that triggers the radiation; it has been constructed to be 
primarily applicable at intermediate $p_t$ and therefore neglects finite-path length 
effects due to gluon formation outside the QGP. These two features are presumably
responsible for the significantly larger energy loss at small path lengths compared to
the other approaches. 
In the CUJET3 framework~\cite{Xu:2014tda,Xu:2015bbz,Shi:2017onf}, the medium contains 
non-perturbative chromo-magnetic degrees of freedom which interact strongly with a 
jet parton, leading to a potential akin to the heavy-quark internal energy. This
also includes a strong running of the coupling at nonperturbative energy scales. 
When the parton energy becomes small at large path lengths, $L\gtrsim$4\,fm, the 
radiative energy loss becomes about a factor of 2 larger than
in the Vitev et al.~and Djordjevic et al.~approaches. However, since such path lengths 
are often smaller than the typical ones in non-central heavy-ion collisions, this
difference may not have marked phenomenological consequences. 
A stronger-than-linear increase in the CUJET energy loss remains until saturation is 
reached when all kinetic energy of the original quark is radiated. Further work is 
required to better understand the relations between the different approaches and 
their results. 

In Fig.~\ref{fig:delE-b}, we display the corresponding fractional energy loss for $b$ 
quarks. As expected, it is substantially smaller than for $c$ quarks, especially at the 
lower energies ($E$=10 and 20\,GeV) due to the well-known dead-cone effect. This implies 
that collisional energy loss play a more important role than for $c$ quarks. In addition,
interference effects are much less pronounced as formation times are generally reduced
to due the larger quark mass. Together, these features create an overall much closer
to linear dependence of the energy loss on path length out to values of 10\,fm. The Nantes 
results are now in better agreement with the Vitev et al.~and Djordjevic et al.~calculations, 
while the non-perturbative interaction encoded in the internal-energy
potential used in CUJET still leads to larger energy loss by a factor of 2-3 (leading
to near saturation at $\Delta E = m_b$ around $L$=10\,fm for a 10\,GeV initial $b$ 
quark). Even for $b$-quark energies as large as 40\,GeV, the deviations from a linear 
behavior are relatively small.

\begin{figure}[!t]
\includegraphics[width=0.95\textwidth]{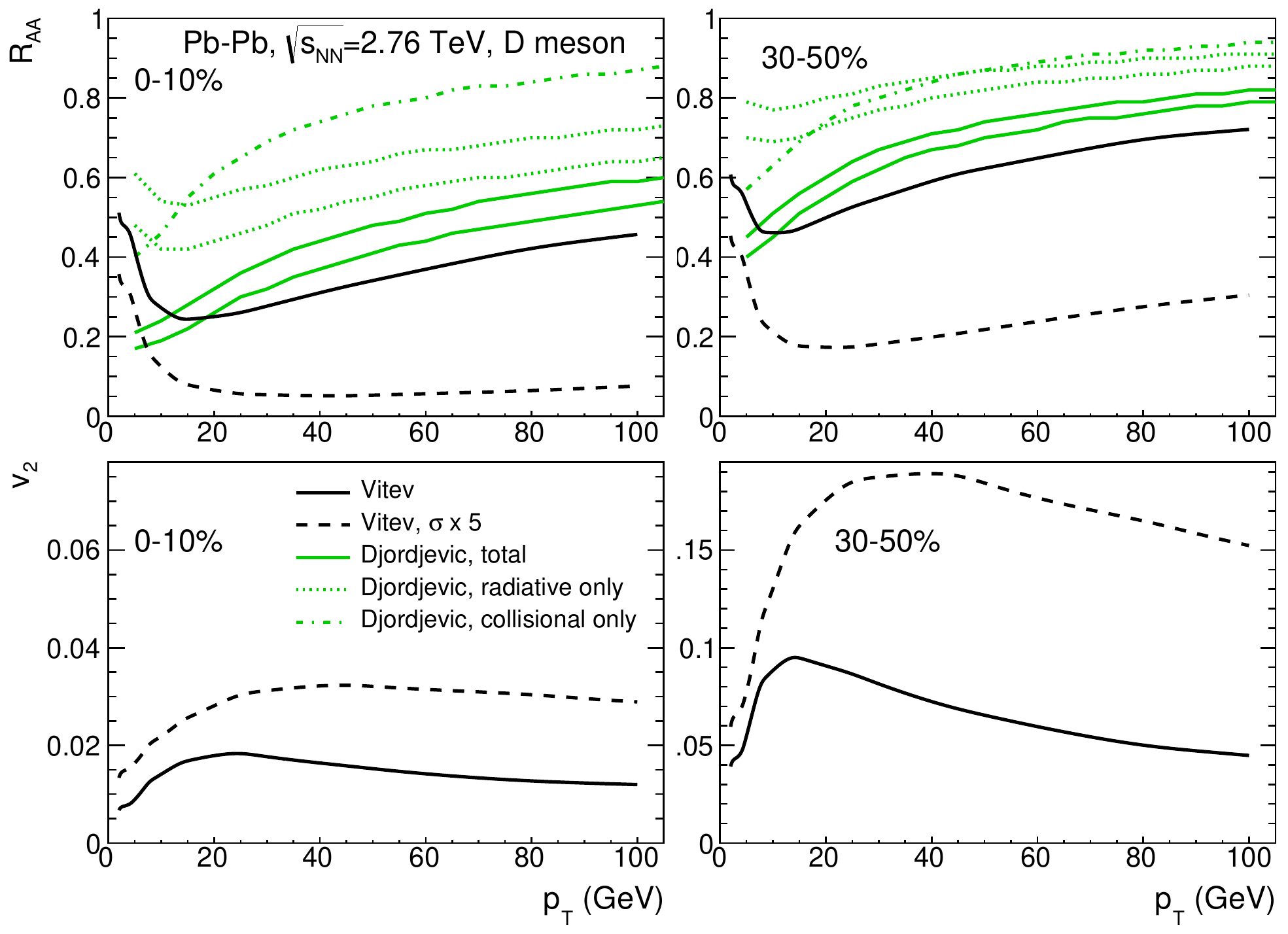}
\caption{Comparison of the nuclear modification factor (upper row) and elliptic flow
(lower row) of $D$-mesons in 0-10\% (left column) and 30-50\% (right column)
Pb-Pb(2.76\,TeV) collisions from two energy loss models (black lines: Vitev et al.,
green lines: Djordjevic et al.) within the QGP phase in these reactions. The theoretical
uncertainty band for the radiative-only (green dotted lines) and total (solid green
lines) results is due to a range of 0.4-0.6 in the ratio of magnetic to electric screening
masses.}
\label{fig:D_eloss}
\end{figure}

Finally, we illustrate for a few cases how the energy loss calculations compare at the 
level of the nuclear modification factor and elliptic flow in Pb-Pb collisions at the
LHC, cf.~Fig.~\ref{fig:D_eloss}. This requires the additional input of a bulk medium 
evolution model. For the Vitev et al. calculations, the QGP fireball has been modelled 
by 1-D Bjorken expansion with Glauber geometry in the transverse plane, while 
for the Djordjevic et al. calculations a static spherically symmetric fireball has been
employed. The baseline calculations in both approaches, including both radiative and
collisional energy loss for Djordjevic et al. and only radiative for Vitev et al., 
agree within 20\% for $p_T$$>10$\,GeV (and generally do a good job in describing 
experimental data at high  $p_T$). For the former, we also show results when only
accounting for either radiative or collisional energy loss. For semi-central collisions,
both contributions are comparable for most of the considered $p_T$ range out to 100\,GeV, 
while for central collision the radiative one becomes dominant for $p_T\gtrsim20$\,GeV.
This can be attributed to the stronger than linear rise in the radiative energy loss 
of charm quarks at high $p_T$, recall the lower panels in Fig.~\ref{fig:delE-c}.
On the other hand, the radiative-only result for the $R_{\rm AA}$ from the Djordjevic 
et al.  calculation shows a factor of $\sim$2 less suppression than the result from 
Vitev et al. which one would not have expected from the charm-quark energy loss
results displayed in Fig.~\ref{fig:delE-c}. This suggests a marked difference in the
underlying bulk evolution models. 
Finally, including a $K$-factor of 5 in the HQ scattering cross section in the Vitev et al. 
calculation leads to a much stronger suppression in the $R_{\rm AA}$ which is well beyond 
what is found in experimental data.

\section{Summary and Perspectives} 
\label{sec_sum}

The characterization of heavy-flavor diffusion in QCD matter remains one of the most
powerful approaches to investigate, both qualitatively and quantitatively, the properties 
of the medium created in high-energy heavy-ion collisions. On the one hand, this pertains 
to determining the temperature and momentum dependence of relevant transport coefficients 
(such as diffusion and energy-loss coefficients), but, maybe more importantly, 
gives the opportunity to unravel underlying microscopic processes 
which reveal the structure of the QCD medium in the strongly coupled regime.
Similarly to other probes, the progress of HF probes hinges on a close connection between 
theory, phenomenology and experimental data, further fueled by dedicated future 
plans~\cite{Abelevetal:2014cna,cms-hl} to improve and extend the current data set.    
The present report is a first attempt to systematically break down the HF probe of QCD 
matter into its main modeling components, with the ultimate goal of understanding and 
quantifying the uncertainties that each component imprints on the final extraction of 
the transport coefficients. These components are the initial heavy-quark 
spectra and their modifications due to nuclear shadowing, the bulk evolution of the
fireball medium, the microscopic description of heavy-quark transport in the QGP and 
through hadronization\footnote{In the present report we have neither explicitely 
addressed the impact of a pre-equilibrium evolution, bridging the (short) time between 
the initial production and the formation of a locally thermalized medium, nor diffusion 
through the hadronic phase.}. 
Another important objective has been the identification of baseline criteria and standard 
inputs that can be broadly agreed upon and channeled into future refinements of the majority 
of the transport approaches. 

As for the initial conditions, a best-fit of state-of-the-art $D$-meson spectra in $pp$ 
collisions at LHC energies has been carried out within the FONLL framework (and associated 
BCFY fragmentation functions), including a systematic error band. The comparison with initial 
$c$-quark $p_t$ spectra currently in use in 6 different approaches showed good agreement 
within this band (with the largest uncertainty at low $p_t\lesssim2$\,GeV). The 
implementation of shadowing is more uncertain, especially if more conservative error bands 
from recent EPPS16 nPDF fits are employed. This uncertainty is larger in the low-$p_t$ region, 
as verified in an explicit transport study for Pb-Pb collisions. The extraction of shadowing 
effects from $p$A data is further complicated by the possible occurence of final-state 
interactions of HF particles.

To test the role of different bulk evolution models, the participating research groups 
delivered the $p_T$ spectra and $v_2$ of direct pions and protons in central and semicentral 
Pb-Pb collisions at $\sqrt{s_{\rm NN}} = 2.76$~TeV at the end of the QGP phase in their 
hydrodynamic or transport model, to benchmark the environment for testing $c$-quark 
diffusion. For the integrated direct-pion yields at $T_{\rm c}$ a rather large spread of 
about $\pm$25\% for central collisions was found within the hydrodynamic models (further
augmented by transport models and in semicentral collisions); the situation improved (with
a couple of outliers) when comparing inclusive spectra (\ie, including resonance feeddown), 
presumably because inclusive pion numbers are closer to observables that the models are 
tuned to in the first place.
Remaining discrepancies should be resolvable in a next iteration through closer inspection 
of, \eg, the hadro-chemistry and more uniform choices of the centrality classes and chemical 
freezeout temperature. The pion and proton radial and 
elliptic ($v_2$) flow at $T_{\rm c}$ exhibited better agreement. Within their evolution 
models the groups carried out charm-quark transport calculations using a {\em common} 
(predefined) $c$-quark interaction with QGP partons (pQCD elastic Born scattering with a 
$K$ factor of 5). Except for 2 outliers, an encouraging degree of agreement of the $c$-quark 
spectra and $v_2$ emerged, with an extracted ``systematic" error of 10-15\%. This 
suggests that the diagnosed spread in the bulk evolution models is in reality smaller 
(presumably because the evolution models are, in principle calibrated to final-state 
hadron spectra), and/or that the results for the $c$-quark observables are 
more robust than light-hadron spectra against details of the medium evolution.         

Concerning the hadronization of heavy quarks at the end of the QGP phase, all approaches 
feature some type of recombination with constituent quarks from the surrounding medium, 
supplemented with independent fragmentation for quarks that are not recombining. 
The employed mechanisms include instantaneous coalescence (both local and global in 
coordinate space), in-medium fragmentation, or heavy-light resonance formation. 
The ensuing spread in the resulting $D$-meson $R_{\rm AA}$ and $v_2$ is appreciably 
increased over the one found at the charm-quark level. We have quantified this by 
introducing a new quantity, $H_{\rm AA}$, the ratio of $D$-meson to $c$-quark spectra 
right after and before hadronization, respectively. The treatment of hadronization has 
therefore been identified as a prime area of future improvements. Quantitative criteria
will have to be applied to benchmark the various approaches, \eg, the compatability
with the equilibrium limit (both chemical and thermal) in the conversion from heavy 
quarks to hadrons.   

We have then performed a model average of $D$-meson $R_{\rm AA}$ and $v_2$ for semi-/central
Pb-Pb collisions at $\sqrt{s_{\rm NN}} = 2.76$~TeV which, without further attempts of 
narrowing down uncertainties, resulted in encouragingly moderate error bands. 
Not withstanding remaining caveats (such as neglecting pre-equilibrium effects and hadronic 
transport), an initial comparison to existing data showed that the pQCD*5 interaction does 
not provide enough interaction strength by an appreciable margin, 
implying that the HF diffusion coefficient, as a measure of low-momentum transport through 
the QCD medium, must be well below ${\cal D}_s(2\pi T)$=6, at least for some temperature 
range (preferentially where the $v_2$ of the fireball is large).

Let us briefly comment on the role of hadronic diffusion, which has not been explicitly
addressed in the present effort. Using different versions of effective hadronic
lagrangians, the interactions of $D$-mesons with light and strange hadrons have been
ultilized to evaluate $D$-meson relaxation time in hot hadronic
matter~\cite{Laine:2011is,He:2011yi,Ghosh:2011bw,Abreu:2011ic,Torres-Rincon:2014ffa,Ozvenchuk:2014rpa}.
After initially rather widely varying results, there is now emerging consensus that the
(scaled) hadronic diffusion coefficient becomes rather small near $T_{\rm pc}$, reaching
down to near 5 or less. This suggests a minimum structure, as well as a possible continuity
with the values for charm quarks in the QGP (as discussed above), in the vicinity of
$T_{\rm pc}$. It also implies that hadronic-diffusion effects in URHICs are quantitatively
significant. Current estimates of the hadronic contribution to the observed $D$-meson
$v_2$ are in the range of 10-40\%~\cite{He:2012df,Song:2015sfa} (also depending on $p_T$)
relative to the QGP contribution, while the $R_{\rm AA}$ tends to be much less affected.
These contributions thus have to be accounted for in future precision extractions of
heavy-flavor transport coefficients from URHIC data.

We have also discussed the microscopic description of heavy-quark diffusion from several
angles. We have emphasized that
the convergence of the perturbative series for the diffusion coefficient is ill-behaved 
even at coupling constants as small as $\alpha_s$=0.1; a key role in this behavior
is played by the fact that the leading contribution already carries a rather large
power in the coupling constant, ${\cal D}_s \sim {\cal O}(1/\alpha_s^2)$. 
Thus, nonperturbative methods are indispensable to develop a credible microscopic 
description of heavy-quark diffusion in the QGP. Due to its
sensitivity to the coupling strength, one may argue
that the heavy-quark diffusion coefficient provides one of the most direct windows
on the nonperturbative many-body physics of the strongly-coupled QGP. Constraints
from lattice QCD, together with nonperturbative many-body methods, will be 
necessary to exploit this opportunity. An increasing interaction strength toward 
$T_{\rm c}$ suggests that the onset of hadronization, expected to be a gradual
process, plays an important role in the interactions of heavy quarks in this
regime, as can be implemented, for example, in a thermodynamic $T$-matrix approach. 
Lattice-QCD calculations of charm-quark susceptibilities, indicative for
an onset of hadronic degrees of freedom above $T_{\rm c}$, support such a picture.
Some care has to be taken in implementing the transport of heavy quarks in heavy-ion
collisions. When the temperature becomes large, the charm-quark mass may no longer be large 
enough to satisfy the parametric hierarchy, $m_Q/T\gg1$, which will ultimately limit the 
ability of the Fokker-Planck/Langevin treatment of charm-quark diffusion at the precision 
frontier. The Boltzmann approach does not require such an approximation. On the other hand, 
the Boltzmann approach will run into issues when the collision rates become so large that 
the medium partons cannot be reliably modeled by an ensemble of (quasiclassical) quasiparticels 
any more. In his case, the Langevin approach is still viable as long as the heavy quarks 
remain good quasiparticles (even if the medium partons are not).  Bottom quarks, due to
their larger mass, thus provide the largest margin for a theoretically accurate 
implementation of heavy-flavor transport and the extraction of the diffusion coefficient. 
  
While the main focus of the working group activities was on low-momentum interactions 
where incoherent elastic collisions dominate, we have also discussed heavy-quark 
interactions at high $p_T$ and radiative energy loss. We have reviewed the definition 
of the usual transport coefficient, $\hat{q}$, along with subtleties in its evaluation 
and discussed recent progress by applying effective theory to perform next-to-leading-order 
calculations. Also in the high-$p_T$ regime, the medium evolution is identified as a 
significant source of uncertainty in current modeling efforts. We have compared results 
within 4 different approaches to energy loss and pertinent manifestations in $R_{\rm AA}$ 
and $v_2$ observables, illustrating the relative role of radiative and elastic contributions. 
Inspection of the results in a fixed-temperature QGP revealed that the energy loss for 
bottom quarks remains essentially linear in the path length, \ie, incoherent, up to energies 
of 20\,GeV, while for charm quarks its magnitude and nonlinearities are significantly more 
pronounced.  Quantitative differences in the magnitude of the radiative energy loss not only 
emerge from different perturbative treatments but also show appreciable sensitivity to
nonperturbative effects of a strongly coupled QGP, both of which deserve further scrutiny.

\vspace{0.5cm}

Based on the insights in this task-force report, and in the interest of a collective
progress in the physics of HF probes of QCD matter, we suggest the following set of 
recommendations for future modeling efforts in heavy-ion collisions:  
\begin{itemize}
\item[1.] 
Adopt FONLL baseline HQ spectra with EPS09 shadowing for the initial conditions in
transport simulations.
\item[2.] 
Employ publicly available hydrodynamic or transport evolution models which have been 
tuned to data, with a maximal range of viable initial conditions and model parameters;
or even a single one with a pre-specified tune as a single point of contact of all 
approaches.
\item[3.] 
Use recombination schemes of heavy quarks with light medium partons which satisfy
4-momentum conservation and recover equilibrium distributions in the long-time limit 
for the resulting hadron distributions.
\item[4.] 
Incorporate nonperturbative interactions in the modeling of heavy-flavor transport
in a QGP at moderate temperatures as established and constrained by information from
lattice QCD; utilize resummed interactions leading to bound-state formation near 
$T_{\rm c}$ to facilitate a seamless transition into coalescence processes.
\item[5.]
Include diffusion through the hadronic phase of heavy-ion collisions.
\end{itemize}

\vspace{0.3cm}

Consequently, to address the question  which particular future measurements could have the 
largest impact on improving our knowledge about the in-medium interactions of heavy flavor,  
we suggest the following observables with associated objectives: 
\begin{itemize}
\item[A.]
Bottom observables as the theoretically cleanest probe of a strongly-coupled QGP, in terms
of the implementation of both microscopic interactions and transport, and as a measure of 
coupling strength without saturation due to thermalization; 
\item[B.]
$v_2$ peak structures and maximal values for $D$ and $B$ mesons to gauge the heavy-flavor
interaction strength and delineate elastic and radiative regimes;
\item[C.]
Precision $R_{\rm AA}$ and $v_2$ of $D$ and $B$ mesons at various beam energies to extract 
temperature and mass dependence of transport coefficients;
\item[D.]
$D_s$ and $\Lambda_c$ hadron observables at low and intermediate $p_T$ to unravel the in-medium 
charm-quark chemistry, specifically its role in hadronization processes and reach in $p_T$; 
\item[E.]
Heavy-flavor (especially bottom) in jets to disentangle gluon vs. heavy-flavor energy loss
and production mechanisms (direct vs. gluon splitting). 
\item[F.] Correlation measurements of heavy-flavor pairs to delineate collisional from 
radiative interactions and test Langevin against Boltzmann transport approaches.
\end{itemize}

\vspace{0.5cm}

Whereas the effort presented here merely constitutes a first step toward a truly systematic 
and broad investigation of heavy-flavor probes of QCD matter, we believe that we have gained 
insights and identified criteria that will prove useful in the future and help to match the
experimental precision of upcoming measurements with a robust theoretical understanding 
and quantitative phenomenology. Concerted theory collaborations will play a critical 
role in achieving this goal.

\vspace{1cm}

{\bf Acknowledgments}\\
We gratefully acknowledge support of the Heavy-Flavor Rapid Reaction Task Force (RRTF)
meetings through the Extreme Matter Institute (EMMI).
This work has been supported by the US National Science Foundation (NSF) under grant 
Nos.~PHY-1614484 (SL, RR), PHY-1352368 (JL, SS) and ACI-1550228 (JETSCAPE) (SC, XNW),
by the US Department of Energy (DOE) under grant Nos.~DE-SC0012704 (PP), DE-AC02-05CH11231 
(SC, XNW), DE-SC0013460 (SC) and the Early Career Program (IV), 
by R\'egion Pays de la Loire (France) under contract no.~2015-08473 (PBG, MN),
by NSF China (NSFC) under grant Nos.~11675079 (MH) and 11221504 (XNW), 
by the European Research Council (ERC) StG under grant No.~259684 (SKD, VG, SP),
by the LANL LDRD program under grant Nos. 2012LANL7033 and 20170073DR (IV), 
by the German Academic Exchange Service (DAAD) (TS), by the 
Deutsche Forschungsgemeinschaft (DFG) under grant Nos.~CRC-TR 211 (EB, HvH, OK, GM, TS)
and SFB-1225 (ISOQUANT) (PBM, SM, JMP, JS),
and by the Major State Basic Research Development Program in China under grant 
No.~2014CB845404 (XNW).

\begin{appendix} 

\section{Overview of Model Approaches Employed in this Work}
\label{app_models}
In Table~\ref{tab_models} we list the acronyms of the model approaches that were involved 
in the various studies reported in this paper, including elastic and radiative interactions 
(both pQCD and non-perturbative) of heavy quarks in the QGP, hadronization mechanisms and 
a wide variety of bulk evolution models (hydrodynamic and transport) for the expanding fireball 
in heavy-ion collisions to make contact with observables. 

\begin{table}[!h]
	\begin{center}
	\begin{tabular}{|l|l|}
		\hline
		Name of the Approach  & References \\
		\hline 
		Catania & \cite{Scardina:2017ipo,Das:2015ana,Das:2013kea}\\
		CUJET & \cite{Xu:2015bbz,Xu:2014tda,Shi:2017onf}\\
	    Djordjevic et al. & \cite{Djordjevic:2009cr,Djordjevic:2013xoa,Djordjevic:2011dd}\\
	    Duke & \cite{Cao:2011et,Cao:2013ita}\\
	    LBL-CCNU & \cite{Cao:2016gvr,Cao:2017hhk} \\
		Nantes (MC@sHQ+EPOS2)& \cite{Gossiaux:2009mk,Aichelin:2013mra,Nahrgang:2013saa} \\
		POWLANG & \cite{Beraudo:2014boa,Beraudo:2015wsd} \\
        PHSD & \cite{Bratkovskaya:2011wp,Song:2015sfa,Song:2015ykw} \\
		SCET & \cite{Kang:2016ofv,Chien:2015hda,Ovanesyan:2011xy} \\	
		TAMU & \cite{Ravagli:2007xx,Riek:2010fk,He:2011zx}\\
		URQMD & \cite{Lang:2012yf,Lang:2012cx} \\
		\hline
	\end{tabular}
\end{center}
\caption{Models of open heavy-flavor ``transport" in hot QCD matter applied to ultra-relativistic 
heavy-ion collisions which participated in the studies reported in the present work, with up to 3 
most pertinent publications describing the approach.}
\label{tab_models}
\end{table}

\end{appendix}

\providecommand{\href}[2]{#2}\begingroup\raggedright\endgroup

\end{document}